\documentclass{aa}

\usepackage{supertabular}
\usepackage{longtable}
\usepackage{txfonts}
\usepackage{natbib}
\usepackage{graphicx}

\newcommand{\ms}{\mbox{m s$^{-1}~$}}

\def\gsim{~\rlap{$>$}{\lower 1.0ex\hbox{$\sim$}}}
\def\lsim{~\rlap{$<$}{\lower 1.0ex\hbox{$\sim$}}}

\def\wpm2{W m$^{-2}$}
\def\eg{{\it e.g.~}}

\def\ie{{\it i.e.\thinspace}}

\def\wpmsq{{W m$^{-2}$}}

\begin{document}
\bibliographystyle{aa}



\title{A dynamically-packed planetary system around GJ
667C with three super-Earths in its habitable zone
\thanks{Based on data obtained from the ESO Science Archive
Facility under request number ANGLADA36104. Such data
had been previously obtained with the HARPS instrument
on the ESO 3.6 m telescope under the programs
183.C-0437, 072.C-0488 and 088.C-0662, and with the UVES spectrograph
at the Very Large Telescopes under the program 087.D-0069.
This study also contains observations obtained at the W.M.
Keck Observatory- which is operated jointly by the
University of California and the California Institute
of Technology- and observations obtained with the
Magellan Telescopes, operated by the Carnegie Institution,
Harvard University, University of Michigan, University of Arizona,
and the Massachusetts Institute of Technology.}
\thanks{
Time-series are available in electronic format at CDS
via anonymous ftp to cdsarc.u-strasbg.fr (130.79.128.5) or via
\texttt{http://cdsweb.u-strabg.fr/cgi-bin/qcat?J/A+A/}
}
}

\subtitle{}

\author{
Guillem Anglada-Escud\'e\inst{1}
\and
Mikko Tuomi\inst{2,3}
\and
Enrico Gerlach \inst{4}
\and
Rory Barnes \inst{5}
\and
Ren\'e Heller \inst{6}
\and
James S. Jenkins \inst{7}
\and
Sebastian Wende \inst{1}
\and
Steven S. Vogt \inst{8}
\and
R. Paul Butler \inst{9}
\and
Ansgar Reiners \inst{1}
\and
Hugh R. A. Jones \inst{2}
}

\offprints{G. Anglada-Escud\'e \email{guillem.anglada@gmail.com}-
Mikko Tuomi, \email{miptuom@utu.fi}}

\institute{
       Universit\"{a}t G\"{o}ttingen,
       Institut f\"ur Astrophysik,
       Friedrich-Hund-Platz 1,
       37077 G\"{o}ttingen, Germany
  \and Centre for Astrophysics, University of Hertfordshire,
       College Lane, Hatfield,
       Hertfordshire AL10 9AB, UK
  \and University of Turku,
       Tuorla Observatory,
       Department of Physics and Astronomy,
       V\"ais\"al\"antie 20, FI-21500, Piikki\"o, Finland
 \and  Technical University of Dresden,
       Institute for Planetary Geodesy,
       Lohrmann-Observatory,
       01062 Dresden, Germany
 \and Astronomy Department, University of Washington,
       Box 951580,
       WA 98195, Seattle, USA
 \and Leibniz Institute for Astrophysics Potsdam (AIP),
       An der Sternwarte 16,
       14482 Potsdam,
       Germany
 \and Departamento de Astronom\'ia, Universidad de Chile,
       Camino El Observatorio 1515, Las Condes,
       Casilla 36-D Santiago, Chile
 \and UCO/Lick Observatory,
       University of California,
       Santa Cruz, CA 95064, USA
 \and Carnegie Institution of Washington,
       Department of Terrestrial
       Magnetism, 5241 Broad Branch Rd. NW,
       20015 Washington D.C., USA
}

\date{submitted Feb 2013}

\begin{abstract}
{
Since low-mass stars have low luminosities,
orbits at which liquid water can exist on Earth-sized
planets are relatively close-in, which produces Doppler signals that
are detectable using state-of-the-art Doppler spectroscopy.
} {
GJ 667C is already known to be orbited by two
super-Earth candidates. We have applied recently developed data analysis
methods to investigate whether the data supports the presence of additional
companions.
} {
We obtain new Doppler measurements from HARPS extracted spectra and
combined them with those obtained from the PFS and HIRES spectrographs.
We used Bayesian and periodogram-based methods to re-assess the number of
candidates and evaluated the confidence of each detection. Among other
tests, we validated the planet candidates by analyzing correlations of
each Doppler signal with measurements of several activity indices and
investigated the possible quasi-periodic nature of signals.
} {
Doppler measurements of GJ~667C are described better
by six (even seven) Keplerian-like signals: the two known candidates
(b and c); three additional few-Earth mass candidates with
periods of 92, 62 and 39 days (d, e and f); a cold
super-Earth in a 260-day orbit (g) and tantalizing evidence
of a $\sim$ 1 M$_\oplus$ object in a close-in orbit of 17 days (h).
We explore whether long-term stable orbits are compatible with
the data by integrating 8$\times$10$^4$
solutions derived from the Bayesian samplings. We
assess their stability using secular frequency analysis.
} {
The system consisting of six planets is compatible with
dynamically stable configurations. As for the solar
system,  the most stable solutions do not contain
mean-motion resonances and are described well by analytic
Laplace-Lagrange solutions. Preliminary analysis also
indicates that masses of the planets cannot be higher than
twice the minimum masses obtained from Doppler
measurements. The presence of a seventh planet (h) is
supported by the fact that it appears squarely centered on
the only island of stability left in the six-planet
solution. Habitability assessments accounting for the
stellar flux, as well as tidal dissipation effects, indicate
that three (maybe four) planets are potentially
habitable. Doppler and space-based transit
surveys indicate that 1) dynamically packed systems of
super-Earths are relatively abundant and 2) M-dwarfs have
more small planets than earlier-type stars. These two
trends together suggest that GJ 667C is one of the first
members of an emerging population of M-stars with multiple
low-mass planets in their habitable zones.
}

\end{abstract}

\keywords{
Techniques : radial velocities --
Methods : data analysis --
Planets and satellites : dynamical evolution and stability --
Astrobiology --
Stars: individual : GJ~667C
}

\titlerunning{Three HZ super-Earths in a seven-planet system}
\maketitle


\section{Introduction}

Since the discovery of the first planets around
other stars, Doppler precision has been steadily
increasing to the point where objects as small as a
few Earth masses can currently be detected around
nearby stars. Of special importance to the exoplanet
searches are low-mass stars (or M-dwarfs) nearest to
the Sun. Since low-mass stars are intrinsically
faint, the orbits at which a rocky planet could
sustain liquid water on its surface \citep[the so-called
habitable zone, ][]{Kasting93} are typically closer
to the star, increasing their Doppler signatures even more.
For this reason, the first super-Earth mass candidates
in the habitable zones of nearby stars have been
detected around M-dwarfs \citep[e.g. GJ 581, ][]{mayor:2009,
vogt:2010}).

Concerning the exoplanet population detected to date, it is
becoming clear that objects between 2 M$_\oplus$ and the mass of
Neptune (also called super-Earths) are very common around all G,
K, and M dwarfs. Moreover, such planets tend to appear in
close in/packed systems around G and K dwarfs (historically
preferred targets for Doppler and transit surveys) with orbits
closer in than the orbit of Mercury around our Sun. These
features combined with a habitable zone closer to the star, point
to the existence of a vast population of habitable worlds in
multiplanet systems around M-dwarfs, especially around old/metal-depleted
stars \citep{jenkins:2012}.

GJ~667C has been reported to host two (possibly three)
super-Earths. GJ~667Cb is a hot super-Earth mass object in an
orbit of 7.2~days and was first announced by
\citet{bonfils:2009:conf}. The second companion has an orbital
period of 28 days, a minimum mass of about 4.5 M$_\oplus$ and,
in principle, orbits well within the liquid water habitable
zone of the star \citep{anglada:2012b, delfosse:2012}. The
third candidate was considered tentative at the time owing to a
strong periodic signal identified in two activity indices. This
third candidate (GJ~667Cd) would have an orbital period between
74 and 105 days and a minimum mass of about 7 M$_\oplus$.
Although there was tentative evidence for more periodic signals
in the data, the data analysis methods used by both
\citet{anglada:2012b} and \citet{delfosse:2012} studies were
not adequate to properly deal with such high multiplicity
planet detection. Recently, \citet{gregory:2012}
presented a Bayesian analysis of the data in
\citet{delfosse:2012} and concluded that several additional
periodic signals were likely present. The proposed solution,
however, contained candidates with overlapping orbits and no
check against activity or dynamics was done, casting serious
doubts on the interpretation of the signals as planet
candidates.

Efficient/confident detection of small amplitude signals requires
more specialized techniques than those necessary to detect single
ones. This was made explicitly obvious in, for example, the
re-analysis of public HARPS data on the M0V star GJ~676A. In
addition to the two signals of gas giant planets reported by
\citet{gj676A}, \citet{gj676A:2012} (AT12 hereafter) identified
the presence of two very significant low-amplitude signals in
closer-in orbits. One of the main conclusions of AT12 was that
correlations of signals already included in the model prevent
detection of additional low-amplitude using techniques based on
the analysis of the residuals only. In AT12, it was also pointed
out that the two additional candidates (GJ~676A d and e) could be
confidently detected with 30\% less measurements using Bayesian
based methods.

In this work, we assess the number of Keplerian-like signals
around GJ~667C using the same analysis methods as in
\citet{gj676A:2012}. The basic data consists of 1) new Doppler
measurements obtained with the HARPS-TERRA software on public
HARPS spectra of GJ~667C \citep[see ][ for a more detailed
description of the dataset]{delfosse:2012}, and 2) Doppler
measurements from PFS/Magellan and HIRES/Keck spectrometers
\citep[available in][]{anglada:2012a}. We give an overview of GJ~667C
as a star and provide updated parameters in Section
\ref{sec:starparam}. The observations and data-products used in
later analyses are described in Section \ref{sec:observations}.
Section \ref{sec:models} describes our statistical tools, models
and the criteria used to quantify the significance of each
detection (Bayesian evidence ratios and log--likelihood
periodograms). The sequence and confidences of the signals in
the Doppler data are given in section \ref{sec:detection} where
up to seven planet-like signals are spotted in the data. To
promote Doppler signals to planets, such signals must be
validated against possible correlations with stellar activity.
In section \ref{sec:activity}, we discuss the impact of stellar
activity on the significance of the signals (especially on the
GJ~667Cd candidate) and we conclude that none of the seven
candidates is likely to be spurious. In section
\ref{sec:subsamples}, we investigate if all signals were
detectable in subsets of the HARPS dataset to rule out spurious
detections from quasi-periodic variability caused by stellar
activity cycles. We find that all signals except the least
significant one are robustly present in both the first and
second-halves of the HARPS observing campaign independently. A
dynamical analysis of the Bayesian posterior samples finds that
a subset of the allowed solutions leads to long-term stable
orbits. This verification steps allows us promoting the first
six signals to  the status of planet candidates. In Section
\ref{sec:dynamics} we also investigate possible mean-motion
resonances (MMR) and mechanisms that guarantee long-term
stability of the system. Given that the proposed system seems
physically viable, we discusses potential habitability of each
candidate in the context of up-to-date climatic models, possible
formation history, and the effect of tides in Section
\ref{sec:habitability}. Concluding remarks and a summary are
given in Section \ref{sec:conclusions}. The appendices describe
additional tests performed on the data to double-check the
significance of the planet candidates.

\section{Properties of GJ~667C} \label{sec:starparam}

\object{GJ~667C} (HR~6426~C), has been classified as an
M1.5V star \citep{Geballe:2002} and is a member of a triple
system, since it is a common proper motion companion
to the K3V+K5V binary pair, GJ~667AB. Assuming the HIPPARCOS
distance to the GJ~667AB binary \citep[$\sim$  6.8
pc][]{HIPPARCOS}, the projected separation between GJ~667C
and GJ~667AB is $\sim$ 230 AU. Spectroscopic measurements 
of the binary have revealed a metallicity significantly 
lower than the Sun \citep[Fe/H =-0.59$\pm$0.10][]{strobel:1981}. The galactic kinematics
of GJ~667 are compatible with both thin and thick disk
populations and there is no clear match to any known
moving group or stream \citep{anglada:2012b}.
Spectrocopic studies of the GJ~667AB pair
\citep{strobel:1981} show that they are on 
the main sequence, indicating an age between 2 and 10~Gyr.
Following the simple models in
\citet{reiners_mohanty:2012}, the low activity and the
estimate of the rotation period of GJ~667C (P$>$ 100~days,
see Section \ref{sec:activity}) also support an age of
$>$ 2~Gyr. In conclusion, while the age of the GJ~667
system is uncertain, all analyses indicate that the system
is old.

We performed a spectroscopic analysis of GJ~667C using
high resolution spectra obtained with the UVES/VLT
spectrograph (program 87.D-0069). Both the HARPS and the
UVES spectra show no $H_{\alpha}$ emission. The value of
the mean S-index measurement (based on the intensity of the
Ca{\sc II} H+K emission lines) is $0.48\pm 0.02$, which puts
the star among the most inactive objects in the HARPS
M-dwarf sample \citep{bonfils:2011}. By comparison, GJ~581(S=$0.45$)
and GJ~876 (S=0.82) are RV-stable enough to
detect multiple low-mass planets around them, while
slightly more active stars like GJ~176 (S=$1.4$), are
stable enough to detect at least one low-mass companion.
Very active and rapidly rotating M-dwarfs, such GJ~388 (AD~Leo)
or GJ~803 (AU~Mic), have S-indices as high as
3.7 and 7.8, respectively. A low activity level allows
one to use a large number of atomic and molecular lines
for the spectral fitting without accounting for magnetic
and/or rotational broadening. UVES observations of
GJ~667C were taken in service mode in three exposures
during the night on August 4th 2011. The high resolution
mode with a slit width of $0.3''$ was used to achieve a
resolving power of $R \sim 100\,000$. The observations
cover a wavelength range from $640$\,nm to $1020$\,nm on
the two red CCDs of UVES.

The spectral extraction and reduction were done using the
ESOREX pipeline for UVES. The wavelength solution is
based, to first order, on the Th-Ar calibration provided
by ESO. All orders were corrected for the blaze function
and also normalized to unity continuum level. Afterwards,
all orders were merged together. For overlapping orders
the redder ends were used due to their better quality. In
a last step, an interactive removal of bad pixels and cosmic
ray hits was performed.

\begin{table}
  \caption{Parameter space covered by the grid of synthetic models.}
  \label{tab:paramspace}
  \centering                                      %
  \begin{tabular}{rcc}
    \hline \hline
		      & Range & Step size \\ \hline
    $T_{\rm eff}$ [K] & 2,300 -- 5,900   & 100 \\
    $\log(g)$         &   0.0 -- +6.0     & 0.5 \\
    $[Fe/H]$          &  -4.0 -- -2.0    & 1.0 \\
		      &  -2.0 -- +1.0    & 0.5 \\
    \hline \hline
  \end{tabular}
\end{table}

The adjustment consists of matching the observed
spectrum to a grid of synthetic spectra from the newest
PHOENIX/ACES grid \citep[see][]{Husser:2013}). The updated
codes use a new equation of state that accounts for the
molecular formation rates at low temperatures. Hence, it
is optimally suited for simulation of spectra of cool
stars. The 1D models are computed in plane parallel
geometry and consist of $64$ layers. Convection is
treated in mixing-length geometry and from the
convective velocity a micro-turbulence velocity
\citep{edmunds:1978} is deduced via $v_{mic}=0.5\cdot
v_{conv}$. The latter is used in the generation of the
synthetic high resolution spectra. An overview of the
model grid parameters is shown in
Table~\ref{tab:paramspace}. Local thermal equilibrium is
assumed in all models.

\begin{figure*} \center
\includegraphics[width=0.8\textwidth,clip]{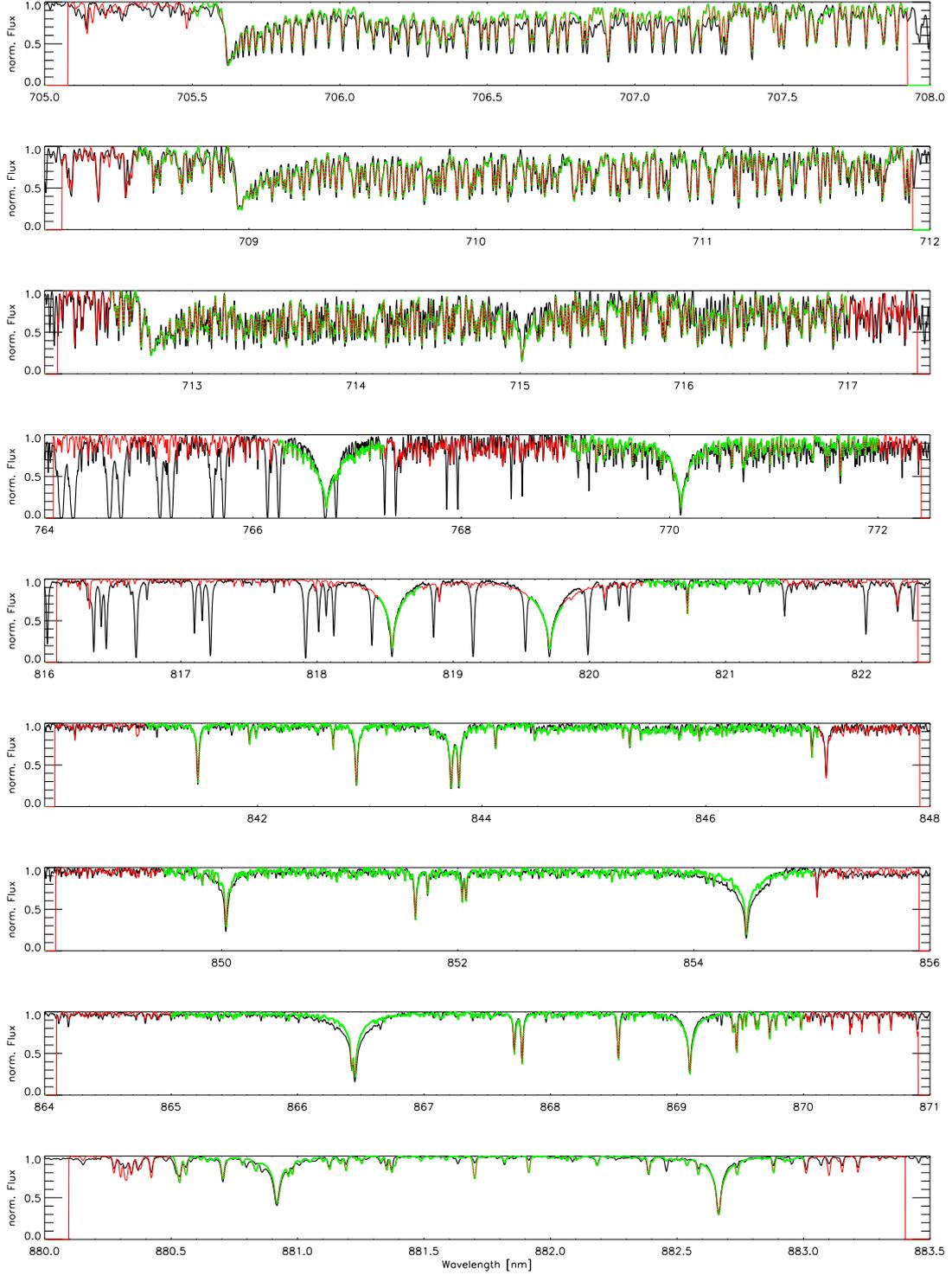}

\caption{Snapshots of the wavelength regions used in the
spectral fit to the UVES spectrum of GJ~667C. The
observed spectrum is represented in black, the green
curves are the parts of the synthetic spectrum used in
the fit. The red lines are also from the synthetic
spectrum that were not used to avoid contamination by
telluric features or because they did not contain
relevant spectroscopic information. Unfitted deep sharp
lines- especially on panels four and five from the top of the
page- are non-removed telluric features.}

\label{fig:GJ667Cmodel}

\end{figure*}

First comparisons of these models with observations show
that the quality of computed spectra is greatly improved
in comparison to older versions. The problem in previous
versions of the PHOENIX models was that observed spectra
in the $\epsilon$- and $\gamma$-TiO bands could not be
reproduced assuming the same effective temperature
parameter \citep{reiners:2005}. The introduction of the
new equation of state apparently resolved this problem.
The new models can consistently reproduce both TiO
absorption bands together with large parts of the visual
spectrum at very high fidelity (see
Fig.~\ref{fig:GJ667Cmodel}).

As for the observed spectra, the models in our grid are
also normalized to the local continuum.  The regions
selected for the fit were chosen as unaffected by
telluric contamination and are marked in green in
Fig.~\ref{fig:GJ667Cmodel}. The molecular TiO bands in
the region between $705$\,nm to $718$\,nm
($\epsilon$-TiO) and $840$\,nm to $848$\,nm
($\gamma$-TiO) are very sensitive to $T_{\rm{eff}}$ but
almost insensitive to $\log{g}$. The alkali lines in the
regions between $764$\,nm to $772$\,nm and $816$\,nm to
$822$\,nm (K- and Na-atomic lines, respectively) are
sensitive to $\log{g}$ and $T_{\rm{eff}}$. All regions
are sensitive to metallicity. The simultaneous fit of
all the regions to all three parameters breaks the
degeneracies, leading to a unique solution of effective
temperature, surface gravity and metallicity.

As the first step, a three dimensional $\chi^2$-map is
produced to determine start values for the fitting
algorithm. Since the model grid for the $\chi^2$-map is
discrete, the real global minimum is likely to lie
between grid points. We use the parameter sets of the
three smallest $\chi^2$-values as starting points for the
adjustment procedure. We use the IDL
\emph{curvefit}-function as the fitting algorithm. Since this
function requires continuous parameters, we use three
dimensional interpolation in the model spectra. As a
fourth free parameter, we allow the resolution of the
spectra to vary in order to account for possible
additional broadening (astrophysical or instrumental).
For this star, the relative broadening is always found to
be $<3\%$ of the assumed resolution of UVES, and is
statistically indistinguishable from 0. More information
on the method and first results on a more representative
sample of stars will be given in a forthcoming
publication.


As already mentioned, the distance to the GJ 667 system
comes from the HIPPARCOS parallax of the GJ 667AB pair
and is rather uncertain (see Table \ref{tab:basic}). This,
combined with the luminosity-mass calibrations in
\citet{delfosse:2000}, propagates into a rather uncertain
mass (0.33$\pm$0.02 $M_\odot$) and luminosity estimates
($0.0137\pm 0.0009 \,L_\odot$) for GJ~667C
\citep{anglada:2012b}. A good trigonometric parallax
measurement and the direct measurement of the size of
GJ~667C using interferometry
\citep[e.g.~][]{vonbraun:2011} are mostly needed to
refine its fundamental parameters. The updated values of
the spectroscopic parameters are slightly changed from
previous estimates. For example, the effective
temperature used in \citet{anglada:2012b} was based on
evolutionary models using the stellar mass as the input
which, in turn, is derived from the rather uncertain
parallax measurement of the GJ~667 system. If the
spectral type were to be understood as a temperature
scale, the star should be classified as an M3V-M4V
instead of the M1.5V type assigned in previous works
\citep[e.g.~][]{Geballe:2002}. This mismatch is a well
known effect on low metallicity M dwarfs (less absorption
in the optical makes them appear of earlier type than solar
metallicity stars with the same effective temperature).
The spectroscopically derived parameters and other basic
properties collected from the literature are listed in
Table \ref{tab:basic}.

\begin{table}
\caption{Stellar parameters of GJ~667C}\label{tab:basic}
\center
\begin{tabular}{lrr}
\hline \hline
Parameters              & Value & Ref.               \\
\hline                                       & \\
R.A.                         &  17 18 57.16  & 1\\
Dec                          & -34 59 23.14  & 1\\
$\mu_{R.A.}$ [mas yr$^{-1}$] &  1129.7(9.7)  & 1\\
$\mu_{Dec.}$ [mas yr$^{-1}$] &  -77.0(4.6)  & 1\\
Parallax [mas]               &   146.3(9.0)  & 1\\
Hel. RV [k\ms]               &     6.5(1.0)  & 2\\
V [mag]                      &    10.22(10)  & 3\\
J [mag]                      &    6.848(21)  & 4\\
H [mag]                      &    6.322(44)  & 4\\
K [mag]                      &    6.036(20)  & 4\\
$T_{eff}$[K]                 &     3350(50)  & 5\\
${\rm[Fe/H]}$                &    -0.55(10)  & 5\\
log $g$ [g in c\ms]          &     4.69(10)  & 5\\
\\
Derived quantities		 \\
\hline				 \\
UVW$_{\rm LSR}$ [k\ms] & (19.5, 29.4,-27.2) & 2 \\
Age estimate	   & $>2$ Gyr	 &  5 \\
Mass [M$_\odot$]  & 0.33(2)   &  5 \\
$L_*$/L$_\odot$   & 0.0137(9) &  2 \\
\hline \hline
\end{tabular}
\tablebib{
(1)~\citet{HIPPARCOS}; (2)\citet{anglada:2012b};
(3)~\citet{mermilliod:1986}; (4)~\citet{twomass};
(5) This work (see text)
}
\end{table}

\section{Observations and Doppler measurements}
\label{sec:observations}

A total of 173 spectra obtained using the HARPS spectrograph \citep{pepe:2002}
have been re-analyzed using the HARPS-TERRA software \citep{anglada:2012a}.
HARPS-TERRA implements a least-squares template matching algorithm to obtain the
final Doppler measurement. This method and is especially well suited to deal
with the highly blended spectra of low mass stars. It only replaces the last
step of a complex spectral reduction procedure as implemented by the HARPS Data
Reduction Software (DRS). Such extraction is automatically done by the HARPS-ESO
services and includes all the necessary steps from 2D extraction of the echelle
orders, flat and dark corrections, and accurate wavelength calibration using
several hundreds of Th Ar lines accross the HARPS wavelength range
\citep{lovis:2007}. Most of the spectra (171) were extracted from the ESO
archives and have been obtained by other groups over the years \citep[e.g.,
][]{bonfils:2011, delfosse:2012} covering from June 2004 to June 2010. To
increase the time baseline and constrain long period trends, two additional
HARPS observations were obtained between March 5th and 8th of 2012. In addition
to this, three activity indices were also extracted from the HARPS spectra.
These are: the S-index (proportional to the chromospheric emission of the star),
the full-width-at-half-maximum of the mean line profile (or FWHM, a measure of
the width of the mean stellar line) and the line bisector (or BIS, a measure of
asymmetry of the mean stellar line). Both the FWHM and BIS are measured by the
HARPS-DRS and were taken from the headers of the corresponding files. All these
quantities might correlate with spurious Doppler offsets caused by stellar
activity. In this sense, any Doppler signal with a periodicity compatible with
any of these signals will be considered suspicious and will require a more
detailed analysis. The choice of these indices is not arbitrary. Each of them is
thought to be related to an underlying physical process that can cause spurious
Doppler offsets. For example, S-index variability typically maps the presence of
active regions on the stellar surface and variability of the stellar magnetic
field (e.g., solar-like cycles). The line bisector and FWHM should have the same
period as spurious Doppler signals induced by spots corotating with the star
(contrast effects combined with stellar rotation, suppression of convection due
to magnetic fields and/or Zeeman splitting in magnetic spots). Some physical
processes induce spurious signals at some particular spectral regions (e.g.,
spots should cause stronger offsets at blue wavelengths). The Doppler signature
of a planet candidate is constant over all wavelengths and, therefore, a signal
that is only present at some wavelengths cannot be considered a credible
candidate. This feature will be explored below to validate the reality of some
of the proposed signals. A more comprehensive description of each index and
their general behavior in response to stellar activity can be found elsewhere
\citep{baliunas:1995, lovis:2011}. In addition to the data products derived from
HARPS observations, we also include 23 Doppler measurements obtained using the
PFS/Magellan spectrograph between June 2011 and October 2011 using the Iodine
cell technique, and 22 HIRES/Keck  Doppler measurements \citep[both RV sets are
provided  in][]{anglada:2012a} that have lower precision but allow extending the
time baseline of the observations. The HARPS-DRS also produces Doppler
measurements using the so--called cross correlation method (or CCF). In the
Appendices, we show that the CCF-Doppler measurements actually contain the same
seven signals providing indirect confirmation and lending further confidence to
the detections.

\section{Statistical and physical models} \label{sec:models}

The basic model of a radial velocity data set from a single
telescope-instrument combination is a sum of $k$ Keplerian
signals, with $k$ = 0, 1, ..., a random variable describing
the instrument noise, and another describing all the
excess noise in the data. The latter noise term,
sometimes referred to as stellar RV jitter
\citep{ford:2005}, includes the noise originating from
the stellar surface due to inhomogeneities, activity-related
phenomena, and can also include instrumental
systematic effects. Following \citet{tuomi:2011}, we
model these noise components as Gaussian random variables
with zero mean and variances of $\sigma^2_i$ and
$\sigma^{2}_{l}$, where the former is the formal
uncertainty in each measurement and the latter is the
jitter that is treated as a free parameter of the model
(one for each instrument $l$). 

Since radial velocity
variations have to be calculated with respect to some
reference velocity, we use a parameter $\gamma_{l}$ that
describes this reference velocity with respect to the
data mean of a given instrument. For several
telescope/instrument combinations, the Keplerian signals
must necessarily be the same but the parameters
$\gamma_{l}$ (reference velocity) and $\sigma^{2}_{l}$
(jitter) cannot be expected to have the same values.
Finally, the model also includes a linear trend
$\dot{\gamma}$ to account for very long period companions
(e.g., the acceleration caused by the nearby GJ 667AB
binary). This model does not include mutual interactions
between planets, which are known to be significant in
some cases \citep[e.g. GJ 876, ][]{LaughlinChambers01}.
In this case, the relatively low masses of the companions
combined with the relatively short time-span of the
observations makes these effects too small to have
noticeable impact on the measured orbits. Long-term
dynamical stability information is incorporated and
discussed later (see Section
\ref{sec:dynamics}). Explicitly, the baseline model for
the RV observations is

\begin{equation}\label{eq:model}
v_{l}(t_i) =
\gamma_{l} +
\dot{\gamma}\, (t_i-t_0) +
\sum_{j=1}^{k} f(t_i, \vec{\beta}_{j}) +
g_l\left[\vec{\psi}; t_{i}, z_{i}, t_{i-1}, r_{i-1}\right]\, ,
\end{equation}

\noindent where $t_0$ is some reference epoch (which we
arbitrarily choose as t$_0$=2450000 JD), $g$ is a function
describing the specific noise properties (instrumental and
stellar) of the $l$-th instrument on
top of the estimated Gaussian uncertainties. We model
this function using first order moving average (MA) terms
\cite{hd40307,tauceti} that and on the residual
r$_{i-1}$ to the previous measurement at
$t_{i-1}$, and using linear correlation terms with
activity indices (denoted as $z_{i}$). This component 
of the model is typically parameterized using one or more 
``nuisance parameters'' $\vec{\psi}$ that are also treated as free
parameters of the model. Function $f$ represents the
Doppler model of a planet candidate with parameters
$\vec{\beta}_{j}$ (Period $P_j$, Doppler semi-amplitude
$K_j$, mean anomaly at reference epoch $M_{0,j}$,
eccentricity $e_j$, and argument of the periastron
$\omega_j$).

The Gaussian white noise component of each measurement
and the Gaussian jitter component associated to each
instrument enter the model in the definition of the
likelihood function $L$ as

\begin{equation}\label{eq:likelihood}
  L(m | \vec{\theta}) =
  \prod_{i=1}^{N}
  \frac{1}{\sqrt{2\pi (\sigma_{i}^{2} + \sigma_{l}^{2})}}
  \exp \Bigg\{ \frac{-\big[m_{i} - v_{l}(t_{i})\big]^{2}}{2(\sigma_{i}^{2} + \sigma_{l}^{2})} \Bigg\} ,
\end{equation}

\noindent where $m$ stands for \textit{data} and $N$ is
the number of individual measurements. With these
definitions, the posterior probability density
$\pi(\vec{\theta} | m)$ of parameters $\vec{\theta}$
given the data $m$ (\vec{\theta} includes the orbital
elements $\vec{\beta}_{j}$, the slope term
$\dot{\gamma}$, the instrument dependent constant
offsets $\gamma_{l}$, the instrument dependent jitter
terms $\sigma_{l}$, and a number of nuisance parameters
$\vec{\psi}$), is derived from the Bayes' theorem as

\begin{equation}\label{eq:posterior}
  \pi(\vec{\theta} | m) = \frac{L(m | \vec{\theta}) \pi(\vec{\theta})}{\int L(m | \vec{\theta}) \pi(\vec{\theta}) d \vec{\theta}} .
\end{equation}

\noindent This equation is where the prior information
enters the model through the choice of the prior density
functions $\pi(\vec{\theta})$. This way, the posterior
density $\pi(\vec{\theta} | m)$ combines the new
information provided by the new data $m$ with our prior
assumptions for the parameters. In a Bayesian sense,
finding the most favored model and allowed confidence
intervals consists of the identification and exploration
of the higher probability regions of the posterior
density. Unless the model of the observations is very
simple (e.g., linear models), a closed form of
$\pi(\vec{\theta} | m)$ cannot be derived analytically
and numerical methods are needed to explore its
properties. The description of the adopted numerical
methods are the topic of the next subsection.

\subsection{Posterior samplings and Bayesian detection criteria}
\label{sec:bayesdetection}

Given a model with $k$ Keplerian signals, we draw
statistically representative samples from the
posterior density of the model parameters (Eq.~\ref{eq:posterior})
using the adaptive Metropolis
algorithm \cite{haario:2001}. This algorithm has been
used successfully in e.g. \cite{tuomi:2011},
\cite{tuomi:2011b} and \cite{gj676A:2012}. The
algorithm appears to be a well suited to the fitting
of Doppler data in terms of its relatively fast
convergence -- even when the posterior is not unimodal
\citep{tuomi:2012} -- and it provides samples that
represent well the posterior densities. We use these
samples to locate the regions of maximum \textit{a
posteriori} probability in the parameter space and to
estimate each parameter confidence interval allowed by
the data. We describe the parameter densities briefly
by using the maximum \emph{a posteriori} probability (MAP)
estimates as the most probable values, i.e. our
preferred solution, and by calculating the 99\%
Bayesian credibility sets (BCSs) surrounding these
estimates. Because of the caveats of point estimates
(e.g., inability to describe the shapes of posterior
densities in cases of multimodality and/or
non-negligible skewness), we also plot marginalized
distributions of the parameters that are more
important from a detection and characterization point
of view, namely, velocity semi-amplitudes $K_j$, and
eccentricities $e_j$.

The availability of samples from the posterior densities
of our statistical models also enables us to take
advantage of the signal detection criteria given in
\cite{tuomi:2012}. To claim that any signal is
significant, we require that 1) its period is
well-constrained from above and below, 2) its RV
amplitude has a density that differs from zero
significantly (excluded from the 99\% credibility
intervals), and 3) the posterior probability of the
model containing $k+1$ signals must be (at least) 150
times greater than that of the model containing only
{\it k} signals.

The threshold of 150 on condition (3) might seem arbitrary, 
and although posterior probabilities also have associated uncertainties
\citep{jenkins:2011}, we consider that such a threshold is a
conservative one. As made explicit in the definition of
the posterior density function $\pi(\theta | m)$, the
likelihood function is not the only source of
information. We take into account the fact that all
parameter values are not equally probable prior to
making the measurements via prior probability densities.
Essentially, our priors are chosen as in
\cite{tuomi:2012}. Of special relevance in the detection
process is the prior choice for the eccentricities. Our
functional choice for it (Gaussian with zero mean and
$\sigma_e$ = 0.3) is based on statistical, dynamical and
population considerations and it is discussed further in
the appendices (Appendix \ref{sec:prior}). For more
details on different prior choices, see the dedicated
discussion in \citet{tuomi:gj163}.

\subsection{Log--Likelihood periodograms}
\label{sec:periodograms}

\begin{figure}
\center
\includegraphics[width=0.4\textwidth,clip]{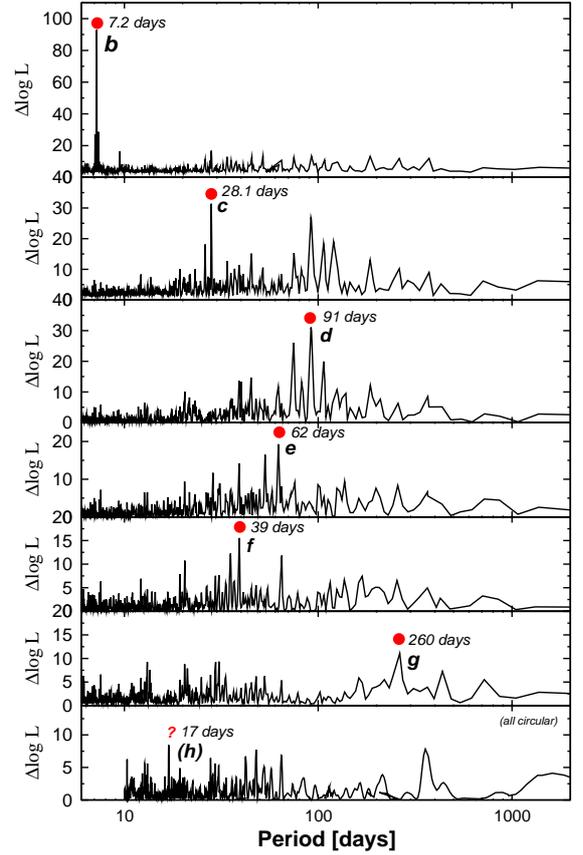}

\caption{Log--likelihood periodograms for the seven
candidate signals sorted by significance. While the
first six signals are easily spotted, the seventh is only
detected with log--L periodograms if all orbits are
assumed to be circular.}

\label{fig:periodograms}

\end{figure}

Because the orbital period (or frequency) is an extremely
non-linear parameter, the orbital solution of a model with
$k+1$ signals typically contains many hundreds or thousands
of local likelihood maxima (also called independent
frequencies). In any method based on stochastic processes,
there is always a chance that the global maxima of the
target function is missed. Our log--likelihood periodogram
(or log--L periodogram) is a tool to systematically identify
periods for new candidate planets of higher probability and
ensure that these areas have been well explored by the
Bayesian samplings (e.g., we always attempt to start the
chains close to the five most significant periodicities left
in the data). A log--L periodogram consists of computing the
improvement of the logarithm of the likelihood (new model
with $k+1$ planets) compared to the logarithm of the
likelihood of the null hypothesis (only $k$ planets) at each
test period. Log--L periodograms are represented as period
versus $\Delta \log L$ plots, where log is always the natural
logarithm. The definition of the likelihood function we use
is shown in Eq.~\ref{eq:likelihood} and typically assumes Gaussian
noise sources only (that is, different jitter parameters are
included for each instrument and g=0 in Eq.~\ref{eq:model}).

$\Delta \log L$ can also be used for estimating the
\textit{frequentist} false alarm probability (FAP) of a
solution using the likelihood-ratio test for nested models.
This FAP estimates what fraction of times one
would recover such a significant solution by an
unfortunate arrangement of Gaussian noise. To compute
this FAP from $\Delta \log L$ we used the up-to-date
recipes provided by \citet{baluev:2009}. We note that
that maximization of the likelihood involves solving for
many parameters simultaneously: orbital parameters of the
new candidate, all orbital parameters of the already
detected signals, a secular acceleration term
$\dot{\gamma}$, a zero-point $\gamma_l$ for each
instrument, and jitter terms $\sigma_{l}$ of each
instrument (see Eq.~\ref{eq:model}). It is, therefore, a
computationally intensive task, especially when several
planets are included and several thousand of test
periods for the new candidate must be explored. 

As discussed in the appendices (see Section
\ref{sec:eccprior}), allowing for full Keplerian
solutions at the period search level makes the method
very prone to false positives. Therefore while a full
Keplerian solution is typically assumed for all the
previously detected $k$-candidates, the orbital model for
the $k+1$-candidate is always assumed to be circular in
our standard setup. This way, our log--L periodograms
represent a natural generalization of more classic
hierarchical periodogram methods. This method
was designed to account for parameter correlations at the
detection level. If such correlations are not accounted
for, the significance of new signals can be strongly biased
causing both false positives and missed detections.
In the study of the planet hosting M-dwarf
GJ~676A \citep{gj676A:2012} and in the more recent manuscript
on GJ~581 \citep{tuomi_jenkins:2012}, we have shown that -while
log--L periodograms represent an improvement with
respect to previous periodogram schemes- the aforementioned
Bayesian approach has a higher sensitivity to lower
amplitude signals and is less prone to false positive
detections. Because of this, the use of log--L
periodograms is not to quantify the significance of a
new signal but to provide visual assessment of
possible aliases or alternative high-likelihood solutions.

Log--L periodograms implicitly assume flat priors for all the
free parameters. As a result, this method provides a quick
way of assessing the sensitivity of a detection against a
choice of prior functions that are different from uniform. As
discussed later, the sixth candidate is only confidently
spotted using log--L periodograms (our detection criteria is
FAP$<$ 1\%) when the orbits of all the candidates are assumed
to be circular. This is the \textit{red line} beyond which
our detection criteria becomes strongly dependent on our
choice of prior on the eccentricity. The same applies to the
seventh tentative candidate signal.

\begin{figure*}
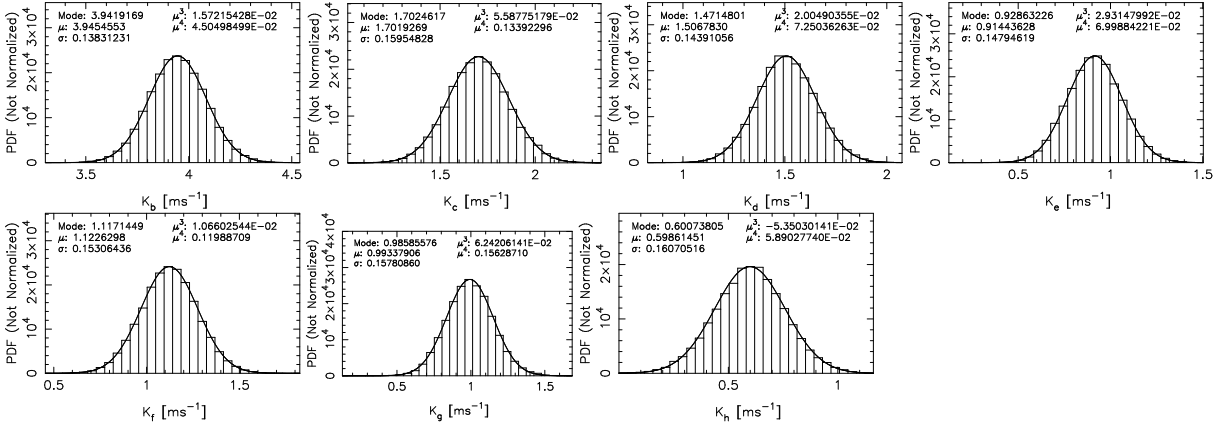

\raggedright
\includegraphics[width=0.15\textwidth,clip, angle=270]{rvdist07_rv_GJ667Cl_dist_Kb.ps}
\includegraphics[width=0.15\textwidth,clip, angle=270]{rvdist07_rv_GJ667Cl_dist_Kc.ps}
\includegraphics[width=0.15\textwidth,clip, angle=270]{rvdist07_rv_GJ667Cl_dist_Kd.ps}
\includegraphics[width=0.15\textwidth,clip, angle=270]{rvdist07_rv_GJ667Cl_dist_Ke.ps}
\includegraphics[width=0.15\textwidth,clip, angle=270]{rvdist07_rv_GJ667Cl_dist_Kf.ps}
\includegraphics[width=0.15\textwidth,clip, angle=270]{rvdist07_rv_GJ667Cl_dist_Kg.ps}
\includegraphics[width=0.15\textwidth,clip, angle=270]{rvdist07_rv_GJ667Cl_dist_Kh.ps}

\caption{Marginalized posterior densities for the Doppler
semi-amplitudes of the seven reported signals.}

\label{fig:densities}
\end{figure*}

\section{Signal detection and confidences} \label{sec:detection}

As opposed to other systems analyzed with the same
techniques \citep[e.g. Tau Ceti or HD
40307, ][]{tauceti,hd40307}, we found that for GJ~667C the
simplest model ($g=0$ in equation \ref{eq:model}) already
provides a sufficient description of the data. For
brevity, we omit here all the tests done with more
sophisticated parameterizations of the noise (see
Appendix \ref{sec:further}) that essentially lead to
unconstrained models for the correlated noise terms and
the same final results. In parallel with the Bayesian
detection sequence, we also illustrate the search using
log--L periodograms. In all that follows we use the three
datasets available at this time : HARPS-TERRA, HIRES and
PFS. We use the HARPS-TERRA Doppler measurements instead
of CCF ones because TERRA velocities have been proven to
be more precise on stable M-dwarfs
\citep{anglada:2012a}.

\begin{table}

\caption{Relative posterior probabilities and log-Bayes
factors of models $\mathcal{M}_{k}$ with $k$ Keplerian
signals given the combined HARPS-TERRA, HIRES, and PFS RV
data of GJ~667C. Factor $\Delta$ indicates how much the
probability increases with respect to the best model with
one less Keplerian and $P$ denotes the MAP period
estimate of the signal added to the solution when
increasing $k$. Only the highest probability sequence is
shown here (reference solution). A complete table with
alternative solutions corresponding to local probability
maxima is given in Appendix \ref{sec:TERRA_analysis}}

\begin{tabular}{lccccc}
\hline \hline
$k$ &
$P(\mathcal{M}_{k} | d)$ &
$\Delta$ &
$\log P(d | \mathcal{M}_{k})$ &
$P$ [days] & ID\\
\hline
0 & 2.7$\times10^{-85}$ & --		     & -602.1 & --		   \\
1 & 3.4$\times10^{-48}$ & 1.3$\times10^{37}$ & -516.0 & 7.2		   \\
2 & 1.3$\times10^{-35}$ & 3.9$\times10^{12}$ & -486.3 & 91		   \\
3 & 8.9$\times10^{-18}$ & 6.7$\times10^{17}$ & -444.5 & 28		   \\
4 & 1.9$\times10^{-14}$ & 2.1$\times10^{3}$  & -436.2 & 53		   \\
4 & 1.2$\times10^{-14}$ & 1.3$\times10^{3}$  & -436.7 & 62		   \\
5 & 1.0$\times10^{-7}$  & 5.5$\times10^{6}$  & -420.0 & 39, 53  	   \\
5 & 1.0$\times10^{-8}$  & 5.3$\times10^{5}$  & -422.3 & 39, 62  	   \\
6 & 4.1$\times10^{-3}$  & 4.0$\times10^{4}$  & -408.7 & 39, 53, 256	   \\
6 & 4.1$\times10^{-4}$  & 4.0$\times10^{3}$  & -411.0 & 39, 62, 256	   \\
7 & 0.057		& 14		     & -405.4 & 17, 39, 53, 256    \\
7 & 0.939		& 230	             & -402.6 & 17, 39, 62, 256    \\
\hline \hline
\end{tabular}
\end{table}

\begin{figure*}
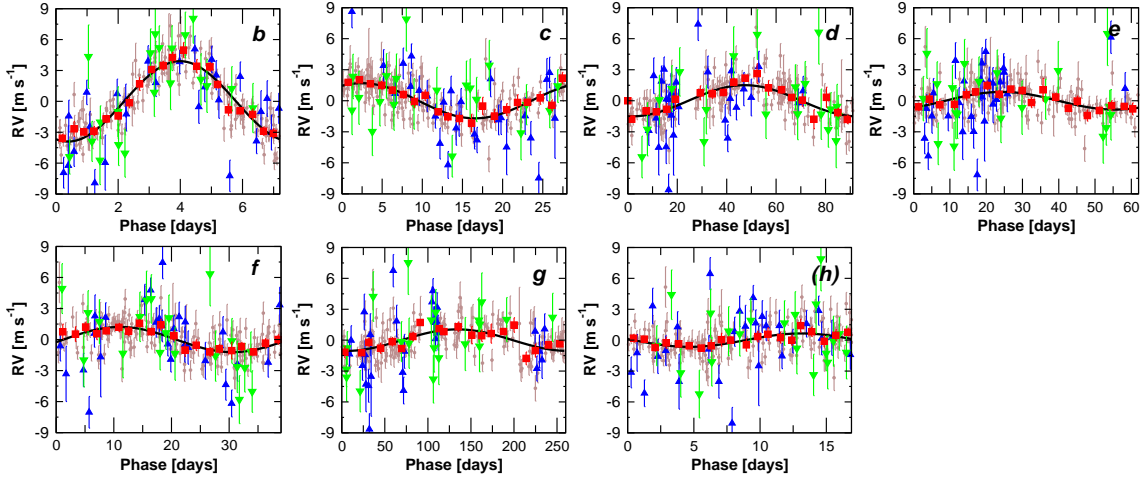

\raggedright
\includegraphics[width=0.20\textwidth,clip]{phase_folded_b.eps}
\includegraphics[width=0.20\textwidth,clip]{phase_folded_c.eps}
\includegraphics[width=0.20\textwidth,clip]{phase_folded_d.eps}
\includegraphics[width=0.20\textwidth,clip]{phase_folded_e.eps}
\includegraphics[width=0.20\textwidth,clip]{phase_folded_f.eps}
\includegraphics[width=0.20\textwidth,clip]{phase_folded_g.eps}
\includegraphics[width=0.20\textwidth,clip]{phase_folded_h.eps}

\caption{RV measurements phase-folded to the best period for
each planet. Brown circles are HARPS-TERRA velocities, PFS
velocities are depicted as blue triangles, and HIRES
velocities are green triangles. Red squares are
averages on 20 phase bins of the HARPS-TERRA velocities.
The black line corresponds to the best circular
orbital fit (visualization purposes only).}

\label{fig:phasefolded}

\end{figure*}

\begin{table*}
\scriptsize
\center

\caption{Reference orbital parameters and their
corresponding 99\% credibility intervals. While the angles
$\omega$ and $M_0$ are unconstrained due to strong
degeneracies at small eccentricities, their sum
$\lambda=M_0 + \omega$ is better behaved and is also
provided here for reference.}
\label{tab:parameters}

\begin{tabular}{lllllll}
\hline \hline
 & b & (h) & c & f & e$^*$\\
\hline
P [days]                & 7.2004 [7.1987, 7.2021] & 16.946 [16.872, 16.997] & 28.140 [28.075, 28.193] & 39.026 [38.815, 39.220]  & 62.24 [61.69, 62.79]      \\
e                       & 0.13 [0.02, 0.23]       & 0.06 [0, 0.38]          &  0.02 [0, 0.17]         & 0.03 [0, 0.19]           & 0.02 [0, 0.24]	     \\
K         [\ms]         & 3.93 [3.55, 4.35]       & 0.61 [0.12, 1.05]       & 1.71 [1.24, 2.18]       & 1.08 [0.62, 1.55]        & 0.92 [0.50, 1.40]	     \\
$\omega$ [rad]          & 0.10 [5.63, 0.85]       & 2.0 [0, 2$\pi$]         & 5.1 [0, $2\pi$]         & 1.8 [0, 2$\pi$]          & 0.5 [0, 2$\pi$]	     \\
M$_0$ [rad]             & 3.42 [2.32, 4.60]       & 5.1 [0, 2$\pi$]         & 0.3 [0, $2\pi$]         & 5.1 [0, 2$\pi$]          & 4.1 [0, 2$\pi$]	     \\
                                                                                                                                 			       \\
$\lambda$ [deg]         & 201[168, 250]           & 45(180)$^\dagger$            & 308(99)$^\dagger$       & 34 (170)$^\dagger$  & 262(150)$^\dagger$	     \\
M $\sin i$ [M$_\oplus$] & 5.6 [4.3, 7.0]          & 1.1 [0.2, 2.1]          & 3.8 [2.6, 5.3]          & 2.7 [1.5, 4.1]           & 2.7 [1.3, 4.3]	     \\
a [AU]                  & 0.0505 [0.0452, 0.0549] & 0.0893 [0.0800, 0.0977] & 0.125 [0.112, 0.137]    & 0.156 [0.139, 0.170]     & 0.213 [0.191, 0.232]      \\
\\
& d & g & Other model parameters\\
\hline
P [days]                & 91.61 [90.72, 92.42]    & 256.2 [248.3, 270.0]  & $\dot\gamma$ [\ms$yr^{-1}$] &  2.07 [1.79, 2.33]   \\
e                       & 0.03 [0, 0.23]	  & 0.08 [0, 0.49]	  & $\gamma_{\rm HARPS}$ [\ms]	 & -30.6 [-34.8, -26.8] \\
K [\ms]                 & 1.52 [1.09, 1.95]	  & 0.95 [0.51, 1.43]	  & $\gamma_{\rm HIRES}$ [\ms]	 & -31.9 [-37.0,, -26.9]\\
$\omega$ [rad]          & 0.7 [0, 2$\pi$]	  & 0.9 [0, 2$\pi$]	  & $\gamma_{\rm PFS}$ [\ms]	 & -25.8 [-28.9, -22.5] \\
M$_{0}$ [rad]           & 3.7 [0, 2$\pi$]	  & 4.1 [0, 2$\pi$]	  & $\sigma_{\rm HARPS}$ [\ms] &  0.92 [0.63, 1.22]   \\
                        &                         &                       & $\sigma_{\rm HIRES}$ [\ms] &  2.56 [0.93, 5.15]   \\
$\lambda$ [deg]         & 251(126)$^\dagger$	  & 285(170)$^\dagger$    & $\sigma_{\rm PFS}$ [\ms]   &  1.31 [0.00, 3.85]   \\
M $\sin i$ [M$_\oplus$] & 5.1 [3.4, 6.9]	  & 4.6 [2.3, 7.2]	  \\
a [AU]                  & 0.276 [0.246, 0.300]    & 0.549 [0.491, 0.601]  \\
\hline \hline

\end{tabular}
\tablefoot{
$\dagger$ Values allowed in the full range of $\lambda$.
Full-width-at-half-maximum of the marginalized posterior is provided
to illustrate the most likely range (see Figure \ref{fig:stab6p}).
$*$ Due to the presence of a strong alias, the orbital
period of this candidate could be 53 days instead. Such an
alternative orbital solution for planet e is given in Table
\ref{tab:CCF_parameters}.
}

\end{table*}

The first three periodicities (7.2 days, 28.1 days and 91
days) were trivially spotted using Bayesian posterior
samplings and the corresponding log--L periodograms. These
three signals were already reported by
\citet{anglada:2012b} and \citet{delfosse:2012}, although
the last one (signal d, at 91 days) remained uncertain due
to the proximity of a characteristic time-scale of the
star's activity. This signal is discussed in the context of
stellar activity in Section \ref{sec:activity}. Signal d
has a MAP period of 91 days and would correspond
to a candidate planet with a minimum mass of $\sim$ 5 M$_\oplus$.

After that, the log--L periodogram search for a fourth
signal indicates a double-peaked likelihood maximum at 53
and 62 days -both candidate periods receiving extremely
low false-alarm probability estimates (see Figure
\ref{fig:periodograms}). Using the recipes in
\citet{dawson:2010}, it is easy to show that the two
peaks are the yearly aliases of each other. Accordingly,
our Bayesian samplings converged to either period equally
well giving slightly higher probability to the 53-day
orbit ($\times 6$). In both cases, we found that
including a fourth signal improved the model probability
by a factor $>$10$^3$. In appendix
\ref{sec:TERRA_analysis} we provide a detailed analysis
and derived orbital properties of both solutions and show
that the precise choice of this fourth period does not
substantially affect the confidence of the rest of
the signals. As will be shown at the end of the
detection sequence, the most likely solution
for this candidate corresponds to a
minimum mass of $2.7$ M$_\oplus$ and a period of 62 days.

After including the fourth signal, a fifth signal at 39.0
days shows up conspicuously in the log--L periodograms. In
this case, the posterior samplings always converged to
the same period of 39.0 days without difficulty (signal
f). Such a planet would have a minimum mass of $\sim$2.7
M$_\oplus$. Given that the model probability improved by a
factor of 5.3$\times$10$^5$ and that the FAP
derived from the log-L periodogram is $0.45\%$, the
presence of this periodicity is also supported by the data
without requiring further assumptions.

The Bayesian sampling search for a sixth signal always
converged to a period of 260 days that also satisfied our
detection criteria and increased the probability of the model
by a factor of $4\times 10^3$. The log--L periodograms did spot
the same signal as the most significant one but assigned a
FAP of $\sim$20\% to it. This apparent contradiction is due
to the prior on the eccentricity. That is, the maximum
likelihood solution favors a very eccentric orbit for the
Keplerian orbit at 62 days ($e_e\sim 0.9$), which is unphysical and
absorbs variability at long timescales through aliases. To
investigate this, we performed a log--L periodogram
search assuming circular orbits for all the candidates. In
this case, the 260-day period received a FAP of $0.5\%$
which would then qualify as a significant detection. Given
that the Bayesian detection criteria are well satisfied and
that the log--L periodograms also provide substantial support
for the signal, we also include it in the model (signal g). Its
amplitude would correspond to a planet with a minimum mass of 4.6 M$_\oplus$.

When performing a search for a seventh signal, the posterior
samplings converged consistently to a global probability
maximum at 17 days (M $\sin i \sim 1.1$ M$_\oplus$) which
improves the model probability by a factor of 230. The
global probability maximum containing seven signals
corresponds to a solution with a period of 62 days for
planet e. This solution has a total probability $\sim$ 16
times larger than the one with $P_e = 53$ days. Although
such a difference is not large enough to make a final decision
on which period is preferred, from now on we will assume
that our reference solution is the one with $P_e = 62.2$
days. The log--L periodogram also spotted the same seventh
period as the next favored one but only when all seven
candidates were assumed to have circular orbits. Given that
this seventh signal is very close to the Bayesian detection
limit, and based on our experience on the analysis of
similar datasets \citep[e.g., GJ 581][]{tuomi_jenkins:2012},
we concede that this candidate requires more measurements to
be securely confirmed. With a minimum mass of only $\sim 1.1$
M$_\oplus$, it would be among the least massive
exoplanets discovered to date.

As a final comment we note that, as in
\citet{anglada:2012b} and \citet{delfosse:2012}, a linear
trend was always included in the model. The most likely
origin of such a trend is gravitational acceleration caused
by the central GJ~667AB binary. Assuming a minimum
separation of 230 AU, the acceleration in the line-of-sight
of the observer can be as large as 3.7 \ms, which is of the
same order of magnitude as the observed trend of $\sim$ 2.2
\ms $yr^{-1}$. We remark that the trend (or part of it)
could also be caused by the presence of a very long period
planet or brown dwarf. Further Doppler follow-up,
astrometric measurements, or direct imaging attempts of
faint companions might help addressing this question.

In summary, the first five signals are easily spotted using
Bayesian criteria and log--L periodograms. The global
solution containing seven-Keplerian signals prefers a period of 62.2 days for
signal e, which we adopt as our reference solution. Still, a
period of 53 days for the same signal cannot be ruled
out at the moment. The statistical significance of a 6th
periodicity depends on the prior choice for the eccentricity, but
the Bayesian odds ratio is high enough to propose it as a
genuine Keplerian signal. The statistical significance of the
seventh candidate (h) is close to our detection limit and
more observations are needed to fully confirm it.

\section{Activity} \label{sec:activity}

\begin{figure}
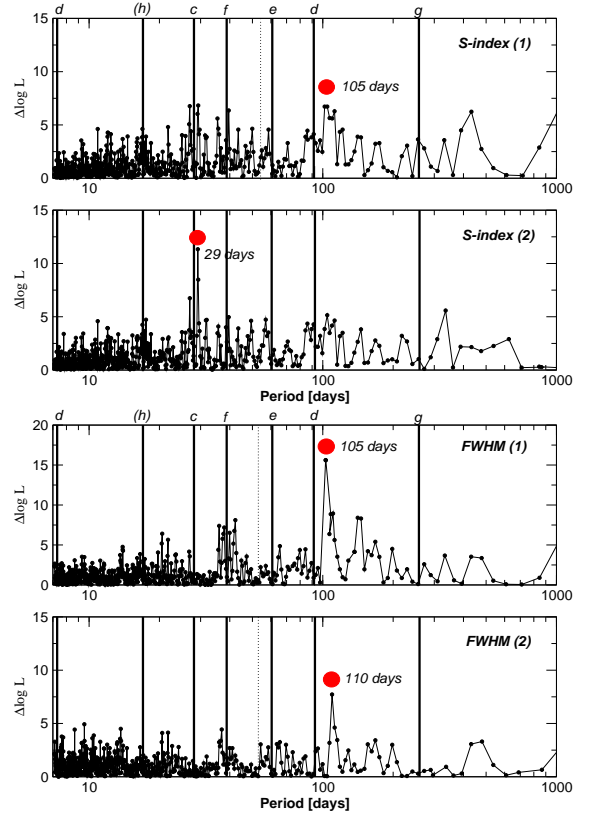

\center
\includegraphics[width=0.4\textwidth,clip]{cahk_periodograms_clean.eps}
\includegraphics[width=0.4\textwidth,clip]{fwhm_periodogram_clean.eps}

\caption{\textbf{Top two panels} Log--L periodograms for
up to 2 signals in the S-index. The most likely periods of
the proposed planet candidates are marked as vertical
lines. \textbf{Bottom two panels.} Log--L periodograms for
up to 2 signals in the FWHM. Given the proximity of these
two signals, it is possible that both of them originate
from the same feature (active region corotating with the
star) that is slowly evolving with time.}

\label{fig:activity}
\end{figure}

In addition to random noise (white or correlated), stellar
activity can also generate spurious Doppler periodicities
that can mimic planetary signals \citep[e.g.,
][]{lovis:2011, reiners:2013}. In this section we
investigate whether there are periodic variations in the
three activity indices of GJ~667C (S-index, BIS and FWHM
are described in Section \ref{sec:observations}). Our
general strategy is the following : if a significant
periodicity is detected in any of the indices and such
periodicity can be related to any of the candidate signals
(same period or aliases), we add a linear correlation term
to the model and compute log--L periodograms and new
samplings of the parameter space. If the data were better
described by the correlation term rather than a genuine
Doppler signal, the overall model probability would
increase and the planet signal in question should decrease
its significance (even disappear).

\begin{figure}
\center
\includegraphics[width=0.45\textwidth,clip]{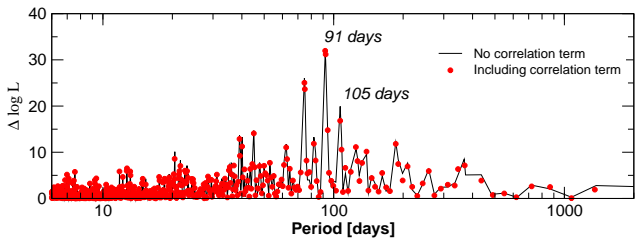}

\caption{Log--likelihood periodograms for planet d (91 days) including the
correlation term (red dots) compared to the original periodogram
without this term (black line). The inclusion of the correlation term increases the contrast
between the peaks at 91 and 105 days, favoring the interpretation of the 91
days signals as a genuine planet candidate.} \label{fig:91days}
\end{figure}

Log--L periodogram analysis of two activity indices
(S-index but specially the FWHM) show a strong periodic
variability at 105 days. As discussed in
\citet{anglada:2012b} and \citet{delfosse:2012}, this cast some doubt
on the candidate at 91 days (d). Despite the fact that the
91-day and 105-day periods are not connected by first
order aliases, the phase sampling is sparse in this period domain
(see phase--folded diagrams of the RV data for the
planet d candidate in Fig.~\ref{fig:phasefolded}) and the
log--L periodogram for this candidate also shows
substantial power at 105 days. From the log--L periodograms
in Figure \ref{fig:periodograms}, one can directly obtain
the probability ratio of a putative solution at 91 days
versus one with a period of 105 days when no correlation
terms are included. This ratio is $6.8\times 10^4$, meaning
that the Doppler period at 91 days is largely favoured over
the 105-day one. All Bayesian samplings starting close to
the 105-day peak ended-up converging on the signal at 91
day. We then applied our validation procedure by inserting
linear correlation terms in the model (g=C$_{F}\times$
FWHM$_i$ or g=C$_S\times$ S$_i$), in eq. \ref{eq:model})
and computed both log--L periodograms and Bayesian
samplings with C$_F$ and C$_S$ as free parameters. In all
cases the $\Delta \log L$ between 91 and 105 days slightly
increased, thus supporting the conclusion that the signal at 91 days is of
planetary origin. For example, the ratio of likelihoods
between the 91 and 105 day signals increased from
6.8$\times$10$^4$ to 3.7$\times$10$^6$ when the
correlation term with the FWHM was included (see Figure
\ref{fig:91days}). The Bayesian samplings including the
correlation term did not improve the model probability (see
Appendix \ref{sec:further}) and still preferred the 91-day
signal without any doubt. We conclude that this signal
is not directly related to the stellar activity and
therefore is a valid planet candidate.

Given that activity might induce higher order harmonics in
the time-series, all seven candidates have been analyzed and
double-checked using the same approach. Some more details on
the results from the samplings are given in the Appendix
\ref{sec:correlation}. All candidates -including the
tentative planet candidate h- passed all these validation
tests without difficulties.

\section{Tests for quasi-periodic signals}
\label{sec:subsamples}

\begin{table*}
\center
\caption{
Most significant periods as extracted using $\log$--L
periodograms on subsamples of the first N${_obs}$
measurements. Boldfaced values indicate coincidence with a
signal of our seven-planet solution (or a first order yearly
alias thereof). A parenthesis in the last period indicates a
preferred period that did not satisfy the frequentist 1\% FAP
threshold but did satisfy the Bayesian detectability
criteria.
}\label{tab:subsamples}
\begin{tabular}{lccccccc}
\hline \hline
N$_{obs}$  & 1          & 2          & 3         & 4             & 5             & 6          & 7       \\
\hline\hline
50         & {\bf7.2}   & {   101.5} & --        & --            & --            & --         & --      \\
75         & {\bf7.2}   & {   103.0} & --        & --            & --            & --         & --      \\
90         & {\bf7.2}   & {\bf28.0}  & {  104.1} & --            & --            & --         & --      \\
100        & {\bf7.2}   & {\bf91.2}  & {\bf28.0} & {\bf54.4}$^a$ & --            & --         & --      \\
120        & {\bf7.2}   & {\bf91.6}  & {\bf28.0} & --            & --            & --         & --      \\
143        & {\bf7.2}   & {\bf91.6}  & {\bf28.0} & {\bf53.6}$^a$ & {\bf35.3}$^a$ & {\bf(260)} & --      \\
160        & {\bf7.2}   & {\bf28.1}  & {\bf91.0} & {\bf38.9}     & {\bf53.4}$^a$ & {\bf275}   & {\bf(16.9)} \\
173        & {\bf7.2}   & {\bf28.1}  & {\bf91.9} & {\bf61.9}     & {\bf38.9}     & {\bf260}   & {\bf(16.9)} \\
\hline
\hline \hline
\end{tabular}
\tablefoot{ $^a$ 1 year$^{-1}$ alias of the preferred
period in Table \ref{tab:parameters}.}
\end{table*}

Activity induced signals and superposition of several
independent signals can be a source of confusion and result in
detections of ``apparent'' false positives. In an attempt to
quantify how much data is necessary to support our preferred
global solution (with seven planets) we applied the log-L
periodogram analysis method to find the solution as a function
of the number of data points. For each dataset, we stopped
searching when no peak above FAP 1\% was found. The process was
fully automated so no human-biased intervention could alter the
detection sequence. The resulting detection sequences are show
in Table \ref{tab:subsamples}. In addition to observing how the
complete seven-planet solution slowly emerges from the data one can
notice that for $N_{\rm obs}<$100 the $k=2$ and $k=3$ solutions
converge to a strong signal at $\sim 100$ days. This period is
dangerously close to the activity one detected in the FWHM and
S-index time-series. To explore what could be the cause of this
feature (perhaps the signature of a quasi-periodic signal), we
examined the $N_{\rm obs}$=75 case in more detail and made a
supervised/visual analysis of that subset.

\begin{figure}
\center
\includegraphics[width=0.4\textwidth,clip]{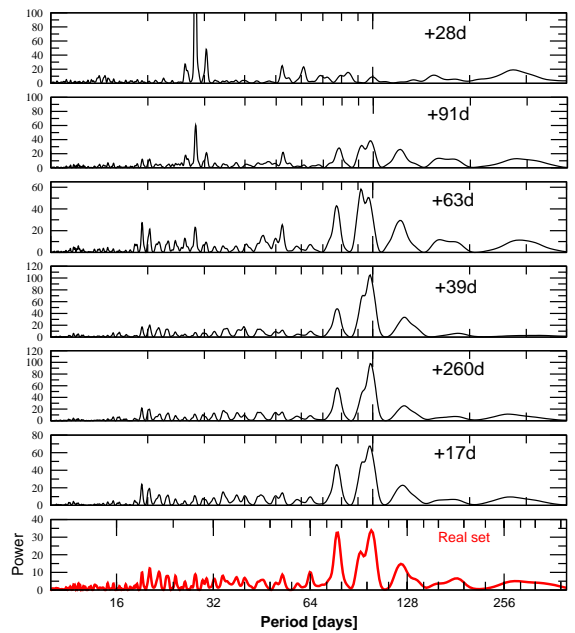}
\caption{Sequence of periodograms obtained from
synthetic noiseless data generated on the first 75
epochs. The signals in Table \ref{tab:parameters} were
sequentially injected from top to bottom. The bottom
panel is the periodogram to the real dataset after
removing the first 7.2 days planet candidate. }
\label{fig:N75}
\end{figure}

The first 7.2 days candidate could be easily extracted.
We then computed a periodogram of the residuals to
figure out if there were additional signals left in the data. In
agreement with the automatic search, the periodogram of
the residuals (bottom of Figure \ref{fig:N75}) show a
very strong peak at $\sim$100 days. The peak was so
strong that we went ahead and assessed its
significance. It had a very low FAP ($<0.01\%$) and
also satisfied our Bayesian detectability criteria. We
could have searched for additional companions, but let
us assume we stopped our analysis here. In this case, we
would have concluded that two signals were strongly
present (7.2 days and 100 days). Because of the
proximity with a periodicity in the FWHM-index (~105
days), we would have expressed our doubts about the
reality of the second signal so only one planet
candidate would have been proposed (GJ 667Cb).

\begin{figure*}
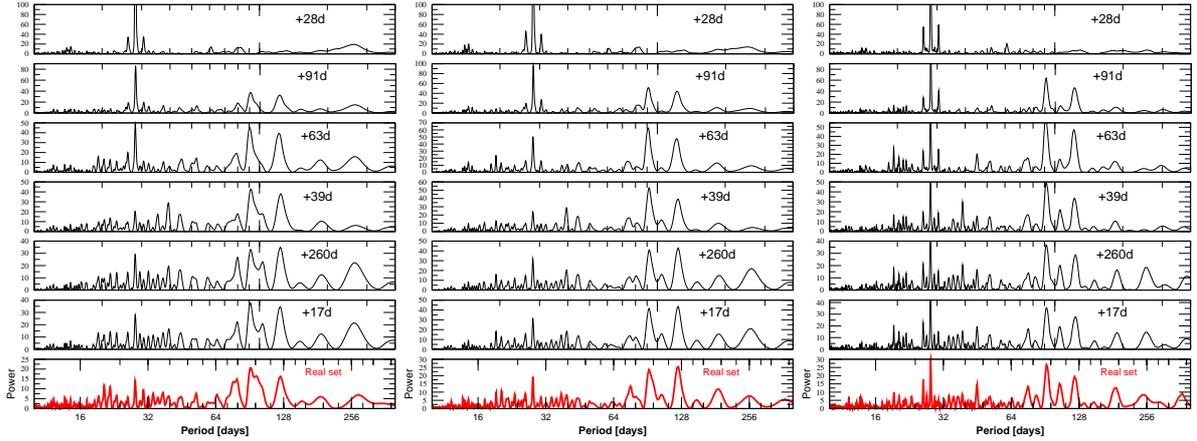

\center
\includegraphics[width=0.28\textwidth,clip]{real_versus_injected_100_sequence.eps}
\includegraphics[width=0.28\textwidth,clip]{real_versus_injected_120_sequence.eps}
\includegraphics[width=0.28\textwidth,clip]{real_versus_injected_173_sequence.eps}
\caption{Same as \ref{fig:N75} but using the first 100 epochs (left), first 120 (center) and all of them (right).}
\label{fig:N100}
\end{figure*}

With 228 RV measurements in hand (173 HARPS-TERRA, 23
PFS and 22 from HIRES) we know that such a conclusion is no
longer compatible with the data. For example, the second and
third planets are very consistently detected at 28 and 91
days. We investigated the nature of that 100 day signal
using synthetic subsets of observations as follows. We took
our preferred seven-planet solution and generated the exact
signal we would expect if we only had planet c (28 days) in
the first 75 HARPS-TERRA measurements (without noise). The
periodogram of such a noiseless time-series (top panel in
Fig.~\ref{fig:N75}) was very different from the real one.
Then, we sequentially added the rest of the signals. As more planets
were added, the periodogram looked closer to the one from the
real data. This illustrates that we would have reached the
same wrong conclusion even with data that had negligible noise.
How well the general structure
of the periodogram was recovered after adding all of the signals 
is rather remarkable (comparing the bottom two panels in Fig.~\ref{fig:N75}).
While this is not a statistical proof of significance, it
shows that the \textit{periodogram} of the residuals from the
1-planet fit (even with only 75 RVs measurements) is indeed
consistent with the proposed seven-planet solution without
invoking the presence of quasi-periodic signals. This
experiment also shows that, until all stronger signals could
be well-decoupled (more detailed investigation showed this
happened at about $N_{\rm obs}\sim 140$), proper and robust
identification of the correct periods was very difficult. We
repeated the same exercise with $N_{\rm obs}$=100, 120 and
173 (all HARPS measurements) and obtained identical behavior
without exception (see panels in Figure \ref{fig:N100}).
Such an effect is not new and the
literature contains several examples that cannot be easily
explained by simplistic aliasing arguments- e.g.,  see GJ
581d \citep{udry:2007, mayor:2009} and HD 125612b/c
\citep{anglada:2010,locurto:2010}. The fact that all signals
detected in the velocity data of GJ 667C have similar
amplitudes -- except perhaps candidate b which has a
considerably higher amplitude -- made this problem especially
severe. In this sense, the currently available set of
observations are a sub--sample of the many more that might be
obtained in the future, so it might happen that one of the
signals we report ``bifurcates'' into other periodicities.
This experiment also suggests that spectral information
beyond the most trivial aliases can be used to verify and
assess the significance of future detections (under
investigation).
\begin{figure*}
\center
\includegraphics[width=0.8\textwidth,clip]{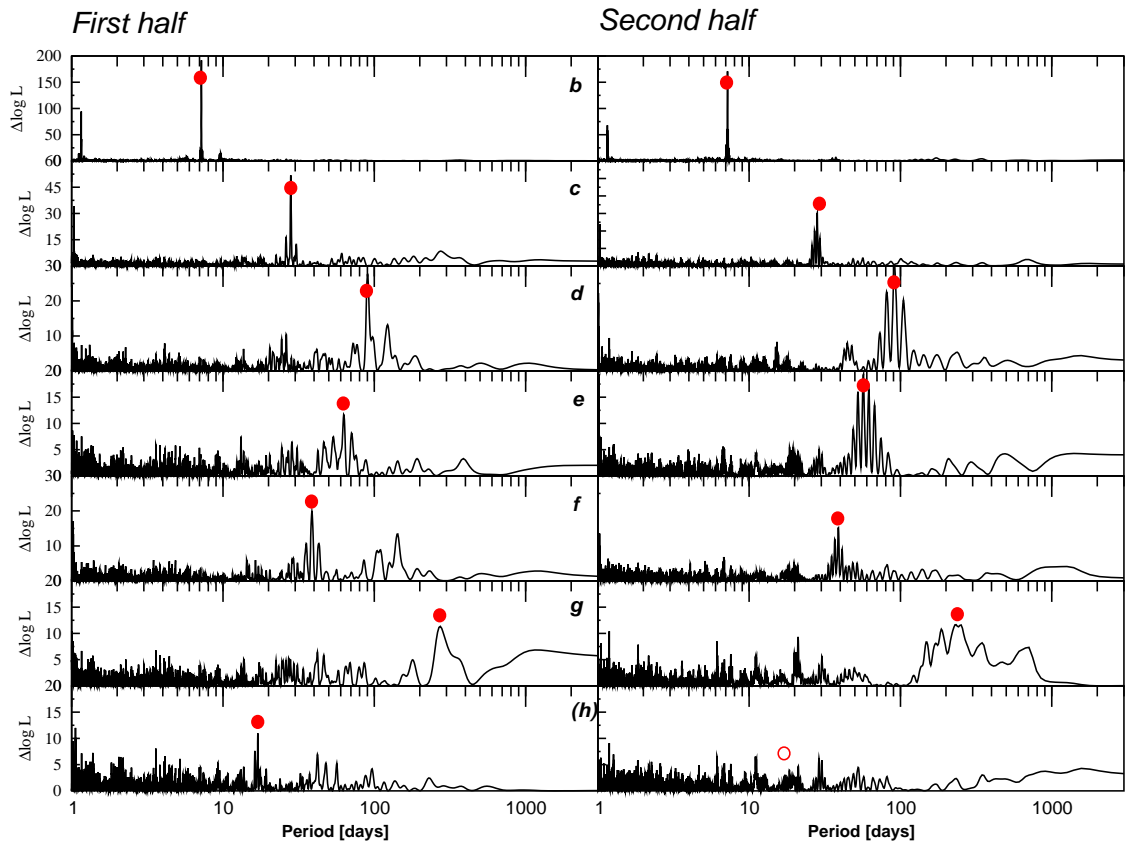}
\caption{Periodograms on first and second half of the time series
as obtained when all signals except one were removed from the data.
Except for signal h, all signals are significantly present in both
halves and could have been recovered using either half if they
had been in single planet systems.}
\label{fig:split}
\end{figure*}

\subsection{Presence of individual signals in one half of the data}

As an additional verification against quasi-periodicity, we investigated
if the signals were present in the data when it was divided into two
halves. The first half corresponds to the first 86 HARPS observations and
the second half covers the remaining data. The data from PFS and HIRES were not
used for this test. The experiment consists of removing all signals
except for one, and then computing the periodogram on those
residuals (first and second halfs separately). If a signal is strongly
present in both halfs, it should, at least, emerge as substantially
significant. All signals except for the seventh one passed this test nicely.
That is, in all cases except for h, the periodograms prominently display
the non-removed signal unambiguously. Besides demonstrating that all
signals are strongly present in the two halves, it also illustrates that
any of the candidates would have been trivially spotted using
periodograms if it had been orbiting alone around the star. The fact
that each signal was not spotted before
\citep{anglada:2012b,delfosse:2012} is a consequence of an inadequate
treatment of signal correlations when dealing with periodograms of the
residuals only. Both the described Bayesian method and the
log-likelihood periodogram technique are able to deal with such
correlations by identifying the combined global solution at the period
search level. As for other multiplanet systems detected using similar
techniques \citep{hd40307,gj676A:2012}, optimal exploration of the
global probability maxima at the signal search level is essential to
properly detect and assess the significance of low mass multiplanet
solutions, especially when several signals have similar amplitudes close
to the noise level of the measurements.

Summarizing these investigations and validation of the signals
against activity, we conclude that

\begin{itemize}
\item
Up to seven periodic signals are detected in the Doppler measurements of
GJ 667C data, with the last (seventh) signal very close to our
detection threshold.

\item
The significance of the signals are not affected by correlations with activity
indices and we could not identify any strong wavelength dependence with any of
them.

\item
The first six signals are strongly present in subsamples of the data. Only the
seventh signal is unconfirmed using half of the data alone. Our analysis
indicates that any of the six stronger signals would had been robustly spotted
with half the available data if each had been orbiting alone around the host
star.

\item
Signal correlations in unevenly sampled data are the reason why \citet{anglada:2012b} and \citet{delfosse:2012} only reported three
of them. This is a known problem when assessing the significance of signals using periodograms of residuals only \citep[see ][as another example]{gj676A:2012}.

\end{itemize}

\noindent
Given the results of these validation tests, we promote
six of the signals (b, c, d, e, f, g) to planet candidates. For
economy of language, we will refer to them as \textit{planets}
but the reader must remember that, unless complementary and
independent verification of a Doppler signal is obtained (e.g.,
transits), they should be called \textit{planet candidates}.
Verifying the proposed planetary system against dynamical
stability is the purpose of the next section.

\section{Dynamical analysis} \label{sec:dynamics}

\begin{figure*}
\center
\includegraphics[width=0.9\textwidth,clip]{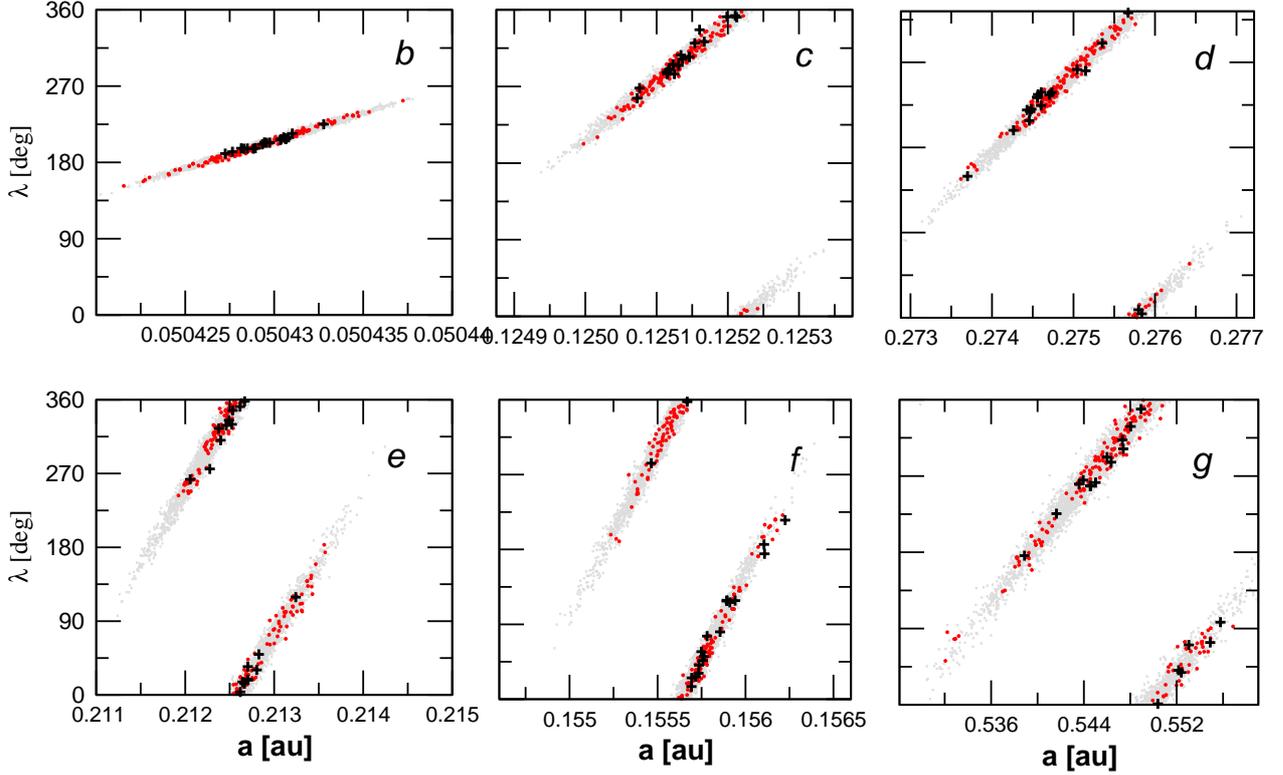}

\caption{Result of the stability analysis of 80,000
six-planet solutions in the plane of initial
semi-major axis $a$ vs. initial mean longitude
$\lambda$ obtained from a numerical integration over
$T\approx 7000$ years. Each initial condition is
represented as a point. Initial conditions leading to
an immediate disintegration of the system are shown as
gray dots. Initial conditions that lead to stable
motion for at least 1 Myr are shown as red points
(D$<10^{-5}$). Black crosses represent the most
stable solutions (D$<10^{-6}$), and can last
over many Myr.}

\label{fig:stab6p}
\end{figure*}

One of the by-products of the Bayesian analysis
described in the previous sections, are numerical
samples of statistically allowed parameter
combinations. The most likely orbital elements and
corresponding confidence levels can be deduced from
these samples. In Table \ref{tab:parameters} we give
the orbital configuration for a planetary system with
seven planets around GJ~667C, which is preferred from a
statistical point of view. To be sure that the
proposed planetary system is physically realistic,
it is necessary to confirm that these parameters not
only correspond to the solution favored by the data,
but also constitute a dynamically stable configuration. Due to the
compactness of the orbits, abundant resonances and
therefore complex interplanetary interactions are
expected within the credibility intervals. To slightly
reduce this complexity and since
evidence for planet h is weak, we split our analysis
and present- in the first part of this section- the
results for the six-planet solution with planets b to g.
The dynamical feasibility of the seven-planet solution
is then assessed by investigating the semi-major axis
that would allow introducing a seventh planet with the
characteristics of planet h.

\begin{table*}
\center
\caption{Astrocentric orbital elements of solution S$_6$.}
\label{tab:mostStable}
\begin{tabular}{crccrrc}
\hline \hline
Planet & $P$ (d) & $a$ (AU) &$e$ & $\omega$ ($^\circ$)
& $M_0$ ($^\circ$) & $M\sin i$ (M$_\oplus$) \\
\hline
b &  7.2006  & 0.05043 &  0.112 &   4.97 &  209.18 &  5.94 \\
c & 28.1231  & 0.12507 &  0.001 & 101.38 &  154.86 &  3.86 \\
f & 39.0819  & 0.15575 &  0.001 &  77.73 &  339.39 &  1.94 \\
e & 62.2657  & 0.21246 &  0.001 & 317.43 &   11.32 &  2.68 \\
d & 92.0926  & 0.27580 &  0.019 & 126.05 &  243.43 &  5.21 \\
g & 251.519  & 0.53888 &  0.107 & 339.48 &  196.53 &  4.41 \\
\hline \hline
\end{tabular}
\end{table*}

\subsection{Finding stable solutions for six planets} \label{sec:stable_six}

A first thing to do is to extract from the Bayesian
samplings those orbital configurations that allow stable
planetary motion over long time scales, if any. Therefore
we tested the stability of each configuration by a separate
numerical integration using the symplectic integrator
SABA$_2$ of \citet{laskar:2001} with a step size
$\tau=0.0625$ days. In the integration, we included a
correction term to account for general relativistic
precession. Tidal effects were neglected for these runs.
Possible effects of tides are discussed separately in
Section \ref{sec:tidal}. The integration was stopped if any
of the planets went beyond 5 AU or planets approached each
other closer than $10^{-4}$ AU.

The stability of those configurations that survived
the integration time span of $10^4$ orbital periods of
planet g (i.~e. $\approx 7000$ years), was then
determined using frequency analysis
\citep{laskar:1993}. For this we computed the
stability index $D_k$ for each planet $k$ as the
relative change in its mean motion $n_k$ over two
consecutive time intervals as was done in
\citet{hd40307}. For stable orbits the computed mean
motion in both intervals will be almost the same and
therefore $D_k$ will be very small. We note that
this also holds true for planets captured inside a
mean-motion resonance (MMR), as long as this resonance
helps to stabilize the system. As an index for the
total stability of a configuration we used
$D=\mathrm{max}(|D_k|)$. The results are summarized in
Figure \ref{fig:stab6p}. To generate Figure
\ref{fig:stab6p}, we extracted a sub-sample of 80,000
initial conditions from the Bayesian samplings. Those
configurations that did not reach the final
integration time are represented as gray dots. By
direct numerical integration of the remaining initial
conditions, we found that almost all configurations
with $D<10^{-5}$ survive a time span of 1~Myr. This
corresponds to $\sim$0.3 percent of the total sample.
The most stable orbits we found ($D< 10^{-6}$) are
depicted as black crosses.

In Figure \ref{fig:stab6p} one can see that the
initial conditions taken from the integrated 80,000
solutions are already confined to a very narrow range
in the parameter space of all possible orbits. This
means that the allowed combinations of initial $a$
and $\lambda$ are already quite restricted by the
statistics. By examining Figure \ref{fig:stab6p} one
can also notice that those initial conditions that
turned out to be long-term stable are quite spread out
along the areas where the density of Bayesian states
is higher. Also, for some of the candidates (d, f and
g), there are regions were no orbit was found
with D$<$10$^{-5}$. The paucity of stable orbits at
certain regions indicate areas of strong chaos within
the statistically allowed ranges (likely disruptive
mean-motion resonances) and illustrate that the
dynamics of the system are far from trivial.

\begin{figure*}
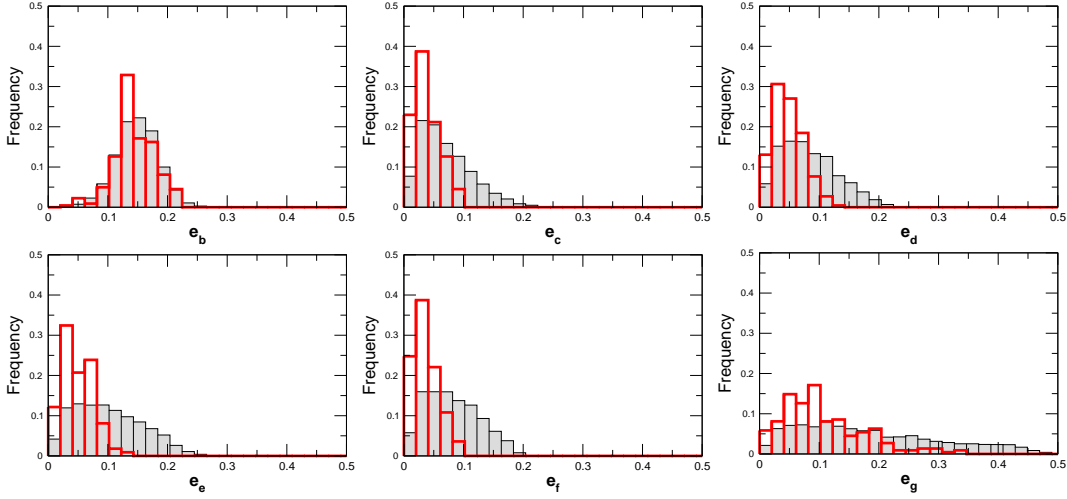

\center
 \includegraphics[width=0.25\textwidth,clip]{eb.eps}
 \includegraphics[width=0.25\textwidth,clip]{ec.eps}
 \includegraphics[width=0.25\textwidth,clip]{ed.eps}

 \includegraphics[width=0.25\textwidth,clip]{ee.eps}
 \includegraphics[width=0.25\textwidth,clip]{ef.eps}
 \includegraphics[width=0.25\textwidth,clip]{eg.eps}

\caption{Marginalized posterior densities for the
orbital eccentricities of the six planet solution (b, c,
d, in the first row; e, f, g in the second) before
(gray histogram) and after (red histogram) ruling out
dynamically unstable configurations.}

\label{fig:eccdist}

\end{figure*}

The distributions of eccentricities are also strongly
affected by the condition of dynamical stability. In
Figure \ref{fig:eccdist} we show the marginalized
distributions of eccentricities for the sample of all
the integrated orbits (gray histograms) and the
distribution restricted to relatively stable orbits
(with D$<10^{-5}$, red histograms). We see that, as
expected, stable motion is only possible with
eccentricities smaller than the mean values allowed by
the statistical samples. The only exceptions are
planets b and g. These two planet candidates are well
separated from the other candidates. As a consequence,
their probability densities are rather unaffected by
the condition of long-term stability. We note here
that the information about the dynamical stability has
been used only \textit{a posteriori}. If we had used
long-term dynamics as a prior (e.g., assign 0
probability to orbits with D$>$10$^{-5}$), moderately
eccentric orbits would have been much more strongly
suppressed than with our choice of prior function (Gaussian
distribution of zero mean and $\sigma=0.3$, see
Appendix \ref{sec:eccprior}). In this sense, our prior
density choice provides a much softer and
uninformative constraint than the dynamical viability
of the system.

In the following we will use the set of initial
conditions that gave the smallest $D$ for a detailed
analysis and will refer to it as S$_6$. In Table
\ref{tab:mostStable}, we present the masses and
orbital parameters of S$_6$, and propose it as the
favored configuration. To double check our dynamical
stability results, we also integrated S$_6$ for $10^8$
years using the HNBody package \citep{RauchHamilton02}
including general relativistic corrections and a
time step of $\tau=10^{-3}$ years.\footnote{Publicly
available at http://janus.astro.umd.edu/HNBody/}

\subsection{Secular evolution} \label{sec:secular}

Although the dynamical analysis of such a complex system
with different, interacting  resonances could be treated
in a separate paper, we present here a basic analysis
of the dynamical architecture of the system. From studies
of the Solar System, we know that, in the absence of mean
motion resonances, the variations in the orbital elements
of the planets are governed by the so-called secular
equations. These equations are obtained after averaging
over the mean longitudes of the planets. Since the
involved eccentricities for GJ~667C are small, the
secular system can be limited here to its linear version,
which is usually called a Laplace-Lagrange solution (for
details see \citet{laskar:1990}). Basically, the solution
is obtained from a transformation of the complex
variables $z_k=e_k \mathrm e^{\imath\varpi_k}$ into the
proper modes $u_k$. Here, $e_k$ are the eccentricities and
$\varpi_k$ the longitudes of the periastron of planet
$k=b,c,\ldots, g$. The proper modes $u_k$ determine the
secular variation of the eccentricities and are given by
$u_k\approx \mathrm e^{i(g_k t + \phi_k)}$.

\begin{figure}
\includegraphics{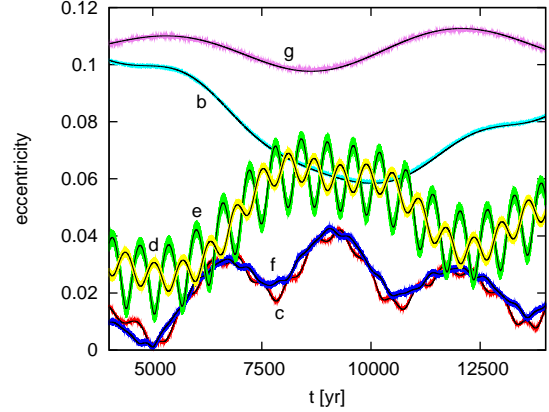}

\caption{Evolution of the eccentricities of solution S$_6$.
Colored lines give  the eccentricity as obtained from a
numerical integration. The thin black lines show the
eccentricity of the respective planet as given by the linear,
secular approximation. Close to each line we give the name
of the corresponding planet.}

\label{fig:secSys}
\end{figure}

Since the transformation into the proper modes depends only
on the masses and semi-major axes of the planets, the secular
frequencies will not change much for the different stable
configurations found in Figure \ref{fig:stab6p}. Here we use
solution S$_6$ to obtain numerically the parameters of the
linear transformation by a frequency analysis of the
numerically integrated orbit. The secular frequencies $g_k$
and the phases $\phi_k$ are given in Table \ref{tab:1}. How
well the secular solution describes the long-term evolution
of the eccentricities can be readily seen in Figure
\ref{fig:secSys}.

\begin{table}[ht]
\caption{\label{tab:1}Fundamental secular frequencies $g_k$, phases $\phi_k$ and
corresponding periods of the six-planet solution.}
\centering
\begin{tabular}{ccrr}
\hline \hline
 $k$ &   $g_k$ & $\phi_k$ &   Period \\
   & [deg/yr]  & [deg]    &   [yr]   \\ \hline
1  & 0.071683  & 356.41 & 5022.09 \\
2  & 0.184816  & 224.04 & 1947.88 \\
3  & 0.991167  & 116.46 &  363.21 \\
4  & 0.050200  &  33.63 & 7171.37 \\
5  & 0.656733  & 135.52 &  548.17 \\
6  & 0.012230  & 340.44 & 29435.80 \\
\hline \hline
\end{tabular}
\end{table}

From Figure \ref{fig:secSys}, it is easy to see that there
exists a strong secular coupling between all the inner
planets. From the Laplace-Lagrange solution, we find that the
long-term variation of the eccentricities of these planets
is determined by the secular frequency $g_1-g_4$ with a
period of $\approx 17000$ years. Here, the variation in
eccentricity of planet b is in anti-phase to that of planets
c to f due to the exchange of orbital angular momentum. On
shorter time scales, we easily spot in Figure
\ref{fig:secSys} the coupling between planets d and e with a
period of $\approx 600$ years ($g_1-g_5$), while the
eccentricities of planets c and f vary with a period of
almost 3000 years ($g_1+g_4$). Such couplings are already
known to prevent close approaches between the planets
\citep{ferraz-mello:2006}. As a result, the periastron of
the planets are locked and the difference $\Delta\varpi$
between any of their $\varpi$ librates around zero.

Although the eccentricities show strong variations, these
changes are very regular and their maximum values remain
bounded. From the facts that 1) the secular solution
agrees so well with numerically integrated orbits, and 2)
at the same time the semi-major axes remain nearly
constant (Table \ref{tab:2}), we can conclude that S$_6$
is not affected by strong MMRs.

Nevertheless, MMRs that can destabilize the whole system
are within the credibility intervals allowed by the
samplings and not far away from the most stable orbits.
Integrating some of the initial conditions marked as
chaotic in Figure \ref{fig:stab6p} one finds that, for
example, planets d and g are in some of these cases
temporarily trapped inside a 3:1 MMR, causing subsequent
disintegration of the system.

\begin{table}[ht]
\caption{\label{tab:2}Minimum and maximum values of the semi-major axes and eccentricities
during a run of S$_6$ over 10 Myr.}
\centering
\begin{tabular}{crrrr}
\hline \hline
$k$ & $a_\mathrm{min}$  & $a_\mathrm{max}$ & $e_\mathrm{min}$  &
$e_\mathrm{max}$ \\ \hline
b  & 0.050431 & 0.050433 & 0.035 & 0.114 \\
c  & 0.125012 & 0.125135 & 0.000 & 0.060 \\
f  & 0.155582 & 0.155922 & 0.000 & 0.061 \\
e  & 0.212219 & 0.212927 & 0.000 & 0.106 \\
d  & 0.275494 & 0.276104 & 0.000 & 0.087 \\
g  & 0.538401 & 0.539456 & 0.098 & 0.116 \\
\hline \hline
\end{tabular}
\end{table}

\subsection{Including planet h} \label{sec:dynamics_seven}

After finding a non-negligible set of stable six-planet
solutions, it is tempting to look for more planets in
the system. From the data analysis, one even knows the
preferred location of such a planet. We first considered
doing an analysis similar to the one for the six-planet case
using the Bayesian samples for the seven-planet solution. As
shown in previous sections, the subset of stable
solutions found by this approach is already small compared
to the statistical samples in the six-planet case ($\sim
0.3\%$). Adding one more planet (five extra dimensions) can
only shrink the relative volume of stable
solutions further. Given the large uncertainties on the orbital
elements of h, we considered this approach too
computationally expensive and inefficient.

As a first approximation to the problem, we checked
whether the distances between neighboring planets are
still large enough to allow stable motion. In
\citet{chambers:1996} the mean lifetime for coplanar
systems with small eccentricities is estimated as a
function of the mutual distance between the planets,
their masses and the number of planets in the system.
From their results, we can estimate the expected
lifetime for the seven-planet solution to be at least
$10^8$ years.

\begin{figure} \centering
\includegraphics[width=0.50\textwidth,clip]{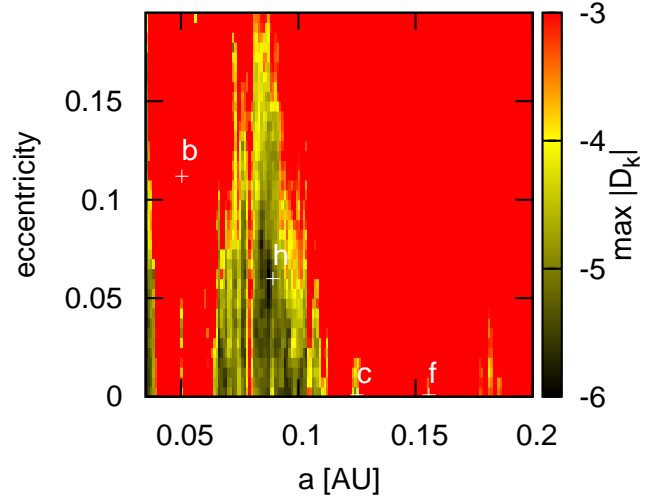}

\caption{Stability plot of the possible location of a 7th
planet in the stable S$_6$ solution (Table 5). We
investigate the stability of an additional planet with
1.1 Earth masses around the location found by the
Bayesian analysis. For these integrations, we varied the
semi-major axis and eccentricity of the putative planet
on a regular grid. The orbital angles $\omega$ and $M_0$
were set to the values of the statistically preferred
solution, while the inclination was fixed to zero. The
nominal positions of the planets as given in Table \ref{tab:mostStable}
are marked with white crosses. }

\label{fig:grid} \end{figure}

Motivated by this result, we explored
the phase space around the proposed orbit for the
seventh planet. To do this, we use solution S$_6$ and
placed a fictitious planet with 1.1 M$_\oplus$  (the
estimated mass of planet h as given in Table
\ref{tab:parameters}) in the semi-major axis range
between 0.035 and 0.2 AU (step size of 0.001 AU)
varying the eccentricity between 0 and 0.2 (step size
of 0.01). The orbital angles $\omega$ and M$_0$ were
set to the values of the statistically preferred
solution for h (see Table \ref{tab:parameters}). For
each of these initial configurations, we integrated the
system for 10$^4$ orbits of planet g and analyzed
stability of the orbits using the same secular
frequency analysis. As a result, we obtained a value
of the chaos index $D$ at each grid point. Figure
\ref{fig:grid} shows that the putative orbit of h
appears right in the middle of the only island of
stability left in the inner region of the system. By
direct numerical integration of solution S$_6$
together with planet h at its nominal position, we
found that such a solution is also stable on Myr
timescales. With this we conclude that the seventh
signal detected by the Bayesian analysis also belongs
to a physically viable planet that might be confirmed
with a few more observations.

\subsection{An upper limit for the masses} \label{sec:upperlimits}

Due to the lack of reported transit, only the minimum
masses  are known for the planet candidates. The true
masses depend on the unknown inclination $i$ of the
system to the line-of-sight. In all the analysis
presented above, we implicitly assume that the GJ~667C
system is observed edge-on ($i=90^\circ$) and that all true
masses are equal to the minimum mass M$\sin{i}$. As
shown in the discussion on  the dynamics, the
stability of the system is fragile in the sense that
dynamically unstable configurations can be found close
in the parameter space to the stable ones. Therefore,
it is likely that a more complete analysis could set
strong limitations on the maximum masses allowed to
each companion. An exploration of the total phase
space including mutual inclinations would require too
much computational resources and is beyond the scope
of this paper. To obtain a basic understanding of the
situation, we only search for a constraint on the
maximum masses of the S$_6$ solution assuming
co-planarity. Decreasing the inclination of the
orbital plane in steps of 10$^\circ$, we applied the
frequency analysis to the resulting system. By making
the planets more massive, the interactions between them
become stronger, with a corresponding shrinking of the
areas of stability. In this experiment, we found that
the system remained stable for at least one Myr for an
inclination down to $i=30^\circ$.  If this result can
be validated by a much more extensive exploration of
the dynamics and longer integration times (in prep.),
it would confirm that the masses of all the candidates
are within a factor of 2 of the minimum masses derived
from Doppler data. Accordingly, c,f and e would
be the first dynamically confirmed super-Earths (true
masses below 10 M$_\oplus$) in the habitable zone of a
nearby star.

\section{Habitability} \label{sec:habitability}

Planets h--d receive 20--200\% of the Earth's current insolation, and
hence should be evaluated in terms of potential
habitability. Traditionally, analyses of planetary habitability begin
with determining if a planet is in the habitable zone \citep{Dole64,
  Hart79, Kasting93, selsis:2007, Kopparapu13}, but many factors are
relevant. Unfortunately, many aspects cannot presently be determined
due to the limited characterization derivable from RV observations.
However, we can explore the issues quantitatively and identify those
situations in which habitability is precluded, and hence determine
which of these planets {\it could} support life. In this
section we provide a preliminary analysis of each potentially
habitable planet in the context of previous results, bearing in mind
that theoretical predictions of the most relevant processes cannot be constrained by existing data.

\subsection{The Habitable Zone}\label{sec:hz}

The HZ is defined at the inner edge by the onset of a
``moist greenhouse,'' and at the outer edge by the
``maximum greenhouse'' \citep{Kasting93}. Both of these
definitions assume that liquid surface water is
maintained under an Earth-like atmosphere. At the inner
edge, the temperature near the surface becomes large
enough that water cannot be confined to the surface and
water vapor saturates the stratosphere. From there,
stellar radiation can dissociate the water and hydrogen
can escape. Moreover, as water vapor is a greenhouse gas,
large quantities in the atmosphere can heat the surface
to temperatures that forbid the liquid phase, rendering
the planet uninhabitable. At the outer edge, the danger
is global ice coverage. While greenhouse gases like
CO$_2$ can warm the surface and mitigate the risk of
global glaciation, CO$_2$ also scatters starlight via
Rayleigh scattering. There is therefore a limit
to the amount of CO$_2$ that can warm a planet as more
CO$_2$ actually cools the planet by increasing its
albedo, assuming a moist or runaway greenhouse was never
triggered.

We use the most recent calculations of the HZ
\citep{Kopparapu13} and find, for a 1 Earth-mass planet, that
the inner and outer boundaries of the habitable zone for GJ~667C
lie between 0.095--0.126 AU and 0.241--0.251 AU respectively. We
will adopt the average of these limits as a working definition
of the HZ: 0.111 -- 0.246\,AU. At the inner edge, larger mass
planets resist the moist greenhouse and the HZ edge is closer
in, but the outer edge is almost independent of mass.
\cite{Kopparapu13} find that a 10~$M_\oplus$ planet can be
habitable 5\% closer to the star than a 1~$M_\oplus$ planet.
However, we must bear in mind that the HZ calculations are based
on 1-dimensional photochemical models that may not apply to
slowly rotating planets, a situation likely for planets c, d, e,
f and h (see below).

From these definitions, we find that planet candidate h ($a =
0.0893$~AU) is too hot to be habitable, but we note its
semi-major axis is consistent with the most optimistic
version of the HZ. Planet c ($a = 0.125$~AU) is close to the
inner edge but is likely to be in the HZ, especially since it
has a large mass. Planets f and e are firmly in the HZ.
Planet d is likely beyond the outer edge of the HZ, but the
uncertainty in its orbit prevents a definitive assessment.
Thus, we conclude that planets c, f, and e are in the HZ, and
planet d might be, \ie there up to four potentially habitable
planets orbiting GJ~667C.

Recently, \cite{Abe11} pointed out that planets with
small, but non-negligible, amounts of water have a larger
HZ than Earth-like planets. From their definition, both h
and d are firmly in the HZ. However, as we discuss below,
these planets are likely water-rich, and hence we do not
use the \cite{Abe11} HZ here.

\subsection{Composition}\label{sec:composition}

Planet formation is a messy process characterized by
scattering, migration, and giant impacts. Hence precise
calculations of planetary composition are presently
impossible, but see \cite{Bond10,Bond12} for some general
trends. For habitability, our first concern is discerning
if a planet is rocky (and potentially habitable) or
gaseous (and uninhabitable). Unfortunately, we cannot
even make this rudimentary evaluation based on available
data and theory. Without radii measurements, we cannot
determine bulk density, which could discriminate between
the two. The least massive planet known to be gaseous is
GJ~1214~b at 6.55~$M_\oplus$~\citep{Charbonneau09}, and
the largest planet known to be rocky is Kepler-10~b at
4.5~$M_\oplus$~\citep{Batalha11}. Modeling of gas
accretion has found that planets smaller than
1~$M_\oplus$ can accrete hydrogen in certain
circumstances \citep{Ikoma01}, but the critical mass is
likely larger \citep{Lissauer09}. The planets in this
system lie near these masses, and hence we cannot
definitively say if any of these planets are gaseous.

Models of rocky planet formation around M dwarfs have found that those
that accrete from primordial material are likely to be
sub-Earth mass~\citep{Raymond07} and
volatile-poor~\citep{Lissauer07}. In contrast, the planets orbiting
GJ~667C are super-Earths in a very packed configuration summing up to
$>25~M_\oplus$ inside 0.5~AU. Therefore, the planets either
formed at larger orbital distances and migrated in
\citep[\eg][]{Lin96}, or additional dust and ice flowed inward
during the protoplanetary disk phase and accumulated into the
planets \cite{HansenMurray12,HansenMurray13}. The large masses
disfavor the first scenario, and we therefore assume that the
planets formed from material that condensed beyond the
snow-line and are volatile rich. If not gaseous, these planets contain
substantial water content, which is a primary requirement for life
(and negates the dry-world HZ discussed above). In conclusion, these
planets could be terrestrial-like with significant water content and
hence are potentially habitable.

\subsection{Stellar Activity and habitability}
\label{sec:hz_activity}

Stellar activity can be detrimental to life as the
planets can be bathed in high energy photons and
protons that could strip the atmosphere or destroy
ozone layers. In quiescence, M dwarfs emit very little
UV light, so the latter is only dangerous if flares
occur frequently enough that ozone does not have time
to be replenished \citep{Segura10}. As already
discussed in Section \ref{sec:starparam}, GJ~667C
appears to be relatively inactive (indeed, we would
not have been able to detect planetary signals
otherwise), and so the threat to life is small
today. If the star was very active in its youth- with
mega-flares like those on the equal mass star AD Leo
\citep{HawleyPettersen91}- any life on  the surface of
planets might have been difficult during those days
\citep{Segura10}. While M dwarfs are likely to be
active for much longer time than the Sun
\citep{west08, reiners_mohanty:2012}, GJ~667C is not
active today and there is no reason to assume that
life could not form after an early phase of strong
stellar activity.

\subsection{Tidal Effects} \label{sec:tidal}

\begin{table*}
\center
\begin{tabular}{c|cccc|cccc}
\hline
\hline
 & \multicolumn{4}{c|}{CPL} & \multicolumn{4}{c}{CTL}\\
\hline
 & \multicolumn{2}{c|}{base} & \multicolumn{2}{c|}{max} &
 \multicolumn{2}{c|}{base} & \multicolumn{2}{c}{max}\\
\hline
 & $t_{lock}$ & $t_{ero}$ & $t_{lock}$ & $t_{ero}$ & $t_{lock}$ &
 $t_{ero}$ & $t_{lock}$ & $t_{ero}$\\
\hline
h & 0.07 & 0.08 & 18.2 & 20.4 & 0.55 & 0.77 & 66.9 & 103\\
c & 0.62 & 0.69 & 177 & 190 & 4.7 & 8.1 & 704 & 1062\\
f & 2.2 & 2.3 & 626 & 660 & 18.5 & 30.1 & 2670 & 3902\\
e & 14.2 & 15.0 & 4082 & 4226 & 129 & 210 & $>10^4$ & $>10^4$\\
d & 70.4 & 73 & $>10^4$ & $>10^4$ & 692 & 1094 & $>10^4$ & $>10^4$\\
\hline\hline
\end{tabular}
\caption{Timescales for the planets' tidal despinning in units of Myr. ``CPL''
denotes the constant-phase-lag model of \citet{FerrazMello08}, ``CTL'' labels
the constant-time-lag model of \citet{Leconte10}.}\label{tab:erosion}
\end{table*}

\begin{figure*}
\center
\includegraphics[width=0.9\textwidth,clip]{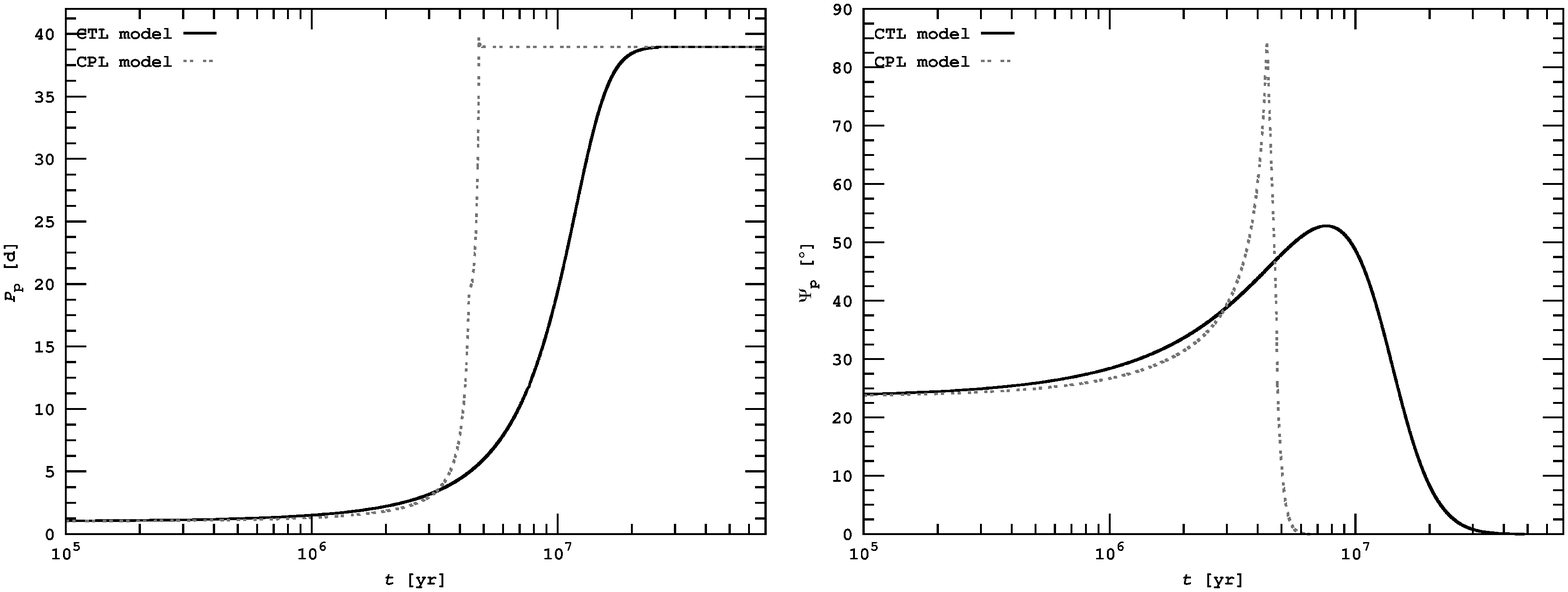}
\caption{
Tidal evolution of the spin properties of planet GJ\,667C\,f. Solid lines depict
predictions from constant-time-lag theory (``CTL''), while dashed lines
illustrate those from a constant-phase-lag model (``CPL''). All tracks assume a
scenario similar to the ``base'' configuration (see text and
Table~\ref{tab:erosion}). \textit{Left}: Despinning for an assumed initial
rotation period of one day. The CPL model yields tidal locking in less than
$5$\,Myr, and CTL theory predicts about $20$\,Myr for tidal locking.
\textit{Right}: Tilt erosion of an assumed initial Earth-like obliquity of
$23.5^\circ$. Time scales for both CPL and CTL models are similar to the locking
time scales.}
\label{fig:tilt}
\end{figure*}

Planets in the HZ of low mass stars may be subject to
strong tidal distortion, which can lead to long-term
changes in orbital and spin properties
\citep{Dole64,Kasting93,Barnes08,Heller11}, and tidal
heating \citep{Jackson08_hab,Barnes09,Barnes12}. Both of
these processes can affect habitability, so we now
consider tidal effects on planets c, d, e, f and h.

Tides will first spin-lock the planetary spin and
drive the obliquity to either 0 or $\pi$. The
timescale for these processes is a complex function of
orbits, masses, radii and spins, \citep[see \eg
][]{Darwin1880,Hut81,FerrazMello08,Leconte10} but for
super-Earths in the HZ of a $\sim 0.3~M_\odot$ star,
\cite{Heller11} found that tidal locking should occur
in $10^6$--$10^9$ years. We have used both the
constant-time-lag and constant-phase-lag tidal models
described in \citet{Heller11} and \citet{Barnes12}
\citep[see also][]{FerrazMello08,Leconte10},  to
calculate how long tidal locking could take for these
planets. We consider two possibilities. Our baseline
case is very similar to that of \citet{Heller11} in
which the planets initially have Earth-like
properties: a 1-day rotation period, an obliquity of
23.5$^\circ$ and the current tidal dissipation of the
Earth (a tidal $Q$ of 12-\cite{Yoder95} or time lag
of 638~s-\cite{Lambeck77,NeronDeSurgyLaskar97}). We
also consider an extreme, but plausible, case that
maximizes the timescale for tidal locking: 8-hour
rotation period, obliquity of 89.9$^\circ$ and a tidal
$Q$ of 1000 or time lag of 6.5~s. In
Table~\ref{tab:erosion} we show the time for the
obliquity to erode to $1^\circ$, $t_\mathrm{ero}$, and
the time to reach the pseudo-synchronous rotation
period, $t_\mathrm{lock}$.

In Figure \ref{fig:tilt}, we depict the tidal
evolution of the rotation period (left panel) and
obliquity (right panel) for planet f as an example.
The assumed initial configuration is similar to the
``base'' scenario. Time scales for rotational
locking and tilt erosion are similar to those shown
in Table~\ref{tab:erosion}.\footnote{Note that
evolution for the CPL model is faster with our
parameterization. In the case of GJ~581\,d, shown in
\citet{Heller11}, the planet was assumed to be less
dissipative in the CPL model ($Q_\mathrm{p}=100$)
and evolution in the CPL model was slower.}

As these planets are on nearly circular orbits, we
expect tidally-locked planets to be synchronously
rotating, although perhaps when the eccentricity is
relatively large pseudo-synchronous rotation could
occur \citep{Goldreich66, MurrayDermott99, Barnes08,
Correia08, FerrazMello08, makarov:2013}. From Table
\ref{tab:erosion} we see that all the planets h--f
are very likely synchronous rotators, planet e is
likely to be, but planet d is uncertain. Should
these planets have tenuous atmospheres ($<0.3$~bar),
then they may not support a habitable surface
\citep{Joshi97}. Considerable work has shown that
thicker atmospheres are able to redistribute dayside
heat to produce clement conditions
\citep{Joshi97,Joshi03,Edson11,Pierrehumbert11,Wordsworth11}.
As we have no atmospheric data, we assert that
tidal locking does not preclude habitability for any
of the HZ planets.

During the tidal despinning, tidal heat can be generated as
dissipation transforms rotational energy into frictional heat. In some
cases, the heating rates can be large enough to trigger a runaway
greenhouse and render a planet uninhabitable \citep{Barnes12}. Tidal
heating is maximized for large radius planets that rotate quickly and
at high obliquity. Using the \citet{Leconte10} model and the Earth's
dissipation, we find that tidal heating of the HZ planets will be
negligible for most cases. Consider an extreme version of planet h,
which is just interior to the HZ. Suppose it has the same tidal
dissipation as the Earth (which is the most dissipative body known), a
rotation period of 10~hr, an eccentricity of 0.1, and an obliquity of
$80^\circ$. The \cite{Leconte10} model predicts such a planet would
have a tidal heat flux of nearly 4000~\wpmsq. However, that model also
predicts the flux would drop to only 0.16~\wpmsq~in just $10^6$
years. The timescale for a runaway greenhouse to sterilize a planet is
on the order of $10^8$ years \citep{Watson81,Barnes12}, so this burst
of tidal heating does not forbid habitability.

\begin{figure*}
\center
\includegraphics[width=0.9\textwidth,clip]{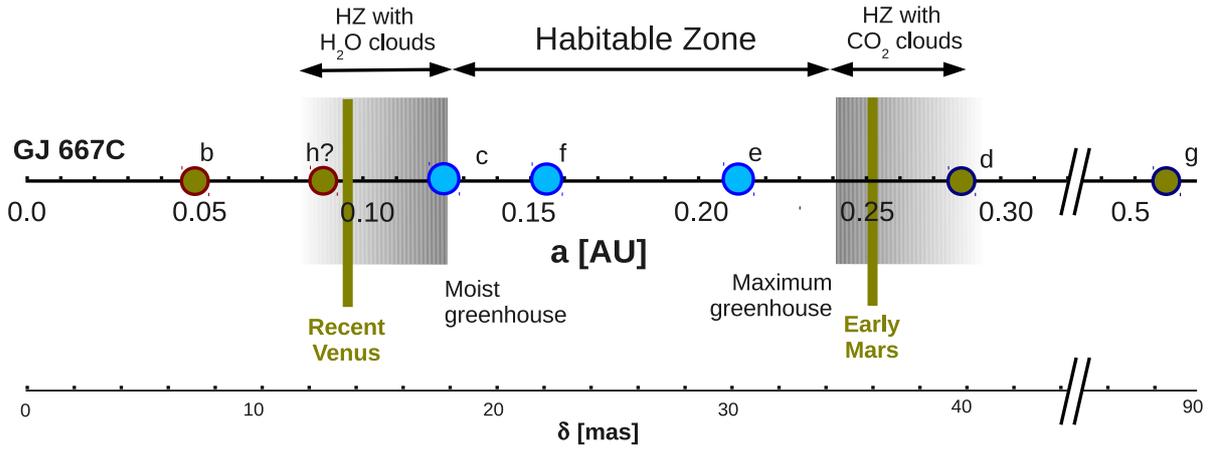}

\caption{Liquid water habitable zone of GJ~667C with the
proposed seven candidates as estimated using the updated
relations in \citet{Kopparapu13}. Three of the reported
planets lie within the HZ. The newly reported planets f
and e are the most comfortably located within it. The
inner edge is set by the moist greenhouse runaway limit
and the outer edge is set by the snow ball runaway
limit. The empirical limits set by a recent uninhabitable
Venus and an early habitable Mars are marked in brown
solid lines. The presence of clouds of water (inner edge)
or CO$_2$ (outer edge) might allow habitable conditions
on planets slightly outside the nominal definition of the
habitable zone \citep{selsis:2007}.
}\label{fig:hz}
\end{figure*}

After tidal locking, the planet would still have about
0.14~\wpmsq~of tidal heating due to the eccentricity
(which, as for the other candidates, can oscillate between
0 and 0.1 due to dynamical interactions). If we
assume an Earth-like planet, then about 90\% of that
heat is generated in the oceans, and 10\% in the rocky
interior. Such a planet would have churning oceans,
and about 0.01~\wpmsq~of tidal heat flux from the
rocky interior. This number should be compared to
0.08~\wpmsq, the heat flux on the Earth due entirely
to other sources. As $e=0.1$ is near the maximum of
the secular cycle, see $\S$~\ref{sec:dynamics}, the
actual heat flux is probably much lower. We conclude
that tidal heating on planet h is likely to be
negligible, with the possibility it could be a minor
contributor to the internal energy budget. As the
other planets are more distant, the tidal heating of
those planets is negligible. The CPL predicts higher
heating rates and planet c could receive $\sim
0.01$~\wpmsq~of internal heating, but otherwise tidal
heating does not affect the HZ planets.

\subsection{The Weather Forecast}\label{sec:weather}

Assuming planets c, f and e have habitable surfaces (see Figure
\ref{fig:hz}), what might their climates be like? To first order we
expect a planet's surface temperature to be cooler as semi-major axis
increases because the incident flux falls off with distance squared.
However, albedo variations can supersede this trend, \eg a closer,
high-albedo planet could absorb less energy than a more distant
low-albedo planet. Furthermore, molecules in the atmosphere can trap
photons near the surface via the greenhouse effect, or scatter stellar
light via Rayleigh scattering, an anti-greenhouse effect. For example,
the equilibrium temperature of Venus is actually lower than the Earth's
due to the former's large albedo, yet the surface temperature of Venus
is much larger than the Earth's due to the greenhouse effect. Here, we
speculate on the climates of each HZ planet based on the our current
understanding of the range of possible climates that HZ planets might
have.

Certain aspects of each planet will be determined by
the redder spectral energy distribution of the host
star. For example, the ``stratosphere'' is expected
to be isothermal as there is negligible UV radiation
\citep{Segura03}. On Earth, the UV light absorbed by
ozone creates a temperature inversion that
delineates the stratosphere. HZ calculations also
assume the albedo of planets orbiting cooler stars
are lower than the Earth's because Rayleigh
scattering is less effective for longer wavelengths,
and because the peak emission in the stellar spectrum
is close to several H$_2$O and CO$_2$ absorption bands
in the near infrared. Therefore, relative to the
Earth's insolation, the HZ is farther from the star.
In other words, if we placed the Earth in orbit
around an M dwarf such that it received the same incident
radiation as the modern Earth, the M dwarf planet would
be hotter as it would have a lower albedo.
 The different character of the light can also impact
plant life, and we might expect less productive
photosynthesis \citep{Kiang07}, perhaps relying on
pigments such as chlorophyll d \citep{Mielke13} or
chlorophyll f \citep{Chen10}.

Planet c is slightly closer to the inner edge of the HZ
than the Earth, and so we expect it to be warmer than the
Earth, too. It receives 1230~\wpmsq~ of stellar
radiation, which is actually less than the Earth's solar
constant of 1360~\wpmsq. Assuming synchronous rotation
and no obliquity, then the global climate depends
strongly on the properties of the atmosphere. If the
atmosphere is thin, then the heat absorbed at the
sub-stellar point cannot be easily transported to the
dark side or the poles. The surface temperature would be
a strong function of the zenith angle of the host star GJ~667C.
For thicker atmospheres, heat redistribution
becomes more significant. With a rotation period
of $\sim$ 28 days, the planet is likely to have Hadley cells
that extend to the poles (at least if Titan, with a
similar rotation period, is a guide), and hence jet
streams and deserts would be unlikely. The location of
land masses is also important. Should land be
concentrated near the sub-stellar point, then silicate
weathering is more effective, and cools the planet by
drawing down CO$_2$ \citep{Edson12}.

Planet f is a prime candidate for habitability and
receives 788~\wpmsq~ of radiation. It likely absorbs less
energy than the Earth, and hence habitability requires
more greenhouse gases, like CO$_2$ or CH$_4$. Therefore a
habitable version of this planet has to have a thicker
atmosphere than the Earth, and we can assume a relatively
uniform surface temperature. Another possibility is an
``eyeball'' world in which the planet is synchronously
rotating and ice-covered except for open ocean at the
sub-stellar point \citep{Pierrehumbert11}. On the other
hand, the lower albedo of ice in the IR may make
near-global ice coverage difficult
\citep{JoshiHaberle12,Shields13}.

Planet e receives only a third the radiation the Earth
does, and lies close to the maximum greenhouse limit. We
therefore expect a habitable version of this planet to
have $>2$ bars of CO$_2$. The planet might not be tidally
locked, and may have an obliquity that evolves
significantly due to perturbations from other planets.
From this perspective planet e might be the most
Earth-like, experiencing a day-night cycle and seasons.

Finally planet d is unlikely to be in the habitable
zone, but it could potentially support sub-surface
life. Internal energy generated by, \eg, radiogenic
heat could support liquid water below an ice layer,
similar to Europa. Presumably the biologically
generated gases could find their way through the ice
and become detectable bio-signatures, but they might
be a very small constituent of a large atmosphere,
hampering  remote detection. While its transit
probability is rather  low ($\sim$0.5\%), its apparent
angular separation from the star is $\sim$ 40
milliarcseconds. This value is the baseline inner
working angle for the Darwin/ESA high-contrast mission
being considered by ESA \citep{darwin} so planet d
could be a primary target for such a mission.

\subsection{Moons in the habitable zone of GJ 667C}
\label{sec:moons}

In addition to planets, extrasolar moons have been
suggested as hosts for life \citep{1987AdSpR...7..125R,
1997Natur.385..234W, tinney:2011, 2013AsBio..13...18H}. In
order to sustain a substantial, long-lived atmosphere under
an Earth-like amount of stellar irradiation
\citep{1997Natur.385..234W,2000ESASP.462..199K}, to drive a
magnetic field over billions of years
\citep{2011ApJ...726...70T}, and to drive tectonic
activity, \citet{2013AsBio..13...18H} concluded that a
satellite in the stellar HZ needs a mass
$\gtrsim~0.25\,M_\oplus$ in order to maintain liquid
surface water.

If potential moons around planets GJ 667C\,c, f, or e formed
in the circumplanetary disk, then they will be much less
massive than the most massive satellites in the Solar System
\citep{2006Natur.441..834C} and thus not be habitable.
However, if one of those planets is indeed terrestrial then
impacts could have created a massive moon as happened on
Earth \citep{1976LPI.....7..120C}. Further possibilities for
the formation of massive satellites are summarized in
\citet[][Sect. 2.1]{2013AsBio..13...18H}.

As the stellar HZ of GJ 667C is very close to this M dwarf
star, moons of planets in the habitable zone would have
slightly eccentric orbits due to stellar perturbations.
These perturbations induce tidal heating and they could be
strong enough to prevent any moon from being habitable
\citep{2012A&A...545L...8H}. Moons around planet d, which
orbits slightly outside the stellar HZ, could offer a more
benign environment to life than the planet itself, if they
experience weak tidal heating of, say, a few watts per
square meter \citep[see Jupiter's moon
Io, ][]{1987AdSpR...7..125R,2000Sci...288.1198S}.

Unless some of these planets are found to transit, there is
no currently available technique to identify satellites
\citep{2009MNRAS.392..181K,2012ApJ...750..115K}. The RV
technique is only sensitive to the combined mass of a planet
plus its satellites so it might be possible that some of the
planets could be somewhat lighter-- but host a massive moon.

\section{Conclusions} \label{sec:conclusions}

We describe and report the statistical methods and tests used
to detect up to seven planet candidates around GJ~667C using
Doppler spectroscopy. The detection of the first five planets
is very robust and independent of any prior choice. In
addition to the first two already reported ones \citep[b and
c][]{anglada:2012a, delfosse:2012} we show that the third
planet also proposed in those papers (planet d) is much
better explained by a Keplerian orbit rather than an
activity-induced periodicity. The next two confidently
detected signals (e and f) both correspond to small
super-Earth mass objects with most likely periods of 62 and
39 days. The detection of the 6th planet is weakly dependent
on the prior choice of the orbital eccentricity. The
statistical evidence for the 7th candidate (planet h) is
tentative and requires further Doppler follow-up for
confirmation. \citet{gregory:2012} proposed a
solution for the system with similar characteristics to the
one we present here but had fundamental differences. In
particular, he also identified the first five stronger signals
but his six-planet solution also included a candidate
periodicity at 30 days- which would be dynamically unstable-
and activity was associated to the signal at 53 days
without further discussion or verification. The difference in our
conclusions are due to a slightly different choice of priors
(especially on the eccentricity), more data was used in our
analysis -only HARPS-CCF data was used by \citet{gregory:2012}-,
and we performed a more thorough investigation of possible
activity-related periodicities.

Numerical integration of orbits compatible with the posterior
density distributions show that there is a subset of
configurations that allow long-term stable configurations.
Except for planets b and g, the condition of dynamical
stability dramatically affects the distribution of allowed
eccentricities indicating that the lower mass planet
candidates (c, e, f) must have quasi-circular orbits. A
system of six planets is rather complex in terms of stabilizing
mechanisms and possible mean-motion resonances. Nonetheless,
we identified that the physically allowed configurations are
those that avoid transient 3:1 MMR between planets d and g.
We also found that the most stable orbital solutions are well
described by the theory of secular frequencies
(Laplace-Lagrange solution). We investigated if the inclusion
of a seventh planet system was dynamically feasible in the
region disclosed by the Bayesian samplings. It is notable
that this preliminary candidate appears around the center of
the region of stability. Additional data should be able to
confirm this signal and provide detectability for longer
period signals.

The closely packed dynamics keeps the eccentricities small
but non-negligible for the lifetime of the system. As a
result, potential habitability of the candidates must account
for tidal dissipation effects among others. Dynamics
essentially affect 1) the total energy budget at the surface
of the planet (tidal heating), 2) synchronization of the
rotation with the orbit (tidal locking), and 3) the timescales
for the erosion of their obliquities. These dynamical
constraints, as well as predictions for potentially habitable
super-Earths around M dwarf stars, suggest that at least
three planet candidates (planets c, e and f) could have
remained habitable for the current life-span of the star.
Assuming a rocky composition, planet d lies slightly outside
the cold edge of the stellar HZ. Still, given the
uncertainties in the planet parameters and in the assumptions
in the climatic models, its potential habitability cannot be
ruled out (e.g., ocean of liquid water under a thick ice
crust, or presence of some strong green-house effect gas).

One of the main results of the Kepler mission is that
high-multiplicity systems of dynamically-packed
super-Earths are quite common around G and K dwarfs
\citep{fabrycky2012}. The putative existence of these kinds
of compact systems around M-dwarfs, combined with a
closer-in habitable zone, suggests the existence of a
numerous population of planetary systems with several
potentially-habitable worlds each. GJ~667C is likely to be
among first of many of such systems that may be discovered
in the coming years.

\begin{acknowledgements}
We acknowledge the constructive discussions with the referees
of this manuscript. The robustness and confidence of the
result greatly improved thanks to such discussions.
G.~Anglada-Escud\'e is supported by the German Federal Ministry of
Education and Research under 05A11MG3.
M.~Tuomi acknowledges D. Pinfield and RoPACS (Rocky Planets Around
Cool Stars), a Marie Curie Initial Training Network
funded by the European Commission's Seventh
Framework Programme.
E.~Gerlach would like to
acknowledge the financial support from the DFG
research unit FOR584.
R.~Barnes is supported by NASA's Virtual Planetary Laboratory
under Cooperative Agreement Number NNH05ZDA001C and
NSF grant AST-1108882.
R.~Heller receives funding from the
Deutsche Forschungsgemeinschaft (reference number
scho394/29-1).
J.S.~Jenkins also acknowledges funding by Fondecyt through
grant 3110004 and partial support from CATA
(PB06, Conicyt), the GEMINI-CONICYT FUND and from
the Comit\'e Mixto ESO-GOBIERNO DE CHILE.
S.~Weende acknowledges DFG funding by SFB-963 and the
GrK-1351
A.~Reiners acknowledges research funding from DFG
grant RE1664/9-1.
S.S.~Vogt gratefully acknowledges support
from NSF grant AST-0307493.
This study contains data obtained from the ESO Science Archive
Facility under request number ANGLADA36104.
%
%
We also acknowledge the efforts of the PFS/Magellan
team in obtaining Doppler measurements.
We thank Sandy Keiser for her efficient setup of
the computer network at Carnegie/DTM. We thank
Dan Fabrycky, Aviv Ofir, Mathias Zechmeister and
Denis Shulyak for useful and constructive discussions.
This research made use of the Magny Cours Cluster
hosted by the GWDG, which is managed by
Georg August University G\"ottingen and the Max
Planck Society.
This research has made extensive use of the SIMBAD database,
operated at CDS, Strasbourg, France; and NASA's Astrophysics 
Data System.
The authors acknowledge the significant
efforts of the HARPS-ESO team in improving the
instrument and its data reduction software that made
this work possible.
We also acknowledge the efforts of the teams and individual 
observers that have been involved in observing the target star 
with HARPS/ESO, HIRES/Keck, PFS/Magellan and UVES/ESO.
\end{acknowledgements}

\bibliography{biblio}

\begin{thebibliography}{115}
\expandafter\ifx\csname natexlab\endcsname\relax\def\natexlab#1{#1}\fi

\bibitem[{{Abe} {et~al.}(2011){Abe}, {Abe-Ouchi}, {Sleep}, \& {Zahnle}}]{Abe11}
{Abe}, Y., {Abe-Ouchi}, A., {Sleep}, N.~H., \& {Zahnle}, K.~J. 2011,
  Astrobiology, 11, 443

\bibitem[{{Anglada-Escud{\'e}} {et~al.}(2012){Anglada-Escud{\'e}}, {Arriagada},
  {Vogt}, {Rivera}, {Butler}, {Crane}, {Shectman}, {Thompson}, {Minniti},
  {Haghighipour}, {Carter}, {Tinney}, {Wittenmyer}, {Bailey}, {O'Toole},
  {Jones}, \& {Jenkins}}]{anglada:2012b}
{Anglada-Escud{\'e}}, G., {Arriagada}, P., {Vogt}, S.~S., {et~al.} 2012, \apjl,
  751, L16

\bibitem[{{Anglada-Escud{\'e}} \& {Butler}(2012)}]{anglada:2012a}
{Anglada-Escud{\'e}}, G. \& {Butler}, R.~P. 2012, \apjs, 200, 15

\bibitem[{{Anglada-Escud{\'e}} {et~al.}(2010){Anglada-Escud{\'e}},
  {L{\'o}pez-Morales}, \& {Chambers}}]{anglada:2010}
{Anglada-Escud{\'e}}, G., {L{\'o}pez-Morales}, M., \& {Chambers}, J.~E. 2010,
  \apj, 709, 168

\bibitem[{{Anglada-Escud{\'e}} \& {Tuomi}(2012)}]{gj676A:2012}
{Anglada-Escud{\'e}}, G. \& {Tuomi}, M. 2012, \aap, 548, A58

\bibitem[{{Baliunas} {et~al.}(1995){Baliunas}, {Donahue}, {Soon}, {Horne},
  {Frazer}, {Woodard-Eklund}, {Bradford}, {Rao}, {Wilson}, {Zhang}, {Bennett},
  {Briggs}, {Carroll}, {Duncan}, {Figueroa}, {Lanning}, {Misch}, {Mueller},
  {Noyes}, {Poppe}, {Porter}, {Robinson}, {Russell}, {Shelton}, {Soyumer},
  {Vaughan}, \& {Whitney}}]{baliunas:1995}
{Baliunas}, S.~L., {Donahue}, R.~A., {Soon}, W.~H., {et~al.} 1995, \apj, 438,
  269

\bibitem[{{Baluev}(2009)}]{baluev:2009}
{Baluev}, R.~V. 2009, \mnras, 393, 969

\bibitem[{{Baluev}(2012)}]{baluev:2012}
{Baluev}, R.~V. 2012, \mnras, 420

\bibitem[{{Barnes} {et~al.}(2009){Barnes}, {Jackson}, {Greenberg}, \&
  {Raymond}}]{Barnes09}
{Barnes}, R., {Jackson}, B., {Greenberg}, R., \& {Raymond}, S.~N. 2009, ApJ.,
  700, L30

\bibitem[{{Barnes} {et~al.}(2013){Barnes}, {Mullins}, {Goldblatt}, {Meadows},
  {Kasting}, \& {Heller}}]{Barnes12}
{Barnes}, R., {Mullins}, K., {Goldblatt}, C., {et~al.} 2013, Astrobiology, 13,
  225

\bibitem[{{Barnes} {et~al.}(2008){Barnes}, {Raymond}, {Jackson}, \&
  {Greenberg}}]{Barnes08}
{Barnes}, R., {Raymond}, S.~N., {Jackson}, B., \& {Greenberg}, R. 2008,
  Astrobiology, 8, 557

\bibitem[{{Batalha} {et~al.}(2011){Batalha}, {Borucki}, {Bryson}, {Buchhave},
  {Caldwell}, {Christensen-Dalsgaard}, {Ciardi}, {Dunham}, {Fressin},
  {Gautier}, {Gilliland}, {Haas}, {Howell}, {Jenkins}, {Kjeldsen}, {Koch},
  {Latham}, {Lissauer}, {Marcy}, {Rowe}, {Sasselov}, {Seager}, {Steffen},
  {Torres}, {Basri}, {Brown}, {Charbonneau}, {Christiansen}, {Clarke},
  {Cochran}, {Dupree}, {Fabrycky}, {Fischer}, {Ford}, {Fortney}, {Girouard},
  {Holman}, {Johnson}, {Isaacson}, {Klaus}, {Machalek}, {Moorehead},
  {Morehead}, {Ragozzine}, {Tenenbaum}, {Twicken}, {Quinn}, {VanCleve},
  {Walkowicz}, {Welsh}, {Devore}, \& {Gould}}]{Batalha11}
{Batalha}, N.~M., {Borucki}, W.~J., {Bryson}, S.~T., {et~al.} 2011, ApJ, 729,
  27

\bibitem[{{Bond} {et~al.}(2010){Bond}, {O'Brien}, \& {Lauretta}}]{Bond10}
{Bond}, J.~C., {O'Brien}, D.~P., \& {Lauretta}, D.~S. 2010, ApJ, 715, 1050

\bibitem[{{Bonfils}(2009)}]{bonfils:2009:conf}
{Bonfils}, X. 2009, in ESO-CAUP conference Series, ed. {N.~Santos}, Vol.~1, --

\bibitem[{{Bonfils} {et~al.}(2013){Bonfils}, {Delfosse}, {Udry}, {Forveille},
  {Mayor}, {Perrier}, {Bouchy}, {Gillon}, {Lovis}, {Pepe}, {Queloz}, {Santos},
  {S{\'e}gransan}, \& {Bertaux}}]{bonfils:2011}
{Bonfils}, X., {Delfosse}, X., {Udry}, S., {et~al.} 2013, \aap, 549, A109

\bibitem[{{Cameron} \& {Ward}(1976)}]{1976LPI.....7..120C}
{Cameron}, A.~G.~W. \& {Ward}, W.~R. 1976, in LPI Science Conference Abstracts,
  Vol.~7, 120

\bibitem[{{Canup} \& {Ward}(2006)}]{2006Natur.441..834C}
{Canup}, R.~M. \& {Ward}, W.~R. 2006, \nat, 441, 834

\bibitem[{{Carter-Bond} {et~al.}(2012){Carter-Bond}, {O'Brien}, \&
  {Raymond}}]{Bond12}
{Carter-Bond}, J.~C., {O'Brien}, D.~P., \& {Raymond}, S.~N. 2012, \apj, 760, 44

\bibitem[{{Cayrel de Strobel}(1981)}]{strobel:1981}
{Cayrel de Strobel}, G. 1981, Bulletin d'Information du Centre de Donnees
  Stellaires, 20, 28

\bibitem[{{Chambers} {et~al.}(1996){Chambers}, {Wetherill}, \&
  {Boss}}]{chambers:1996}
{Chambers}, J.~E., {Wetherill}, G.~W., \& {Boss}, A.~P. 1996, \icarus, 119, 261

\bibitem[{{Charbonneau} {et~al.}(2009){Charbonneau}, {Berta}, {Irwin}, {Burke},
  {Nutzman}, {Buchhave}, {Lovis}, {Bonfils}, {Latham}, {Udry}, {Murray-Clay},
  {Holman}, {Falco}, {Winn}, {Queloz}, {Pepe}, {Mayor}, {Delfosse}, \&
  {Forveille}}]{Charbonneau09}
{Charbonneau}, D., {Berta}, Z.~K., {Irwin}, J., {et~al.} 2009, Nature, 462, 891

\bibitem[{Chen {et~al.}(2010)Chen, Schliep, Willows, Cai, Neilan, \&
  Scheer}]{Chen10}
Chen, M., Schliep, M., Willows, R.~D., {et~al.} 2010, Science, 329, 1318

\bibitem[{{Cockell} {et~al.}(2009){Cockell}, {L{\'e}ger}, {Fridlund}, {Herbst},
  {Kaltenegger}, {Absil}, {Beichman}, {Benz}, {Blanc}, {Brack}, {Chelli},
  {Colangeli}, {Cottin}, {Coud{\'e} du Foresto}, {Danchi}, {Defr{\`e}re}, {den
  Herder}, {Eiroa}, {Greaves}, {Henning}, {Johnston}, {Jones}, {Labadie},
  {Lammer}, {Launhardt}, {Lawson}, {Lay}, {LeDuigou}, {Liseau}, {Malbet},
  {Martin}, {Mawet}, {Mourard}, {Moutou}, {Mugnier}, {Ollivier}, {Paresce},
  {Quirrenbach}, {Rabbia}, {Raven}, {Rottgering}, {Rouan}, {Santos}, {Selsis},
  {Serabyn}, {Shibai}, {Tamura}, {Thi{\'e}baut}, {Westall}, \&
  {White}}]{darwin}
{Cockell}, C.~S., {L{\'e}ger}, A., {Fridlund}, M., {et~al.} 2009, Astrobiology,
  9, 1

\bibitem[{{Correia} {et~al.}(2008){Correia}, {Levrard}, \&
  {Laskar}}]{Correia08}
{Correia}, A.~C.~M., {Levrard}, B., \& {Laskar}, J. 2008, \aap, 488, L63

\bibitem[{{Darwin}(1880)}]{Darwin1880}
{Darwin}, G.~H. 1880, Royal Society of London Philosophical Transactions Series
  I, 171, 713

\bibitem[{{Dawson} \& {Fabrycky}(2010)}]{dawson:2010}
{Dawson}, R.~I. \& {Fabrycky}, D.~C. 2010, \apj, 722, 937

\bibitem[{{Delfosse} {et~al.}(2012){Delfosse}, {Bonfils}, {Forveille}, {Udry},
  {Mayor}, {Bouchy}, {Gillon}, {Lovis}, {Neves}, {Pepe}, {Perrier}, {Queloz},
  {Santos}, \& {S{\'e}gransan}}]{delfosse:2012}
{Delfosse}, X., {Bonfils}, X., {Forveille}, T., {et~al.} 2012, arXiv:1202.2467

\bibitem[{{Delfosse} {et~al.}(2000){Delfosse}, {Forveille}, {S{\'e}gransan},
  {Beuzit}, {Udry}, {Perrier}, \& {Mayor}}]{delfosse:2000}
{Delfosse}, X., {Forveille}, T., {S{\'e}gransan}, D., {et~al.} 2000, \aap, 364,
  217

\bibitem[{{Dole}(1964)}]{Dole64}
{Dole}, S.~H. 1964, {Habitable planets for man} ({Blaisdell Pub.~Co.})

\bibitem[{{Edmunds}(1978)}]{edmunds:1978}
{Edmunds}, M.~G. 1978, \aap, 64, 103

\bibitem[{{Edson} {et~al.}(2011){Edson}, {Lee}, {Bannon}, {Kasting}, \&
  {Pollard}}]{Edson11}
{Edson}, A., {Lee}, S., {Bannon}, P., {Kasting}, J.~F., \& {Pollard}, D. 2011,
  Icarus, 212, 1

\bibitem[{{Edson} {et~al.}(2012){Edson}, {Kasting}, {Pollard}, {Lee}, \&
  {Bannon}}]{Edson12}
{Edson}, A.~R., {Kasting}, J.~F., {Pollard}, D., {Lee}, S., \& {Bannon}, P.~R.
  2012, Astrobiology, 12, 562

\bibitem[{{Fabrycky} {et~al.}(2012){Fabrycky}, {Lissauer}, {Ragozzine}, {Rowe},
  {Agol}, {Barclay}, {Batalha}, {Borucki}, {Ciardi}, {Ford}, {Geary}, {Holman},
  {Jenkins}, {Li}, {Morehead}, {Shporer}, {Smith}, {Steffen}, \&
  {Still}}]{fabrycky2012}
{Fabrycky}, D.~C., {Lissauer}, J.~J., {Ragozzine}, D., {et~al.} 2012,
  arXiv:1202.6328

\bibitem[{{Ferraz-Mello} {et~al.}(2006){Ferraz-Mello}, {Micchtchenko}, \&
  {Beaug{\'e}}}]{ferraz-mello:2006}
{Ferraz-Mello}, S., {Micchtchenko}, T.~A., \& {Beaug{\'e}}, C. 2006, {Regular
  motions in extra-solar planetary systems} (Springer), 255

\bibitem[{{Ferraz-Mello} {et~al.}(2008){Ferraz-Mello}, {Rodr{\'{\i}}guez}, \&
  {Hussmann}}]{FerrazMello08}
{Ferraz-Mello}, S., {Rodr{\'{\i}}guez}, A., \& {Hussmann}, H. 2008, Celestial
  Mechanics and Dynamical Astronomy, 101, 171

\bibitem[{{Ford}(2005)}]{ford:2005}
{Ford}, E.~B. 2005, \aj, 129, 1706

\bibitem[{{Forveille} {et~al.}(2011){Forveille}, {Bonfils}, {Lo Curto},
  {Delfosse}, {Udry}, {Bouchy}, {Lovis}, {Mayor}, {Moutou}, {Naef}, {Pepe},
  {Perrier}, {Queloz}, \& {Santos}}]{gj676A}
{Forveille}, T., {Bonfils}, X., {Lo Curto}, G., {et~al.} 2011, \aap, 526, A141+

\bibitem[{{Geballe} {et~al.}(2002){Geballe}, {Knapp}, {Leggett}, {Fan},
  {Golimowski}, {Anderson}, {Brinkmann}, {Csabai}, {Gunn}, {Hawley},
  {Hennessy}, {Henry}, {Hill}, {Hindsley}, {Ivezi{\'c}}, {Lupton}, {McDaniel},
  {Munn}, {Narayanan}, {Peng}, {Pier}, {Rockosi}, {Schneider}, {Smith},
  {Strauss}, {Tsvetanov}, {Uomoto}, {York}, \& {Zheng}}]{Geballe:2002}
{Geballe}, T.~R., {Knapp}, G.~R., {Leggett}, S.~K., {et~al.} 2002, \apj, 564,
  466

\bibitem[{{Goldreich}(1966)}]{Goldreich66}
{Goldreich}, P. 1966, Astron.~J., 71, 1

\bibitem[{{Gregory}(2012)}]{gregory:2012}
{Gregory}, P.~C. 2012, arXiv:1212.4058

\bibitem[{{Haario} {et~al.}(2001){Haario}, {Saksman}, \&
  {Tamminen}}]{haario:2001}
{Haario}, H., {Saksman}, E., \& {Tamminen}, J. 2001, Bernouilli, 7, 223

\bibitem[{{Hansen} \& {Murray}(2013)}]{HansenMurray13}
{Hansen}, B. \& {Murray}, N. 2013, arXiv:1301.7431

\bibitem[{{Hansen} \& {Murray}(2012)}]{HansenMurray12}
{Hansen}, B.~M.~S. \& {Murray}, N. 2012, \apj, 751, 158

\bibitem[{{Hart}(1979)}]{Hart79}
{Hart}, M.~H. 1979, Icarus, 37, 351

\bibitem[{{Hawley} \& {Pettersen}(1991)}]{HawleyPettersen91}
{Hawley}, S.~L. \& {Pettersen}, B.~R. 1991, ApJ, 378, 725

\bibitem[{{Heller}(2012)}]{2012A&A...545L...8H}
{Heller}, R. 2012, \aap, 545, L8

\bibitem[{{Heller} \& {Barnes}(2013)}]{2013AsBio..13...18H}
{Heller}, R. \& {Barnes}, R. 2013, Astrobiology, 13, 18

\bibitem[{{Heller} {et~al.}(2011){Heller}, {Leconte}, \& {Barnes}}]{Heller11}
{Heller}, R., {Leconte}, J., \& {Barnes}, R. 2011, A\&A, 528, A27+

\bibitem[{{Husser} {et~al.}(2013){Husser}, {Wende-von Berg}, {Dreizler},
  {Homeier}, {Reiners}, {Barman}, \& {Hauschildt}}]{Husser:2013}
{Husser}, T.-O., {Wende-von Berg}, S., {Dreizler}, S., {et~al.} 2013, \aap,
  553, A6

\bibitem[{{Hut}(1981)}]{Hut81}
{Hut}, P. 1981, A\&A, 99, 126

\bibitem[{{Ikoma} {et~al.}(2001){Ikoma}, {Emori}, \& {Nakazawa}}]{Ikoma01}
{Ikoma}, M., {Emori}, H., \& {Nakazawa}, K. 2001, \apj, 553, 999

\bibitem[{{Jackson} {et~al.}(2008){Jackson}, {Barnes}, \&
  {Greenberg}}]{Jackson08_hab}
{Jackson}, B., {Barnes}, R., \& {Greenberg}, R. 2008, MNRAS, 391, 237

\bibitem[{{Jenkins} \& {Peacock}(2011)}]{jenkins:2011}
{Jenkins}, C.~R. \& {Peacock}, J.~A. 2011, \mnras, 413, 2895

\bibitem[{{Jenkins} {et~al.}(2013){Jenkins}, {Jones}, {Tuomi}, {Murgas},
  {Hoyer}, {Jones}, {Barnes}, {Pavlenko}, {Ivanyuk}, {Rojo}, {Jord{\'a}n},
  {Day-Jones}, {Ruiz}, \& {Pinfield}}]{jenkins:2012}
{Jenkins}, J.~S., {Jones}, H.~R.~A., {Tuomi}, M., {et~al.} 2013, \apj, 766, 67

\bibitem[{{Joshi}(2003)}]{Joshi03}
{Joshi}, M. 2003, Astrobiology, 3, 415

\bibitem[{{Joshi} \& {Haberle}(2012)}]{JoshiHaberle12}
{Joshi}, M.~M. \& {Haberle}, R.~M. 2012, Astrobiology, 12, 3

\bibitem[{{Joshi} {et~al.}(1997){Joshi}, {Haberle}, \& {Reynolds}}]{Joshi97}
{Joshi}, M.~M., {Haberle}, R.~M., \& {Reynolds}, R.~T. 1997, Icarus, 129, 450

\bibitem[{{Kaltenegger}(2000)}]{2000ESASP.462..199K}
{Kaltenegger}, L. 2000, in ESA Special Publication, Vol. 462, Exploration and
  Utilisation of the Moon, ed. B.~H. {Foing} \& M.~{Perry}, 199

\bibitem[{{Kasting} {et~al.}(1993){Kasting}, {Whitmire}, \&
  {Reynolds}}]{Kasting93}
{Kasting}, J.~F., {Whitmire}, D.~P., \& {Reynolds}, R.~T. 1993, Icarus, 101,
  108

\bibitem[{{Kiang} {et~al.}(2007){Kiang}, {Segura}, {Tinetti}, {Govindjee},
  {Blankenship}, {Cohen}, {Siefert}, {Crisp}, \& {Meadows}}]{Kiang07}
{Kiang}, N.~Y., {Segura}, A., {Tinetti}, G., {et~al.} 2007, Astrobiology, 7,
  252

\bibitem[{{Kipping}(2009)}]{2009MNRAS.392..181K}
{Kipping}, D.~M. 2009, \mnras, 392, 181

\bibitem[{{Kipping} {et~al.}(2012){Kipping}, {Bakos}, {Buchhave},
  {Nesvorn{\'y}}, \& {Schmitt}}]{2012ApJ...750..115K}
{Kipping}, D.~M., {Bakos}, G.~{\'A}., {Buchhave}, L., {Nesvorn{\'y}}, D., \&
  {Schmitt}, A. 2012, \apj, 750, 115

\bibitem[{{Kopparapu} {et~al.}(2013){Kopparapu}, {Ramirez}, {Kasting}, {Eymet},
  {Robinson}, {Mahadevan}, {Terrien}, {Domagal-Goldman}, {Meadows}, \&
  {Deshpande}}]{Kopparapu13}
{Kopparapu}, R.~K., {Ramirez}, R., {Kasting}, J.~F., {et~al.} 2013, \apj, 765,
  131

\bibitem[{{Lambeck}(1977)}]{Lambeck77}
{Lambeck}, K. 1977, Royal Society of London Philosophical Transactions Series
  A, 287, 545

\bibitem[{{Laskar}(1990)}]{laskar:1990}
{Laskar}, J. 1990, \icarus, 88, 266

\bibitem[{{Laskar}(1993)}]{laskar:1993}
{Laskar}, J. 1993, Celestial Mechanics and Dynamical Astronomy, 56, 191

\bibitem[{{Laskar} \& {Robutel}(2001)}]{laskar:2001}
{Laskar}, J. \& {Robutel}, P. 2001, Celestial Mechanics and Dynamical
  Astronomy, 80, 39

\bibitem[{{Laughlin} \& {Chambers}(2001)}]{LaughlinChambers01}
{Laughlin}, G. \& {Chambers}, J.~E. 2001, \apjl, 551, L109

\bibitem[{{Leconte} {et~al.}(2010){Leconte}, {Chabrier}, {Baraffe}, \&
  {Levrard}}]{Leconte10}
{Leconte}, J., {Chabrier}, G., {Baraffe}, I., \& {Levrard}, B. 2010, A\&A, 516,
  A64+

\bibitem[{{Lin} {et~al.}(1996){Lin}, {Bodenheimer}, \& {Richardson}}]{Lin96}
{Lin}, D.~N.~C., {Bodenheimer}, P., \& {Richardson}, D.~C. 1996, \nat, 380, 606

\bibitem[{{Lissauer}(2007)}]{Lissauer07}
{Lissauer}, J.~J. 2007, ApJ, 660, L149

\bibitem[{{Lissauer} {et~al.}(2009){Lissauer}, {Hubickyj}, {D'Angelo}, \&
  {Bodenheimer}}]{Lissauer09}
{Lissauer}, J.~J., {Hubickyj}, O., {D'Angelo}, G., \& {Bodenheimer}, P. 2009,
  \icarus, 199, 338

\bibitem[{{Lo Curto} {et~al.}(2010){Lo Curto}, {Mayor}, {Benz}, {Bouchy},
  {Lovis}, {Moutou}, {Naef}, {Pepe}, {Queloz}, {Santos}, {Segransan}, \&
  {Udry}}]{locurto:2010}
{Lo Curto}, G., {Mayor}, M., {Benz}, W., {et~al.} 2010, \aap, 512, A48

\bibitem[{{Lovis} {et~al.}(2011){Lovis}, {Dumusque}, {Santos}, {Bouchy},
  {Mayor}, {Pepe}, {Queloz}, {S{\'e}gransan}, \& {Udry}}]{lovis:2011}
{Lovis}, C., {Dumusque}, X., {Santos}, N.~C., {et~al.} 2011, arXiv:1107.5325

\bibitem[{{Lovis} \& {Pepe}(2007)}]{lovis:2007}
{Lovis}, C. \& {Pepe}, F. 2007, \aap, 468, 1115

\bibitem[{{Makarov} \& {Efroimsky}(2013)}]{makarov:2013}
{Makarov}, V.~V. \& {Efroimsky}, M. 2013, \apj, 764, 27

\bibitem[{{Mayor} {et~al.}(2009){Mayor}, {Bonfils}, {Forveille}, {Delfosse},
  {Udry}, {Bertaux}, {Beust}, {Bouchy}, {Lovis}, {Pepe}, {Perrier}, {Queloz},
  \& {Santos}}]{mayor:2009}
{Mayor}, M., {Bonfils}, X., {Forveille}, T., {et~al.} 2009, \aap, 507, 487

\bibitem[{{Mermilliod}(1986)}]{mermilliod:1986}
{Mermilliod}, J.-C. 1986, Catalogue of Eggen's UBV data., 0 (1986),

\bibitem[{Mielke {et~al.}(2013)Mielke, Kiang, Blankenship, \&
  Mauzerall}]{Mielke13}
Mielke, S.~P., Kiang, N.~Y., Blankenship, R.~E., \& Mauzerall, D. 2013, BBA -
  Bioenergetics, 1827, 255

\bibitem[{{Murray} \& {Dermott}(1999)}]{MurrayDermott99}
{Murray}, C.~D. \& {Dermott}, S.~F. 1999, {Solar system dynamics} (Cambridge
  University Press)

\bibitem[{{Neron de Surgy} \& {Laskar}(1997)}]{NeronDeSurgyLaskar97}
{Neron de Surgy}, O. \& {Laskar}, J. 1997, A\&A, 318, 975

\bibitem[{{Pepe} {et~al.}(2002){Pepe}, {Mayor}, {Galland}, \&
  et~al.}]{pepe:2002}
{Pepe}, F., {Mayor}, M., {Galland}, \& et~al. 2002, \aap, 388, 632

\bibitem[{{Pierrehumbert}(2011)}]{Pierrehumbert11}
{Pierrehumbert}, R.~T. 2011, ApJ, 726, L8+

\bibitem[{{Rauch} \& {Hamilton}(2002)}]{RauchHamilton02}
{Rauch}, K.~P. \& {Hamilton}, D.~P. 2002, in BAAS, Vol.~34, AAS/Division of
  Dynamical Astronomy Meeting \#33, 938

\bibitem[{{Raymond} {et~al.}(2007){Raymond}, {Scalo}, \& {Meadows}}]{Raymond07}
{Raymond}, S.~N., {Scalo}, J., \& {Meadows}, V.~S. 2007, ApJ, 669, 606

\bibitem[{{Reiners}(2005)}]{reiners:2005}
{Reiners}, A. 2005, Astronomische Nachrichten, 326, 930

\bibitem[{{Reiners} \& {Mohanty}(2012)}]{reiners_mohanty:2012}
{Reiners}, A. \& {Mohanty}, S. 2012, \apj, 746, 43

\bibitem[{{Reiners} {et~al.}(2013){Reiners}, {Shulyak}, {Anglada-Escud{\'e}},
  {Jeffers}, {Morin}, {Zechmeister}, {Kochukhov}, \& {Piskunov}}]{reiners:2013}
{Reiners}, A., {Shulyak}, D., {Anglada-Escud{\'e}}, G., {et~al.} 2013, \aap,
  552, A103

\bibitem[{{Reynolds} {et~al.}(1987){Reynolds}, {McKay}, \&
  {Kasting}}]{1987AdSpR...7..125R}
{Reynolds}, R.~T., {McKay}, C.~P., \& {Kasting}, J.~F. 1987, Advances in Space
  Research, 7, 125

\bibitem[{{Segura} {et~al.}(2003){Segura}, {Krelove}, {Kasting}, {Sommerlatt},
  {Meadows}, {Crisp}, {Cohen}, \& {Mlawer}}]{Segura03}
{Segura}, A., {Krelove}, K., {Kasting}, J.~F., {et~al.} 2003, Astrobiology, 3,
  689

\bibitem[{{Segura} {et~al.}(2010){Segura}, {Walkowicz}, {Meadows}, {Kasting},
  \& {Hawley}}]{Segura10}
{Segura}, A., {Walkowicz}, L.~M., {Meadows}, V., {Kasting}, J., \& {Hawley}, S.
  2010, Astrobiology, 10, 751

\bibitem[{{Selsis} {et~al.}(2007){Selsis}, {Kasting}, {Levrard}, \&
  et~al.}]{selsis:2007}
{Selsis}, F., {Kasting}, J.~F., {Levrard}, B., \& et~al. 2007, \aap, 476, 1373

\bibitem[{{Shields} {et~al.}(2013){Shields}, {Meadows}, {Bitz},
  {Pierrehumbert}, {Joshi}, \& {Robinson}}]{Shields13}
{Shields}, A.~L., {Meadows}, V.~S., {Bitz}, C.~M., {et~al.} 2013,
  arXiv:1305.6926

\bibitem[{{Skrutskie} {et~al.}(2006){Skrutskie}, {Cutri}, {Stiening},
  {Weinberg}, {Schneider}, {Carpenter}, {Beichman}, {Capps}, {Chester},
  {Elias}, {Huchra}, {Liebert}, {Lonsdale}, {Monet}, {Price}, {Seitzer},
  {Jarrett}, {Kirkpatrick}, {Gizis}, {Howard}, {Evans}, {Fowler}, {Fullmer},
  {Hurt}, {Light}, {Kopan}, {Marsh}, {McCallon}, {Tam}, {Van Dyk}, \&
  {Wheelock}}]{twomass}
{Skrutskie}, M.~F., {Cutri}, R.~M., {Stiening}, R., {et~al.} 2006, \aj, 131,
  1163

\bibitem[{{Spencer} {et~al.}(2000){Spencer}, {Rathbun}, {Travis}, {Tamppari},
  {Barnard}, {Martin}, \& {McEwen}}]{2000Sci...288.1198S}
{Spencer}, J.~R., {Rathbun}, J.~A., {Travis}, L.~D., {et~al.} 2000, Science,
  288, 1198

\bibitem[{{Tachinami} {et~al.}(2011){Tachinami}, {Senshu}, \&
  {Ida}}]{2011ApJ...726...70T}
{Tachinami}, C., {Senshu}, H., \& {Ida}, S. 2011, \apj, 726, 70

\bibitem[{{Tinney} {et~al.}(2011){Tinney}, {Wittenmyer}, {Butler}, {Jones},
  {O'Toole}, {Bailey}, {Carter}, \& {Horner}}]{tinney:2011}
{Tinney}, C.~G., {Wittenmyer}, R.~A., {Butler}, R.~P., {et~al.} 2011, \apj,
  732, 31

\bibitem[{{Tuomi}(2011)}]{tuomi:2011}
{Tuomi}, M. 2011, \aap, 528, L5

\bibitem[{{Tuomi}(2012)}]{tuomi:2012}
{Tuomi}, M. 2012, \aap, 543, A52

\bibitem[{{Tuomi} \& {Anglada-Escude}(2013)}]{tuomi:gj163}
{Tuomi}, M. \& {Anglada-Escude}, G. 2013, arXiv:1306.1717, A\&A accepted

\bibitem[{{Tuomi} {et~al.}(2013){Tuomi}, {Anglada-Escud{\'e}}, {Gerlach},
  {Jones}, {Reiners}, {Rivera}, {Vogt}, \& {Butler}}]{hd40307}
{Tuomi}, M., {Anglada-Escud{\'e}}, G., {Gerlach}, E., {et~al.} 2013, \aap, 549,
  A48

\bibitem[{{Tuomi} \& {Jenkins}(2012)}]{tuomi_jenkins:2012}
{Tuomi}, M. \& {Jenkins}, J.~S. 2012, arXiv:1211.1280

\bibitem[{{Tuomi} {et~al.}(2012){Tuomi}, {Jones}, {Jenkins}, {Tinney},
  {Butler}, {Vogt}, {Barnes}, {Wittenmyer}, {O'Toole}, {Horner}, {Bailey},
  {Carter}, {Wright}, {Salter}, \& {Pinfield}}]{tauceti}
{Tuomi}, M., {Jones}, H.~R.~A., {Jenkins}, J.~S., {et~al.} 2012,
  arXiv:1212.4277

\bibitem[{{Tuomi} {et~al.}(2011){Tuomi}, {Pinfield}, \& {Jones}}]{tuomi:2011b}
{Tuomi}, M., {Pinfield}, D., \& {Jones}, H.~R.~A. 2011, \aap, 532, A116

\bibitem[{{Udry} {et~al.}(2007){Udry}, {Bonfils}, {Delfosse}, {Forveille},
  {Mayor}, {Perrier}, {Bouchy}, {Lovis}, {Pepe}, {Queloz}, \&
  {Bertaux}}]{udry:2007}
{Udry}, S., {Bonfils}, X., {Delfosse}, X., {et~al.} 2007, \aap, 469, L43

\bibitem[{{van Leeuwen}(2007)}]{HIPPARCOS}
{van Leeuwen}, F. 2007, \aap, 474, 653

\bibitem[{{Vogt} {et~al.}(2010){Vogt}, {Butler}, {Rivera}, {Haghighipour},
  {Henry}, \& {Williamson}}]{vogt:2010}
{Vogt}, S.~S., {Butler}, R.~P., {Rivera}, E.~J., {et~al.} 2010, \apj, 723, 954

\bibitem[{{von Braun} {et~al.}(2011){von Braun}, {Boyajian}, {Kane}, {van
  Belle}, {Ciardi}, {L{\'o}pez-Morales}, {McAlister}, {Henry}, {Jao}, {Riedel},
  {Subasavage}, {Schaefer}, {ten Brummelaar}, {Ridgway}, {Sturmann},
  {Sturmann}, {Mazingue}, {Turner}, {Farrington}, {Goldfinger}, \&
  {Boden}}]{vonbraun:2011}
{von Braun}, K., {Boyajian}, T.~S., {Kane}, S.~R., {et~al.} 2011, \apjl, 729,
  L26

\bibitem[{{Watson} {et~al.}(1981){Watson}, {Donahue}, \& {Walker}}]{Watson81}
{Watson}, A.~J., {Donahue}, T.~M., \& {Walker}, J.~C.~G. 1981, Icarus, 48, 150

\bibitem[{{West} {et~al.}(2008){West}, {Hawley}, {Bochanski}, {Covey}, {Reid},
  {Dhital}, {Hilton}, \& {Masuda}}]{west08}
{West}, A.~A., {Hawley}, S.~L., {Bochanski}, J.~J., {et~al.} 2008, Astron.~J.,
  135, 785

\bibitem[{{Williams} {et~al.}(1997){Williams}, {Kasting}, \&
  {Wade}}]{1997Natur.385..234W}
{Williams}, D.~M., {Kasting}, J.~F., \& {Wade}, R.~A. 1997, \nat, 385, 234

\bibitem[{{Wordsworth} {et~al.}(2011){Wordsworth}, {Forget}, {Selsis},
  {Millour}, {Charnay}, \& {Madeleine}}]{Wordsworth11}
{Wordsworth}, R.~D., {Forget}, F., {Selsis}, F., {et~al.} 2011, ApJ, 733, L48+

\bibitem[{{Yoder}(1995)}]{Yoder95}
{Yoder}, C.~F. 1995, in Global Earth Physics: A Handbook of Physical Constants,
  ed. {T.~J.~Ahrens}, 1--+

\bibitem[{{Zakamska} {et~al.}(2011){Zakamska}, {Pan}, \&
  {Ford}}]{zakamska:2011}
{Zakamska}, N.~L., {Pan}, M., \& {Ford}, E.~B. 2011, \mnras, 410, 1895

\bibitem[{{Zechmeister} {et~al.}(2013){Zechmeister}, {K{\"u}rster}, {Endl}, {Lo
  Curto}, {Hartman}, {Nilsson}, {Henning}, {Hatzes}, \&
  {Cochran}}]{zechmeister:2013}
{Zechmeister}, M., {K{\"u}rster}, M., {Endl}, M., {et~al.} 2013, \aap, 552, A78

\end{thebibliography}

\clearpage

\newpage
\appendix

\section{Priors} \label{sec:prior}

The choice of uninformative and adequate priors plays
a central role in Bayesian statistics. More classic
methods, such as weighted least-squares solvers, can
be derived from Bayes theorem by assuming uniform
prior distributions for all free parameters. Using the
definition of Eq.~\ref{eq:posterior}, one can 
note that, under coordinate transformations in the
parameter space (e.g., change from $P$ to $P^{-1}$ as
the free parameter) the posterior probability
distribution will change its shape through the
Jacobian determinant of this transformation. This
means that the posterior distributions are
substantially different under changes of
parameterizations and, even in the case of
least-square statistics, one must be very aware of the
prior choices made (explicit, or implicit through the
choice of parameterization). This discussion is
addressed in more detail in \citet{tuomi:gj163}. For
the Doppler data of GJ~667C, our reference prior
choices are summarized in Table \ref{tab:priors}. The
basic rationale on each prior choice can also be found
in \citet{tuomi:2012}, \citet{gj676A:2012} and
\citet{tuomi:gj163}. The prior choice for the eccentricity can be decisive
in detection of weak signals. Our choice for this
prior ($\mathcal{N}(0, \sigma_{e}^{2})$) is justified
using statistical, dynamical and population arguments.

\begin{table}
\center

\caption{Reference prior probability densities and
ranges of the model parameters.\label{tab:priors}}

\begin{tabular}{lccc}
\hline\hline
Parameter & $\pi(\theta)$ & Interval & Hyper-parameter \\
          &               &          & values \\
\hline
$K$          & Uniform & $[0, K_{max}]$ & $K_{max}=5$ \ms\\
$\omega$     & Uniform & [0, $2 \pi$]   & -- \\
$e$          & $\propto \mathcal{N}(0, \sigma_{e}^{2})$
                       & [0,1] & $\sigma_{e}=0.3$ \\
$M_{0}$      & Uniform & [0, $2 \pi$] & --\\
$\sigma_{J}$ & Uniform & $[0, K_{max}]$ & (*) \\
$\gamma$     & Uniform & $[-K_{max}, K_{max}]$ & (*) \\
$\phi$       & Uniform & [-1, 1] \\
$\log P$     & Uniform & $[\log P_{0}, \log P_{max}]$ & $P_0 = 1.0$ days\\
             &         &                              & $P_{max}=3000$ days \\
\hline\hline
\end{tabular}
\tablefoot{
* Same $K_{max}$ as for the $K$ parameter in first row.
}
\end{table}

\subsection{Eccentricity prior : statistical argument} \label{sec:eccprior}

Our first argument is based on statistical
considerations to minimize the risk of false
positives. That is, since $e$ is a strongly non-linear
parameter in the Keplerian model of Doppler signals
(especially if $e>$0.5), the likelihood function has
many local maxima with high eccentricities. Although
such solutions might appear very significant, it can
be shown that, when the detected amplitudes approach
the uncertainty in the measurements, these
high-eccentricity solutions are mostly spurious.

\begin{figure*}
\center
\includegraphics[width=0.6\textwidth,clip]{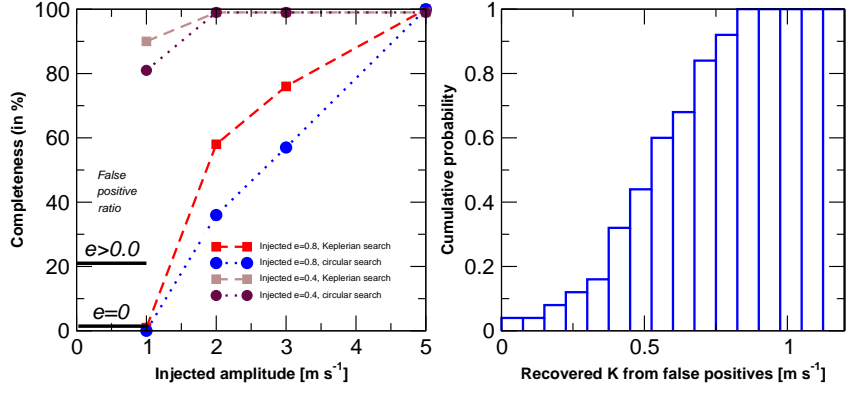}

\caption{Left panel. Detection completeness as a
function of the injected signal using a fully
Keplerian search versus a circular orbit search. Red
and blue lines are for an injected eccentricity of
0.8, and the brown and purple ones are for injected
signals with eccentricity of 0.4. Black horizontal
lines  on the left show the fraction of false-positive
detections satisfying the FAP threshold of 1\% using
both methods (Keplerian versus circular). While the
completeness is slightly enhanced in the low $K$
regime, the fraction of false positives is
unacceptable and, therefore, the implicit assumptions
of the method (e.g., uniform prior for $e$) are not
correct. \textbf{Right panel}. Distribution of
semi-amplitudes $K$ for these false positive detections.
Given that the injected noise level is 1.4 \ms (1 \ms
nominal uncertainty, 1 \ms jitter), it is clear that
signals detected with fully Keplerian log--L
periodograms with $K$ below the noise level cannot be
trusted.}\label{fig:completeness}

\end{figure*}

To illustrate this, we generate simulated sets of
observations (same observing epochs, no signal, Gaussian
white noise injected, 1\ms jitter level), and search for the
maximum likelihood solution using the log--L periodograms
approach (assuming a fully Keplerian solution at the period
search level, see Section \ref{sec:periodograms}). Although
no signal is injected, solutions with a formal false-alarm
probability (FAP) smaller than 1\% are found in 20\% of the
sample. On the contrary, our log--L periodogram search for
circular orbits found 1.2\% of false positives, matching the
expectations given the imposed $1\%$ threshold. We performed
an additional test to assess the impact of the eccentricity
prior on the detection completeness. That is, we injected one
Keplerian signal ($e=0.8$) at the same observing epochs with
amplitudes of 1.0 \ms and white Gaussian noise of 1 \ms. We
then performed the log--L periodogram search on a large
number of these datasets (10$^3$). When the search model was
allowed to be a fully Keplerian orbit, the correct solution
was only recovered 2.5\% of the time, and no signals at the
right period were spotted assuming a  circular orbit. With a
$K=2.0$ \ms, the situation improved and $60\%$ of the orbits
were identified in the full Keplerian case, against $40\%$ of
them in the purely circular one. More tests are illustrated
in the left panel of Fig. \ref{fig:completeness}. When an
eccentricity of 0.4 and K=1 \ms signal was injected, the
completeness of the fully Keplerian search increased to 91\%
and the completeness of the circular orbit search increased
to 80\%. When a $K>2$ \ms signal was injected, the
orbits were identified in $>99\%$ of the cases. We also
obtained a histogram of the semi-amplitudes of the false
positive detections obtained when no signal was injected. The
histogram shows that these amplitudes were smaller than 1.0
\ms with a 99\% confidence level (see right panel of
Fig.~\ref{fig:completeness}). This illustrates that
statistical tests based on point estimates below the noise
level do not produce reliable assessments on the significance
of a fully Keplerian signal. For a given dataset, information
based on simulations (e.g., Fig. A.1) or a physically
motivated prior is necessary to correct such detection bias
\citep{zakamska:2011}.

In summary, while a uniform prior on eccentricity only looses a few very
eccentric solutions in the low amplitude regime, the rate of false
positive detections ($\sim$ 20\%) is unacceptable. On the other hand, only a small fraction of
highly eccentric orbits are \textit{missed} if the search is
done assuming strictly circular orbits. This implies that,
for log--likelihood periodogram searches, circular orbits
should always be assumed when searching for a new
low-amplitude signals and that, when approaching amplitudes
comparable to the measurement uncertainties, assuming
circular orbits is a reasonable strategy to avoid false
positives. In a Bayesian sense, this means that we have to be
very skeptic of highly eccentric orbits when searching for
signals close to the noise level. The natural way to address
this self-consistently is by imposing a prior on the
eccentricity that suppresses the likelihood of highly
eccentric orbits. The log--Likelihood periodograms indicate
that the strongest possible prior (force $e=0$), already does a
very good job so, in general, any function less informative
than a delta function ($\pi(e) = \delta(0)$) and a bit more
constraining than a uniform prior ($\pi(e)=1$) can provide a
more optimal compromise between sensitivity and robustness.
Our particular functional choice of the prior (Gaussian
distribution with zero-mean and $\sigma=0.3$) is based on
dynamical and population analysis considerations.

\subsection{Eccentricity prior : dynamical argument}

From a physical point of view, we expect \textit{a priori}
that eccentricities closer to zero are more probable than
those close to unity when multiple planets are involved.
That is, when one or two planets are clearly present (e.g.
GJ~667Cb and GJ~667Cc are solidly detected even with a flat
prior in $e$), high eccentricities in the remaining lower
amplitude candidates would correspond to unstable and
therefore physically impossible systems.

Our prior for $e$ takes this feature into account
(reducing the likelihood of highly eccentric solutions)
but still allows high eccentricities if the data insists
so \citep{tuomi:2012}. At the sampling level, the only
orbital configurations that we explicitly forbid is that
we do not allow solutions with orbital crossings. While a
rather weak limitation, this requirement essentially
removes all extremely eccentric multiplanet solutions
($e>$0.8) from our MCMC samples. This requirement does
not, for example, remove orbital configurations with
close encounters between planets, and the solutions we
receive still have to be analyzed by numerical
integration to make sure that they correspond to stable
systems. As shown in Section \ref{sec:dynamics},
posterior numerical integration of the samplings
illustrate that our prior function was, after all, rather
conservative.

\subsection{Eccentricity prior : population argument}

To investigate how realistic our prior choice is compared
to the statistical properties of the known exoplanet
populations, we obtained the parameters of all planet
candidates with $M \sin i$ smaller than 0.1 M$_{jup}$ as
listed in The Extrasolar Planets Encyclopaedia
\footnote{\texttt{http:\\exoplanet.eu}} as at 2012 December 1.
We then produced a histogram in eccentricity bins of
$0.1$. The obtained distribution follows very nicely a
Gaussian function with zero mean and $\sigma_e = 0.2$,
meaning that our prior choice is more uninformative (and
therefore, more conservative) than the current
distribution of detections. This issue is the central
topic of \citet{tuomi:gj163}, and a more detailed
discussion (plus some illustrative plots) can be found in
there.

\section{Detailed Bayesian detection sequences} \label{sec:detailed}

In this section, we provide a more detailed description of
detections of seven signals in the combined HARPS-TERRA, PFS,
and HIRES data. We also show that the same seven signals
(with some small differences due to aliases) are also
detected independently when using HARPS-CCF
velocities instead of HARPS TERRA ones. The PFS and HIRES
measurements used are again those provided in
\citet{anglada:2012a}.

\subsection{HARPS-CCF, PFS and HIRES} \label{sec:ccf_analysis}

First, we perform a detailed analysis with the CCF values
provided by \citet{delfosse:2012} in combination with the
PFS and HIRES velocities. When increasing the number of
planets, parameter $k$, in our statistical model, we were
able to determine the relative probabilities of models
with $k=0, 1, 2, ...$ rather easily. The parameter spaces
of all models could be sampled with our Markov chains
relatively rapidly and the parameters of the signals
converged to the same periodicities and RV amplitudes
regardless of the exact initial states of the parameter
vectors.

Unlike in \citet{anglada:2012b} and \citet{delfosse:2012},
we observed immediately that $k=2$ was not the number of
signals favored by the data. While the obvious signals at
7.2  and 28.1 days were easy to detect using samplings of
the parameter space, we also quickly discovered a third
signal at a period of 91 days. These signals correspond to
the planets GJ~667C b, c, and d of \citet{anglada:2012b}
and \citet{delfosse:2012}, respectively, though the
orbital period of companion d was found to be 75 days by
\citet{anglada:2012b} and 106 days by
\citet{delfosse:2012}. The periodograms
also show considerable power at 106 days corresponding
to the solution of \citet{delfosse:2012}. Again, it did
not provide a solution as probable as the signal at
91 days and the inclusion of linear correlation terms
with activity did not affect its significance
(see also Sec.~\ref{sec:activity}).

\begin{table*}
\center

\caption{Relative posterior probabilities and log-Bayes
factors of models $\mathcal{M}_{k}$ with $k$ Keplerian
signals derived from the combined HARPS-CCF, HIRES, and PFS RV
data on the left and HARPS-TERRA HIRES, PFS on the right.
Factor $\Delta$ indicates how much the probability
increases with respect to the best model with one less
Keplerian signal. $P_{s}$ denotes the MAP period
estimate of the signal added to the solution when
increasing the number $k$. For $k=$4, 5, 6, and 7, we
denote all the signals on top of the three most
significant ones at 7.2, 28, and 91 days because the 53
and 62 day periods are each other's yearly aliases and
the relative significance of these two and the signal
with a period of 39 days are rather
similar.\label{tab:total_probabilities}}

\begin{tabular}{l|cccc|cccc|}
\hline \hline
 & \multicolumn{4}{|c|}{HARPS-CCF, PFS, HIRES}  &\multicolumn{4}{|c|}{HARPS-TERRA, PFS, HIRES}\\
\hline
$k$ & $P(\mathcal{M}_{k} | d)$ & $\Delta$ & $\log P(d | \mathcal{M}_{k})$ & $P_{s}$ [days] & $P(\mathcal{M}_{k} | d)$ & $\Delta$ & $\log P(d | \mathcal{M}_{k})$ & $P_{s}$ [days] \\
\hline
0 & 2.2$\times10^{-74}$ & -- & -629.1                 & --               & 2.7$\times10^{-85}$ & --		    & -602.1 & --                \\
1 & 2.4$\times10^{-40}$ & 1.1$\times10^{34}$ & -550.0 & 7.2              & 3.4$\times10^{-48}$ & 1.3$\times10^{37}$ & -516.0 & 7.2               \\
2 & 1.3$\times10^{-30}$ & 5.6$\times10^{9}$ & -526.9  & 28               & 1.3$\times10^{-35}$ & 3.9$\times10^{12}$ & -486.3 & 91                \\
3 & 8.7$\times10^{-21}$ & 6.5$\times10^{9}$ & -503.6  & 91               & 8.9$\times10^{-18}$ & 6.7$\times10^{17}$ & -444.5 & 28                \\
4 & 5.1$\times10^{-17}$ & 5.9$\times10^{3}$ & -494.2  & 39               & 1.5$\times10^{-14}$ & 1.7$\times10^{3}$  & -436.4 & 39                \\
4 & 1.0$\times10^{-14}$ & 1.2$\times10^{6}$ & -488.9 & 53               & 1.9$\times10^{-14}$ & 2.1$\times10^{3}$  & -436.2 & 53                \\
4 & 2.0$\times10^{-17}$ & 2.3$\times10^{3}$ & -495.2 & 62               & 1.2$\times10^{-14}$ & 1.3$\times10^{3}$  & -436.7 & 62                \\
5 & 8.0$\times10^{-9}$ & 7.6$\times10^{5}$ & -474.7   & 39, 53           & 1.0$\times10^{-7}$  & 5.5$\times10^{6}$  & -420.0 & 39, 53            \\
5 & 5.4$\times10^{-12}$ & 5.2$\times10^{2}$ & -482.0 & 39, 62           & 1.0$\times10^{-8}$  & 5.3$\times10^{5}$  & -422.3 & 39, 62            \\
6 & 3.4$\times10^{-4}$ & 4.3$\times10^{4}$ & -463.3   & 39, 53, 256      & 4.1$\times10^{-3}$  & 4.0$\times10^{4}$  & -408.7 & 39, 53, 256       \\
6 & 1.3$\times10^{-7}$ & 16 & -471.2                 & 39, 62, 256      & 4.1$\times10^{-4}$  & 4.0$\times10^{3}$  & -411.0 & 39, 62, 256       \\
7 & 0.998 & 2.9$\times10^{3}$ & -454.6                & 17, 39, 53, 256  & 0.057	       & 14		    & -405.4 & 17, 39, 53, 256   \\
7 & 1.5$\times10^{-3}$ & 4.3 & -461.2                & 17, 39, 62, 256  & 0.939	       & 230		    & -402.6 & 17, 39, 62, 256   \\
\hline \hline
\end{tabular}
\end{table*}

\begin{table*}
\center

\caption{Seven-Keplerian solution of the combined RVs of
GJ 667C with HARPS-CCF data. MAP estimates of the
parameters and their 99\% BCSs. The corresponding solution
derived from HARPS-TERRA data is given in Table
\ref{tab:parameters}. Note that each dataset prefers a
different alias for planet f (53 versus 62
days).}\label{tab:CCF_parameters}

\begin{tabular}{lcccc}
\hline \hline
Parameter & b & h & c & f \\
\hline
$P$ [days] & 7.1998 [7.1977, 7.2015] & 16.955 [16.903, 17.011] & 28.147 [28.084, 28.204] & 39.083 [38.892, 39.293] \\
$e$ & 0.10 [0, 0.25] & 0.16 [0, 0.39] & 0.02 [0, 0.20] & 0.03 [0, 0.18] \\
$K$ [ms$^{-1}$] & 3.90 [3.39, 4.37] & 0.80 [0.20, 1.34] & 1.60 [1.09, 2.17] & 1.31 [0.78, 1.85] \\
$\omega$ [rad] & 0.2 [0, 2$\pi$] & 2.3 [0, 2$\pi$] & 2.3 [0, 2$\pi$] & 3.6 [0, 2$\pi$] \\
$M_{0}$ [rad] & 3.2 [0, 2$\pi$] & 6.0 [0, 2$\pi$] & 2.9 [0, 2$\pi$] & 2.8 [0, 2$\pi$] \\
\hline
& e & d & g \\
\hline
$P$ [days] & 53.19 [52.73, 53.64] & 91.45 [90.81, 92.23] & 256.4 [248.6, 265.8] \\
$e$ & 0.13 [0, 0.19] & 0.12 [0, 0.29] & 0.18 [0, 0.49] \\
$K$ [ms$^{-1}$] & 0.96 [0.48, 1.49] & 1.56 [1.11, 2.06] & 0.97 [0.41, 1.53] \\
$\omega$ [rad] & 0.8 [0, 2$\pi$] & 3.0 [0, 2$\pi$] & 6.2 [0, 2$\pi$] \\
$M_{0}$ [rad] & 5.9 [0, 2$\pi$] & 5.4 [0, 2$\pi$] & 1.0 [0, 2$\pi$] \\
\hline
$\gamma_{1}$ [ms$^{-1}$] (HARPS) & -32.6 [-37.3, -28.2] \\
$\gamma_{2}$ [ms$^{-1}$] (HIRES) & -33.3 [-38.9,, -28.2] \\
$\gamma_{3}$ [ms$^{-1}$] (PFS) & -27.7 [-31.0, -24.0] \\
$\dot{\gamma}$ [ms$^{-1}$a$^{-1}$] & 2.19 [1.90, 2.48] \\
$\sigma_{J,1}$ [ms$^{-1}$] (HARPS) & 0.80 [0.20, 1.29] \\
$\sigma_{J,2}$ [ms$^{-1}$] (HIRES) & 2.08 [0.54, 4.15] \\
$\sigma_{J,3}$ [ms$^{-1}$] (PFS) & 1.96 [0.00, 4.96] \\
\hline \hline
\end{tabular}
\end{table*}

We could identify three more signals in the data at
39, 53, and 260 days with low RV amplitudes of 1.31,
0.96, and 0.97 ms$^{-1}$, respectively.
The 53-day signal had a strong alias at 62 days
and so we treated these as alternative models and calculated
their probabilities as well. The inclusion
of $k=6$ and $k=7$ signals at 260 and 17 days
improved the models irrespective of the preferred
choice of the period of planet e (see Table
\ref{tab:total_probabilities}). To assess the robustness
of the detection of the 7-th signal, we started
alternative chains at random periods. All the
chains that show good signs of convergence (bound
period) unequivocally suggested 17 days for the
last candidate. Since all these signals satisfied
our Bayesian detection criteria, we denoted 
all of them to planet candidates.

We performed samplings of the parameter space of the
model with $k=8$ but could not spot a clear 8-th
periodicity. The period of
such 8th signal converged to a broad probability
maximum between 1200 and 2000 days but the
corresponding RV amplitude received a posterior
density that did not differ significantly from zero.
The probability of the model with $k=8$
only exceeded the probability of $k=7$ by a factor of 60.

We therefore conclude that the combined data
set with HARPS-CCF RVs was in favor of $k=7$. The
corresponding orbital parameters of these seven
Keplerian signals are shown in Table
\ref{tab:CCF_parameters}. Whether there is a weak
signal at roughly 1200-2000 days or not remains to be
seen when future data become available. The naming of
the seven candidate planets (b to h) follows the
significance of detection.

\subsection{HARPS-TERRA, PFS and HIRES (reference
solution)}\label{sec:TERRA_analysis}

The HARPS-TERRA velocities combined with PFS and
HIRES velocities contained the signals of GJ~667C b, c, and d
very clearly. We could extract these signals from the data
with ease and obtained estimates for orbital periods that
were consistent with the estimates obtained using the CCF
data in the previous subsection. Unlike for the HARPS-CCF
data, however, the 91 day signal was more significantly
present in the HARPS-TERRA data and it corresponded to the
second most significant periodicity instead of the 28 day
one. Also, increasing $k$ improved the statistical model
significantly and we could again detect all the additional
signals in the RVs.

As for the CCF data, we branched the best fit solution
into the two preferred periods for the planet e
(53/62 days). The solution and model probabilities
are listed on the right-side of Table \ref{tab:total_probabilities}.
The only significant
difference compared to the HARPS-CCF one is that the
62-day period for
planet e is now preferred. Still, the model with 53 days
is only seventeen times less probable than the one with a 62
days signal, so we cannot confidently rule out that
the 53 days one is the real one. For simplicity and to
avoid confusion, we use the 62-day signal in our
reference solution and is the one used to analyze
dynamical stability and habitability for the system.
As an additional check, preliminary dynamical analysis
of solutions with a period of 53 days for planet e
showed very similar behaviour to the reference
solution in terms of long-term stability (similar
fraction of dynamically stable orbits and similar
distribution for the $D$ chaos indices).
Finally, we made several efforts at sampling the
eight-Keplerian model with different initial states.
As for the CCF data, there are hints of a signal
in the $\sim 2000$ day period domain that could not
be constrained.

\section{Further activity-related tests}
\label{sec:further}

\subsection{Correlated noise}\label{sec:rednoise}

The possible effect of correlated noise must
be investigated together with the significance of the
reported detections \citep[e.g. GJ
581;][]{baluev:2012,tuomi_jenkins:2012}. We studied
the impact of correlated noise by adding a first order
Moving Average term \citep[MA(1), see ][]{tauceti} to
the model in Eq.~\ref{eq:model} and repeated the
Bayesian search for all the signals. The MA(1) model
proposed in \citep{tauceti} contains two free
parameters: a characteristic time-scale $\tau$ and a
correlation coefficient $c$. Even for the HARPS data
set with the greatest number of measurements, the
characteristic time-scale could not be constrained.
Similarly, the correlation coefficient \citep[see
e.g.][]{hd40307,tauceti} had a posteriori density that
was not different from its prior, i.e. it was found to
be roughly uniform in the interval [-1,1], which is a
natural range for this parameter because it makes the
corresponding MA model stationary. While the $k=7$
searches lead to the same seven planet solution, models
containing such noise terms produced lower integrated
probabilities, which suggests over-parameterization.
When this happens, one should use the simplest model
(principle of parsimony) and accept that the noise is
better explained by the white noise components only.
Finally, very low levels of correlated noise are also
consistent with the good match between synthetic and
real periodograms of subsamples in Section
\ref{sec:subsamples}.

\subsection{Including activity correlation in the model}
\label{sec:correlation}

Because the HARPS activity-indices (S-index, FWHM, and
BIS) were available, we analyzed the data by taking
into account possible correlations with any of these
indices. We added an extra component into the
statistical model taking into account linear
correlation with each of these parameters as $F(t_i,
C) = C~ z_i$ (see Eq.~\ref{eq:model}), where $z_i$ is
the any of the three indices at epoch $t_i$.

\begin{figure}
\center
\includegraphics[width=0.3\textwidth,clip, angle=270]{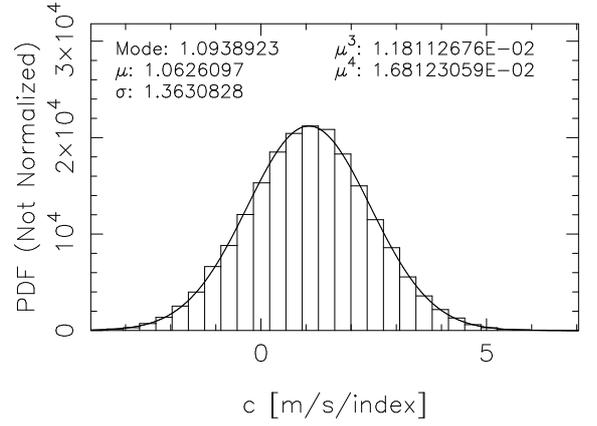}

\caption{Value of the linear correlation parameter of
the S-index ($C_S$) with the radial velocity data  for
a model with $k=3$ Keplerians (detection of planet d).}
\label{fig:S-index_corr}

\end{figure}

In Section \ref{sec:activity}, we found that the log--L of
the solution for planet d slightly improved when adding
the correlation term. The slight improvement of the
likelihood is compatible with a consistently positive
estimate of $C$ for both FWHM and the S-index obtained
with the MC samplings (see Fig. \ref{fig:S-index_corr},
for an example distribution of $C_{\rm S-index}$ as
obtained with a $k=3$ model). While the correlation terms
were observed to be mostly positive in all cases, the $0$
value for the coefficient was always within the 95\%
credibility interval. Moreover, we found that the
integrated model probabilities decreased when compared to
the model with the same number of planets but no
correlation term included. This means that such models are
over-parameterized and, therefore, they are penalized by
the principle of parsimony.


\subsection{Wavelength dependence of the signals}

Stellar activity might cause spurious Doppler
variability that is wavelength dependent (e.g., cool
spots). Using the HARPS-TERRA software on the HARPS
spectra only, we extracted one time-series for each
echelle aperture (72 of them). This process results
is $N_{\rm obs} = 72 \times 173 = 12456$ almost
independent RV measurements with corresponding
uncertainties. Except for calibration related
systematic effects (unknown at this level of
precision), each echelle aperture can be treated as
an independent instrument. Therefore, the reference
velocities $\gamma_l$ and jitter terms $\sigma_l$ of
each aperture were considered as free parameters. To
assess the wavelength dependence of each signal, the
Keplerian model of the i-th planet (one planet at a
time) also contained one semi-amplitude $K_{i,l}$ per
echelle aperture and all the other parameters ($e_i$,
$\omega_i$, $M_{0,i}$ and $P_{i}$) were common to all
apertures. The resulting statistical model has $72
\times 3 + 5 \times k - 1$ parameters when $k$
Keplerian signals are included and one is investigated
(250 free parameters for $k=7$). We started the
posterior samplings in the vicinity of the solutions
of interest because searching the period space would
have been computationally too demanding. Neglecting
the searches for periodicities, we could obtain
relatively well converged samples from the posterior
densities in a reasonable time-scale (few days of
computer time).

The result is the semi-amplitude $K$ of each signal
measured as a function of wavelength. Plotting this $K$
against central wavelength of each echelle order
enabled us to visually assess if signals had strong
wavelength dependencies (i.e. whether there were
signals only present in a few echelle orders) and
check if the obtained solution was consistent with the
one derived from the combined Doppler measurements. By
visual inspection and applying basic statistical
tests, we observed that the scatter in the amplitudes
for all the candidates was consistent within the error
bars and no significant systematic trend (e.g.
increasing $K$ towards the blue or the red) was found in
any case. Also, the weighted means of the derived
amplitudes were fully consistent with the values in
Table \ref{tab:parameters}. We are developing a more
quantitative version of these tests by studying reported
activity-induced signals on a larger sample of stars.
As examples of low--amplitude wavelength-dependent
signals ruled out using similar tests in the HARPS
wavelength range see : \citet{hd40307} on HD 40307
(K3V), \citet{anglada:2012a} on HD 69830 (G8V) and
\citet{reiners:2013} on the very active M dwarf AD Leo
(M3V).

\longtab{2}{
\begin{longtable}{lcccccccc}
\caption{HARPS-TERRA Doppler measurements of GJ~667C.
Measurements are in the barycenter of the Solar System
and corrected for perspective acceleration. The median
has been velocity has been subtracted for cosmetic
purposes. Nominal uncertainties in FWHM
and BIS are 2.0 and 2.35 times the corresponding $\sigma_{\rm ccf}$
\citep[see Section 4.5 in ][]{zechmeister:2013}. No consistent
CCF measurement could be obtained for the last two spectra
because of conflicting HARPS-DRS software versions (different
binary masks) used to produce them. Except for
those two spectra and according to the ESO archive documentation, all
CCF measurements were generated using the default M2 binary mask.
}\label{tab:rvs}
\\
\hline\hline
BJD     & RV$_{TERRA}$ & $\sigma_{\rm TERRA}$ & RV$_{\rm ccf}$ & $\sigma_{\rm ccf}$ & FWHM  & BIS  & S-index & $\sigma_S$\\
(days)  & (\ms)        & (\ms)                & (\ms)          & (\ms)              & (k\ms)& (\ms)& (--)    & (--)      \\ \hline
\endfirsthead
\caption{continued.}\\
\hline\hline
BJD     & RV$_{TERRA}$ & $\sigma_{\rm TERRA}$ & RV$_{\rm ccf}$ & $\sigma_{\rm ccf}$ & FWHM  & BIS  & S-index & $\sigma_S$\\
(days)  & (\ms)        & (\ms)                & (\ms)          & (\ms)              & (k\ms)& (\ms)& (--)    & (--)      \\ \hline
\endhead
\hline\hline
\endfoot
 2453158.764366  &    -3.10  &     0.95   & -3.11  & 1.05 & 3.0514 & -7.93  & 0.4667 & 0.0095 \\
 2453201.586794  &   -11.88  &     1.25   & -11.8  & 1.09 & 3.0666 & -9.61  & 0.4119 & 0.0074 \\
 2453511.798846  &    -7.61  &     0.89   & -9.22  & 1.07 & 3.0742 & -7.42  & 0.5915 & 0.0088 \\
 2453520.781048  &    -3.92  &     1.17   & -0.37  & 1.23 & 3.0701 & -11.99 & 0.4547 & 0.0082 \\
 2453783.863348  &     0.25  &     0.61   & 0.34   & 0.65 & 3.0743 & -13.31 & 0.4245 & 0.0053 \\
 2453810.852282  &    -3.48  &     0.55   & -3.00  & 0.54 & 3.0689 & -10.62 & 0.4233 & 0.0044 \\
 2453811.891816  &     2.20  &     1.08   & 0.24   & 1.02 & 3.0700 & -9.37  & 0.4221 & 0.0066 \\
 2453812.865858  &    -0.34  &     0.71   & -0.56  & 0.72 & 3.0716 & -9.78  & 0.4125 & 0.0054 \\
 2453814.849082  &   -10.16  &     0.49   & -10.06 & 0.47 & 3.0697 & -10.63 & 0.4848 & 0.0042 \\
 2453816.857459  &    -9.15  &     0.52   & -9.89  & 0.65 & 3.0698 & -12.20 & 0.4205 & 0.0051 \\
 2453830.860468  &    -6.96  &     0.56   & -7.29  & 0.59 & 3.0694 & -11.59 & 0.4729 & 0.0052 \\
 2453832.903068  &    -0.49  &     0.64   & -0.35  & 0.68 & 3.0706 & -13.33 & 0.4930 & 0.0058 \\
 2453834.884977  &    -1.50  &     0.72   & -1.68  & 0.57 & 3.0734 & -8.20  & 0.4456 & 0.0049 \\
 2453836.887788  &    -6.99  &     0.48   & -6.24  & 0.48 & 3.0723 & -8.27  & 0.4864 & 0.0044 \\
 2453861.796371  &     6.38  &     0.59   & 7.84   & 0.59 & 3.0780 & -11.47 & 0.6347 & 0.0060 \\
 2453862.772051  &     6.69  &     0.76   & 8.00   & 0.74 & 3.0768 & -12.54 & 0.5534 & 0.0065 \\
 2453863.797178  &     4.57  &     0.59   & 4.58   & 0.56 & 3.0759 & -10.71 & 0.4891 & 0.0051 \\
 2453864.753954  &     1.21  &     0.68   & 2.52   & 0.65 & 3.0783 & -9.21  & 0.4854 & 0.0055 \\
 2453865.785606  &    -1.85  &     0.61   & -2.55  & 0.55 & 3.0752 & -7.73  & 0.4815 & 0.0050 \\
 2453866.743120  &    -1.36  &     0.58   & -2.32  & 0.49 & 3.0770 & -7.49  & 0.5277 & 0.0045 \\
 2453867.835652  &    -0.48  &     0.66   & -0.05  & 0.64 & 3.0816 & -10.55 & 0.4708 & 0.0055 \\
 2453868.813512  &     2.34  &     0.56   & 0.62   & 0.61 & 3.0754 & -10.01 & 0.4641 & 0.0053 \\
 2453869.789495  &     3.85  &     0.63   & 4.73   & 0.65 & 3.0795 & -12.71 & 0.4837 & 0.0055 \\
 2453870.810097  &     2.37  &     0.88   & 2.82   & 0.81 & 3.0813 & -10.48 & 0.4567 & 0.0062 \\
 2453871.815952  &    -1.11  &     0.61   & -3.03  & 0.81 & 3.0790 & -9.16  & 0.5244 & 0.0068 \\
 2453882.732970  &    -2.96  &     0.52   & -4.17  & 0.51 & 3.0795 & -8.09  & 0.5121 & 0.0047 \\
 2453886.703550  &    -4.54  &     0.58   & -3.78  & 0.48 & 3.0757 & -10.11 & 0.4607 & 0.0042 \\
 2453887.773514  &    -5.97  &     0.48   & -3.98  & 0.44 & 3.0700 & -10.94 & 0.4490 & 0.0041 \\
 2453917.737524  &    -4.12  &     0.88   & -2.44  & 1.14 & 3.0666 & -10.91 & 0.5176 & 0.0084 \\
 2453919.712544  &     0.98  &     0.99   & 0.69   & 1.17 & 3.0774 & -8.01  & 0.4324 & 0.0073 \\
 2453921.615825  &    -1.67  &     0.49   & -1.24  & 0.51 & 3.0671 & -9.87  & 0.4305 & 0.0043 \\
 2453944.566259  &    -2.02  &     0.98   & -2.16  & 1.00 & 3.0776 & -9.25  & 0.6143 & 0.0079 \\
 2453947.578821  &     3.89  &     1.68   & 5.83   & 2.43 & 3.0806 & -8.54  & 0.7079 & 0.0134 \\
 2453950.601834  &    -1.01  &     0.89   & 1.65   & 0.92 & 3.0780 & -11.80 & 0.5612 & 0.0071 \\
 2453976.497106  &     2.40  &     0.61   & 3.52   & 0.60 & 3.0791 & -12.74 & 0.5365 & 0.0054 \\
 2453979.594316  &    -2.67  &     0.95   & -0.48  & 1.19 & 3.0776 & -9.20  & 0.5517 & 0.0091 \\
 2453981.555311  &    -4.77  &     0.64   & -4.29  & 0.57 & 3.0749 & -13.12 & 0.5339 & 0.0055 \\
 2453982.526504  &    -4.36  &     0.81   & -2.88  & 0.69 & 3.0717 & -11.84 & 0.4953 & 0.0061 \\
 2454167.866839  &    -1.87  &     0.62   & -2.51  & 0.61 & 3.0798 & -10.14 & 0.5141 & 0.0053 \\
 2454169.864835  &    -0.10  &     0.59   & -0.04  & 0.63 & 3.0793 & -11.94 & 0.4729 & 0.0052 \\
 2454171.876906  &     5.17  &     0.71   & 6.08   & 0.58 & 3.0744 & -7.24  & 0.4893 & 0.0050 \\
 2454173.856452  &    -1.18  &     0.83   & -1.44  & 0.61 & 3.0746 & -10.33 & 0.4809 & 0.0052 \\
 2454194.847290  &     1.27  &     0.59   & 0.85   & 0.69 & 3.0756 & -8.43  & 0.4586 & 0.0054 \\
 2454196.819157  &    -3.57  &     0.79   & -3.06  & 0.79 & 3.0759 & -12.33 & 0.4809 & 0.0061 \\
 2454197.797125  &    -3.83  &     0.86   & -4.71  & 0.97 & 3.0726 & -9.12  & 0.4584 & 0.0069 \\
 2454198.803823  &    -4.06  &     0.76   & -4.99  & 0.79 & 3.0708 & -9.33  & 0.5685 & 0.0068 \\
 2454199.854238  &     0.18  &     0.55   & 0.97   & 0.51 & 3.0714 & -10.66 & 0.4652 & 0.0044 \\
 2454200.815699  &     1.30  &     0.60   & 2.55   & 0.57 & 3.0708 & -10.26 & 0.4468 & 0.0047 \\
 2454201.918397  &     0.54  &     0.79   & 2.31   & 0.63 & 3.0681 & -11.27 & 0.4690 & 0.0056 \\
 2454202.802697  &    -2.96  &     0.69   & -3.23  & 0.66 & 3.0696 & -8.49  & 0.4954 & 0.0056 \\
 2454227.831743  &    -1.26  &     0.84   & 0.47   & 0.95 & 3.0619 & -9.96  & 0.4819 & 0.0071 \\
 2454228.805860  &     3.35  &     0.68   & 5.19   & 0.65 & 3.0651 & -15.03 & 0.4603 & 0.0055 \\
 2454229.773888  &     7.44  &     1.29   & 7.23   & 1.28 & 3.0708 & -6.34  & 0.5213 & 0.0082 \\
 2454230.845843  &     1.51  &     0.58   & 1.97   & 0.62 & 3.0631 & -8.92  & 0.4409 & 0.0053 \\
 2454231.801726  &    -0.57  &     0.62   & -1.15  & 0.55 & 3.0704 & -8.86  & 0.5993 & 0.0055 \\
 2454232.721251  &    -0.63  &     1.15   & -2.17  & 1.41 & 3.0719 & -9.70  & 0.3737 & 0.0079 \\
 2454233.910349  &    -1.27  &     1.29   & -2.10  & 1.68 & 3.0687 & -12.12 & 0.5629 & 0.0112 \\
 2454234.790981  &    -1.89  &     0.74   & -1.48  & 0.66 & 3.0672 & -8.39  & 1.2169 & 0.0093 \\
 2454253.728334  &     0.99  &     0.79   & 1.65   & 0.84 & 3.0773 & -10.30 & 0.4509 & 0.0062 \\
 2454254.755898  &    -2.64  &     0.54   & -3.25  & 0.52 & 3.0779 & -7.99  & 0.4426 & 0.0046 \\
 2454255.709350  &    -2.92  &     0.74   & -2.83  & 0.72 & 3.0775 & -7.36  & 0.4829 & 0.0059 \\
 2454256.697674  &    -0.21  &     0.97   & -0.45  & 0.84 & 3.0775 & -9.19  & 0.4608 & 0.0063 \\
 2454257.704446  &     2.93  &     0.66   & 2.39   & 0.70 & 3.0766 & -11.09 & 0.4549 & 0.0055 \\
 2454258.698322  &     4.19  &     0.83   & 5.19   & 0.63 & 3.0799 & -9.57  & 0.4760 & 0.0052 \\
 2454291.675565  &    -5.58  &     1.16   & -4.45  & 1.35 & 3.0802 & -9.95  & 0.4298 & 0.0086 \\
 2454292.655662  &    -4.37  &     0.75   & -1.25  & 0.76 & 3.0820 & -11.83 & 0.4487 & 0.0056 \\
 2454293.708786  &     0.89  &     0.63   & 2.84   & 0.59 & 3.0732 & -11.52 & 0.5344 & 0.0056 \\
 2454295.628628  &     3.05  &     0.92   & 3.67   & 1.03 & 3.0786 & -6.85  & 0.4975 & 0.0072 \\
 2454296.670395  &    -4.68  &     0.75   & -3.99  & 0.74 & 3.0703 & -7.79  & 0.5453 & 0.0067 \\
 2454297.631678  &    -5.53  &     0.63   & -4.81  & 0.55 & 3.0725 & -10.38 & 0.5212 & 0.0053 \\
 2454298.654206  &    -5.39  &     0.67   & -6.73  & 0.71 & 3.0743 & -5.18  & 0.5718 & 0.0066 \\
 2454299.678909  &    -1.46  &     0.85   & -2.26  & 0.92 & 3.0785 & -6.46  & 0.5299 & 0.0070 \\
 2454300.764649  &     0.14  &     0.74   & -0.07  & 0.63 & 3.0693 & -12.07 & 0.4803 & 0.0057 \\
 2454314.691809  &    -0.53  &     1.88   & -2.89  & 2.22 & 3.0756 & -12.48 & 0.3823 & 0.0102 \\
 2454315.637551  &     3.41  &     1.12   & 2.31   & 1.47 & 3.0701 & -10.42 & 0.4835 & 0.0091 \\
 2454316.554926  &     5.78  &     0.96   & 6.61   & 1.12 & 3.0746 & -6.02  & 0.4402 & 0.0069 \\
 2454319.604048  &    -6.64  &     0.79   & -7.01  & 0.59 & 3.0694 & -7.54  & 0.4643 & 0.0052 \\
 2454320.616852  &    -5.58  &     0.65   & -6.49  & 0.69 & 3.0698 & -3.94  & 0.4611 & 0.0057 \\
 2454340.596942  &    -1.52  &     0.60   & -0.55  & 0.55 & 3.0691 & -10.11 & 0.4480 & 0.0048 \\
 2454342.531820  &    -2.39  &     0.66   & -1.74  & 0.54 & 3.0667 & -9.95  & 0.4573 & 0.0048 \\
 2454343.530662  &     0.55  &     0.64   & 1.39   & 0.61 & 3.0669 & -7.25  & 0.4900 & 0.0055 \\
 2454346.551084  &    -0.17  &     1.01   & -0.82  & 1.14 & 3.0677 & -5.48  & 0.5628 & 0.0086 \\
 2454349.569500  &    -5.24  &     0.65   & -4.02  & 0.77 & 3.0658 & -11.12 & 0.3809 & 0.0058 \\
 2454522.886464  &    -1.68  &     0.70   & -1.11  & 0.61 & 3.0688 & -9.85  & 0.5582 & 0.0056 \\
 2454524.883089  &     4.38  &     0.69   & 3.05   & 0.69 & 3.0668 & -8.66  & 0.4779 & 0.0057 \\
 2454525.892144  &     1.96  &     0.72   & 0.69   & 0.58 & 3.0692 & -8.77  & 0.4202 & 0.0047 \\
 2454526.871196  &    -1.08  &     0.54   & 0.30   & 0.52 & 3.0717 & -9.58  & 0.4898 & 0.0046 \\
 2454527.897962  &    -2.69  &     0.64   & -3.31  & 0.65 & 3.0689 & -8.46  & 0.4406 & 0.0052 \\
 2454528.903672  &    -2.80  &     0.71   & -5.04  & 0.74 & 3.0679 & -8.40  & 0.4666 & 0.0058 \\
 2454529.869217  &     0.48  &     0.63   & -0.10  & 0.62 & 3.0664 & -8.23  & 0.4255 & 0.0050 \\
 2454530.878876  &     1.40  &     0.68   & 1.38   & 0.53 & 3.0667 & -7.31  & 0.4331 & 0.0044 \\
 2454550.901932  &    -6.95  &     0.70   & -6.46  & 0.58 & 3.0680 & -5.44  & 0.4330 & 0.0047 \\
 2454551.868783  &    -3.73  &     0.65   & -3.44  & 0.53 & 3.0654 & -7.36  & 0.4287 & 0.0045 \\
 2454552.880221  &     0.24  &     0.59   & -0.25  & 0.50 & 3.0665 & -9.12  & 0.4342 & 0.0042 \\
 2454554.846366  &     2.14  &     0.57   & 1.73   & 0.68 & 3.0699 & -5.49  & 0.4116 & 0.0052 \\
 2454555.870790  &    -2.84  &     0.58   & -2.26  & 0.58 & 3.0663 & -7.66  & 0.4704 & 0.0050 \\
 2454556.838936  &    -4.14  &     0.59   & -3.47  & 0.51 & 3.0686 & -11.20 & 0.4261 & 0.0043 \\
 2454557.804592  &    -4.56  &     0.66   & -4.37  & 0.60 & 3.0650 & -8.87  & 0.4306 & 0.0049 \\
 2454562.905075  &     0.67  &     0.70   & 0.80   & 0.57 & 3.0668 & -7.11  & 0.4709 & 0.0051 \\
 2454563.898808  &    -1.37  &     0.71   & -1.39  & 0.53 & 3.0656 & -11.93 & 0.4127 & 0.0046 \\
 2454564.895759  &    -2.63  &     0.85   & -2.82  & 0.71 & 3.0680 & -8.67  & 0.5068 & 0.0061 \\
 2454568.891702  &     3.27  &     0.87   & 4.85   & 1.02 & 3.0735 & -11.20 & 0.4682 & 0.0069 \\
 2454569.881078  &    -0.46  &     0.83   & 0.46   & 0.78 & 3.0720 & -16.02 & 0.4939 & 0.0061 \\
 2454570.870766  &    -1.70  &     0.72   & -1.21  & 0.88 & 3.0715 & -10.68 & 0.4606 & 0.0063 \\
 2454583.933324  &     0.44  &     1.00   & 1.56   & 1.11 & 3.0711 & -17.51 & 0.5177 & 0.0087 \\
 2454587.919825  &    -0.50  &     0.90   & -1.42  & 1.10 & 3.0824 & -7.86  & 0.4602 & 0.0078 \\
 2454588.909632  &     4.05  &     0.98   & 3.18   & 1.05 & 3.0828 & -6.95  & 0.5501 & 0.0080 \\
 2454590.901964  &     4.22  &     0.93   & 4.19   & 0.93 & 3.0758 & -9.05  & 0.4707 & 0.0073 \\
 2454591.900611  &     1.69  &     0.91   & -1.27  & 0.96 & 3.0753 & -7.39  & 0.5139 & 0.0075 \\
 2454592.897751  &    -2.50  &     0.68   & -2.50  & 0.63 & 3.0757 & -8.84  & 0.4741 & 0.0057 \\
 2454593.919961  &    -2.30  &     0.74   & -2.58  & 0.65 & 3.0680 & -12.41 & 0.5039 & 0.0063 \\
 2454610.878230  &     9.08  &     0.88   & 10.36  & 0.95 & 3.0671 & -9.46  & 0.4037 & 0.0069 \\
 2454611.856581  &     5.49  &     0.56   & 6.40   & 0.54 & 3.0650 & -8.37  & 0.4296 & 0.0050 \\
 2454616.841719  &     4.81  &     0.91   & 5.15   & 0.88 & 3.0713 & -8.09  & 0.3999 & 0.0065 \\
 2454617.806576  &     8.12  &     0.93   & 7.30   & 1.33 & 3.0753 & -14.38 & 0.4948 & 0.0086 \\
 2454618.664475  &    10.67  &     1.76   & 7.01   & 2.51 & 3.0854 & -7.21  & 0.6755 & 0.0135 \\
 2454639.867730  &     3.14  &     1.06   & 4.26   & 1.10 & 3.0588 & -8.27  & 0.4083 & 0.0083 \\
 2454640.723804  &     5.06  &     0.64   & 7.07   & 0.66 & 3.0705 & -13.61 & 0.4387 & 0.0055 \\
 2454642.676950  &    -0.81  &     0.47   & 1.56   & 0.61 & 3.0704 & -10.27 & 0.4720 & 0.0053 \\
 2454643.686130  &    -2.06  &     0.72   & -4.52  & 0.76 & 3.0709 & -9.26  & 0.4809 & 0.0064 \\
 2454644.732044  &    -1.19  &     0.46   & -1.85  & 0.56 & 3.0680 & -8.64  & 0.5097 & 0.0054 \\
 2454646.639658  &     5.74  &     1.11   & 5.01   & 0.95 & 3.0737 & -10.14 & 0.4316 & 0.0066 \\
 2454647.630210  &     5.37  &     0.68   & 3.28   & 0.72 & 3.0693 & -6.35  & 0.4938 & 0.0062 \\
 2454648.657090  &     2.58  &     0.92   & 0.96   & 0.94 & 3.0720 & -8.85  & 0.4597 & 0.0068 \\
 2454658.650838  &    -4.20  &     0.97   & -3.30  & 0.88 & 3.0714 & -13.06 & 0.4193 & 0.0065 \\
 2454660.650214  &    -0.82  &     1.13   & -0.40  & 1.06 & 3.0728 & -10.20 & 0.4224 & 0.0074 \\
 2454661.760056  &     1.72  &     0.73   & 1.76   & 0.84 & 3.0737 & -11.56 & 0.4238 & 0.0065 \\
 2454662.664144  &     3.30  &     0.72   & 2.58   & 0.97 & 3.0713 & -8.43  & 0.4675 & 0.0070 \\
 2454663.784376  &    -1.92  &     0.93   & -1.14  & 0.78 & 3.0643 & -11.52 & 0.3811 & 0.0061 \\
 2454664.766558  &    -1.00  &     1.51   & 0.0    & 1.85 & 3.0765 & -9.85  & 0.4702 & 0.0106 \\
 2454665.774513  &    -1.88  &     0.87   & -2.51  & 0.85 & 3.0695 & -9.89  & 0.4183 & 0.0065 \\
 2454666.683607  &    -0.37  &     0.87   & 0.36   & 0.79 & 3.0717 & -9.64  & 0.4098 & 0.0060 \\
 2454674.576462  &     4.82  &     1.01   & 6.41   & 1.39 & 3.0901 & -6.38  & 0.4226 & 0.0083 \\
 2454677.663487  &     7.37  &     1.78   & 8.63   & 3.11 & 3.1226 & -4.66  & 0.4452 & 0.0117 \\
 2454679.572671  &     2.94  &     1.26   & -1.23  & 1.48 & 3.0822 & -5.41  & 0.5622 & 0.0103 \\
 2454681.573996  &     2.51  &     0.89   & 2.86   & 1.10 & 3.0780 & -6.26  & 0.4443 & 0.0075 \\
 2454701.523392  &    -0.50  &     0.68   & -0.28  & 0.67 & 3.0719 & -4.48  & 0.5141 & 0.0058 \\
 2454708.564794  &    -0.12  &     0.86   & -0.67  & 0.79 & 3.0803 & -12.80 & 0.5160 & 0.0062 \\
 2454733.487290  &     8.06  &     3.51   & 10.75  & 3.89 & 3.0734 & -3.79  & 0.5017 & 0.0146 \\
 2454735.499425  &     0.00  &     1.04   & -2.22  & 1.19 & 3.0720 & -11.44 & 0.4337 & 0.0072 \\
 2454736.550865  &    -3.28  &     0.91   & -4.99  & 1.05 & 3.0671 & -8.70  & 0.4647 & 0.0075 \\
 2454746.485935  &    -4.49  &     0.58   & -5.00  & 0.53 & 3.0611 & -13.01 & 0.4259 & 0.0045 \\
 2454992.721062  &     6.84  &     0.79   & 7.80   & 0.65 & 3.0748 & -10.71 & 0.4826 & 0.0053 \\
 2455053.694541  &    -3.20  &     0.84   & -3.32  & 1.09 & 3.0741 & -11.73 & 0.4427 & 0.0078 \\
 2455276.882590  &     0.27  &     0.74   & 1.86   & 0.77 & 3.0732 & -11.03 & 0.4699 & 0.0061 \\
 2455278.827303  &     1.84  &     0.92   & 1.02   & 0.85 & 3.0760 & -8.41  & 0.5883 & 0.0074 \\
 2455280.854800  &     5.26  &     0.76   & 4.41   & 0.87 & 3.0793 & -12.50 & 0.4817 & 0.0065 \\
 2455283.868014  &    -0.69  &     0.68   & -0.09  & 0.61 & 3.0793 & -8.23  & 0.5411 & 0.0054 \\
 2455287.860052  &     4.99  &     0.72   & 5.42   & 0.64 & 3.0779 & -11.76 & 0.5366 & 0.0056 \\
 2455294.882720  &     8.56  &     0.69   & 6.81   & 0.58 & 3.0775 & -8.71  & 0.5201 & 0.0051 \\
 2455295.754277  &    10.15  &     1.06   & 8.21   & 0.98 & 3.0743 & -8.94  & 0.5805 & 0.0076 \\
 2455297.805750  &     4.95  &     0.64   & 3.95   & 0.70 & 3.0779 & -10.19 & 0.4614 & 0.0057 \\
 2455298.813775  &     2.52  &     0.75   & 2.95   & 0.67 & 3.0807 & -7.72  & 0.5828 & 0.0061 \\
 2455299.785905  &     3.74  &     1.60   & 4.62   & 2.19 & 3.0793 & -10.36 & 0.4187 & 0.0106 \\
 2455300.876852  &     5.07  &     0.60   & 5.78   & 0.75 & 3.0792 & -9.53  & 0.5104 & 0.0060 \\
 2455301.896438  &     9.54  &     0.99   & 8.40   & 1.32 & 3.0774 & -12.07 & 0.4395 & 0.0085 \\
 2455323.705436  &     8.56  &     0.86   & 8.82   & 0.78 & 3.0702 & -7.78  & 0.4349 & 0.0067 \\
 2455326.717047  &     2.17  &     1.05   & 0.67   & 1.27 & 3.0649 & -9.48  & 0.5955 & 0.0103 \\
 2455328.702599  &     1.56  &     1.02   & 1.83   & 1.01 & 3.0658 & -8.48  & 0.5077 & 0.0089 \\
 2455335.651717  &     1.01  &     0.92   & 3.02   & 1.22 & 3.0593 & -10.79 & 0.4685 & 0.0092 \\
 2455337.704618  &     7.58  &     1.03   & 6.13   & 1.24 & 3.0725 & -11.55 & 0.4859 & 0.0090 \\
 2455338.649293  &    13.01  &     1.97   & 12.32  & 2.54 & 3.0687 & -5.74  & 0.4969 & 0.0134 \\
 2455339.713716  &     6.70  &     1.03   & 5.13   & 1.57 & 3.0700 & -11.07 & 0.4760 & 0.0097 \\
 2455341.789626  &    -0.40  &     0.63   & 0.01   & 0.71 & 3.0812 & -11.27 & 0.4916 & 0.0061 \\
 2455342.720036  &     4.80  &     0.91   & 5.68   & 1.13 & 3.0718 & -7.25  & 0.4674 & 0.0076 \\
 2455349.682257  &     6.55  &     0.78   & 4.00   & 0.97 & 3.0685 & -4.609 & 0.4787 & 0.0073 \\
 2455352.601155  &    12.92  &     1.11   & 14.24  & 1.35 & 3.0700 & -7.48  & 0.4530 & 0.0093 \\
 2455354.642822  &     7.52  &     0.60   & 8.38   & 0.65 & 3.0663 & -10.63 & 0.4347 & 0.0057 \\
 2455355.576777  &     6.41  &     0.97   & 4.76   & 0.96 & 3.0681 & -9.41  & 0.4278 & 0.0074 \\
 2455358.754723  &     8.54  &     1.17   & 10.00  & 1.30 & 3.0619 & -11.04 & 0.3527 & 0.0095 \\
 2455359.599377  &     6.89  &     0.90   & 6.90   & 0.97 & 3.0724 & -11.39 & 0.3649 & 0.0070 \\
 2455993.879754  &    12.28  &     1.04   & --     & --   & --	   & --     & 0.5179 & 0.0086 \\
 2455994.848576  &    15.43  &     1.30   & --     & --   & --     & --     & 0.6712 & 0.0120
\end{longtable}
}

\end{document}